\documentclass[a4paper,11pt]{article}
\usepackage{jcappub}
\usepackage{lineno}
\usepackage{amsmath}	
\usepackage{bm}
\usepackage{color}

\usepackage{amssymb}
\usepackage{amsmath}
\usepackage{graphics}
\usepackage{graphicx}
\usepackage{hyperref}
\usepackage{siunitx}
\usepackage{booktabs}                                                                                                 

\usepackage[normalem]{ulem}

\newcommand{\be}{\begin{equation}}
\newcommand{\ee}{\end{equation}}
\newcommand{\bea}{\begin{eqnarray}}
\newcommand{\eea}{\end{eqnarray}}
\def\ba#1\ea{\begin{align}#1\end{align}}

\newcommand{\refeq}[1]{Eq.~(\ref{eq:#1})}          
\newcommand{\refeqs}[2]{Eqs.~(\ref{eq:#1})--(\ref{eq:#2})}          
\newcommand{\reffig}[1]{Fig.~\ref{fig:#1}}          
\newcommand{\reftab}[1]{Table~\ref{tab:#1}}          
          
\newcommand{\refsec}[1]{Sec.~\ref{sec:#1}}
\newcommand{\refapp}[1]{App.~\ref{app:#1}}
\newcommand{\vs}{\nonumber\\}

\def\bfx{\bm x}
\def\bfv{\bm v}
\def\bfs{\bm s}
\def\bfk{\bm k}

\def\twofast{{\sc TwoFAST}}
\def\powerfull{{\sc PowerFull}}

\def\JM{{JM}}
\def\JMp{{(JM)}}
\def\lm{{\ell m}}

\def\bt{b_e}

\def\dd{{\rm d}}

\def\nhat{\hat{\bm {n}}}
\def\khat{\hat{\bm {k}}}

\newcommand{\TF}{\mathrm{TF}}
\newcommand{\Luc}{\mathrm{Luc}}

\definecolor{royalblue}{rgb}{.25,.41,.88}

\newcommand{\fixed}[1]{\textcolor{magenta}{{#1}}}

\title{PowerFull: fast and accurate computation of the relativistic angular galaxy power spectrum including local primordial non-Gaussianity}

\author[a,b]{Gregory Lukens}
\author[a,b,c]{Donghui Jeong}
\affiliation[a]{Department of Astronomy and Astrophysics, The Pennsylvania State University, University Park, PA 16802, USA}
\affiliation[b]{Institute for Gravitation and the Cosmos, The Pennsylvania State University, University Park, PA 16802, USA}
\affiliation[c]{School of Physics, Korea Institute for Advanced Study, 85 Hoegiro, Dongdaemun-gu, Seoul, 02455, Republic of Korea}
\emailAdd{gmlukens@psu.edu}
\emailAdd{djeong@psu.edu}

\abstract{
Current and forthcoming wide-field galaxy surveys, such as SPHEREx and Euclid, provide a unique opportunity to probe the ultra-large scales of structure formation, where primordial non-Gaussianity, often parameterized by $f_\text{NL}$, is expected to leave its most detectable imprint. At the same time, these surveys inevitably enter regimes where relativistic and wide-angle effects become significant, requiring careful modeling to extract unbiased cosmological information. We address these challenges by applying the total-angular-momentum (TAM) formalism to describe redshift-space clustering beyond the flat-sky approximation. The standard Fourier-mode description, which characterizes distortions by the angle between a mode and a line-of-sight, becomes ill-defined over wide fields, whereas the TAM basis naturally separates radial and angular contributions and incorporates the relevant relativistic effects. Within this framework, we provide a new parameterization of wide-angle predictions that can be compared directly with survey data. We then introduce \textsc{PowerFull}, a modification of the \texttt{Julia}-based \textsc{2-FAST} package, which enables efficient computation of the full relativistic angular power spectrum. Finally, using Fisher information matrix–based analyses, we show that a survey like SPHEREx reaches $\sigma(f_\text{NL}) \sim 1$, but that neglecting the relativistic and wide-angle contributions shifts the recovered $f_\text{NL}$ by an amount of order its own uncertainty: at this precision the inferred value is biased unless the clustering is modeled consistently.}

\begin{document}
\maketitle
\flushbottom

\section{Introduction}

The large-scale structure (LSS) of the universe contains a plethora of cosmological information, offering insights into the physics of the early universe and the nature of cosmic evolution. 
By measuring the clustering of galaxies across cosmic time, we can infer the statistical properties of the primordial density field, test models of structure formation, and constrain fundamental cosmological parameters. Current and forthcoming wide-field surveys, such as SPHEREx \cite{Dore::2014spherex}, LSST \cite{Ivezi::2019lsst}, Roman \cite{Eifler::2019roman}, DESI \cite{DESI::2016desi}, and Euclid \cite{Amendola::2018euclid}, observe LSS over increasingly large volumes by measuring the on-sky positions and redshifts of hundreds of millions of galaxies. The unprecedented levels of precision and vast reach of these surveys present an exciting opportunity to probe fundamental questions in physics, such as the nature of dark matter, dark energy, and inflation.

In order to take full advantage of the opportunity offered by these surveys, it is important to model the observed galaxy clustering accurately. In practice, the redshift measured for a galaxy is a result of the Hubble flow (the expansion of the universe) and the line-of-sight peculiar velocity (the motion of the galaxy induced by the underlying gravitational potential). The combination is degenerate, so we cannot isolate the peculiar velocity component from the redshift contribution due to the Hubble flow unless other distance information is supplied. If the peculiar velocity field was truly random, we could ignore this effect, as the peculiar velocity ($\sim 100\,\,\mathrm{km/s}$) is typically much smaller than the recessional velocity ($H_0d \gg 1000\,\,\mathrm{km/s}$ for $d\gg 10~{\rm Mpc}/h$). However, the velocity field is directly correlated to the underlying density field, so this effect, however small, induces redshift-space distortions (RSDs) in the observed clustering of galaxies \cite{Kaiser:1987qv}. Therefore, an essential aspect of extracting cosmological information from LSS is properly modeling RSDs.

In the Newtonian framework, RSDs arise purely as a result of the peculiar velocities of galaxies, introducing an anisotropy in the clustering statistics, as mentioned in the previous paragraph. However, this treatment neglects relativistic corrections that become important as the survey volume approaches the Hubble scale. The full general relativistic (GR) treatment of RSDs \cite{Sasaki::1987rsd,Yoo::2010gr,Challinor::2011gr,Bonvin::2011gr,Jeong::2012gr,Jeong_Schmidt::2015review} accounts for additional effects that alter the photons as their geodesics traverse vast swaths of the cosmos; the integrated Sachs-Wolfe effect (ISW) (due to time-dependent gravitational potentials), Shapiro time delay (the additional light-travel time induced by gravitational potentials along the line-of-sight), and gravitational lensing from foreground LSS (which distorts the apparent position and magnification of galaxies). Since these corrections scale with $(k/k_{\rm H})^{-n}$ for integer powers $n$, they can be ignored for surveys with volumes much smaller than the horizon $k_{\rm H}\simeq aH$. However, these relativistic effects become non-negligible in the current generation of surveys and must be included in the treatment of RSDs in order to precisely extract the cosmological information contained in them.

Among the key imprints within LSS is primordial non-Gaussianity (PNG), a signature of inflation that provides a powerful probe of the initial conditions of the universe \cite{Maldacena::2003png,Acquaviva::2003png,Creminelli::2004png}. While the standard inflationary paradigm predicts nearly Gaussian primordial density fluctuations, deviations from Gaussianity arise as a result of multi-field inflation, non-slow-roll dynamics, and higher-order interactions in the inflaton field. PNG of the local variety (i.e., impacting the primordial gravitational potential $\Phi$ locally) \cite{Salopek::1990png,Komatsu::2001} induces a characteristic scale-dependent bias ($\propto k^{-2}$) in the observed galaxy density field through scale-dependent galaxy bias \cite{Dalal::2008png,Matarrese::2008png}. This scale dependence ensures that precise constraints on PNG (in the form of $f_\text{NL}$) must involve constraining galaxy clustering statistics at very large scales. At present, the most stringent constraints on $f_\text{NL}$ are from measurements of temperature anisotropies in the CMB conducted by the Planck satellite: $f_\text{NL} = -0.9 \pm 5.1$ \cite{Planck::2020fnl}. In order to distinguish between different inflation models, galaxy surveys aim to constrain PNG down to $\sigma(f_\text{NL}) \sim 1$ \cite{Alvarez::2014png,Alnoso::2015fnl_2,Camera2015SKA,Foglieni::2023fnl,Noorikuhani::2023fnl}, and therefore, must constrain clustering power at scales where the GR terms become non-trivial.

To accurately extract cosmological information from LSS, we rely on statistical measures of the clustering of galaxies. The two-point correlation function (2PCF) and its Fourier space counterpart, the power spectrum, are widely used tools to quantify the statistical properties of the density field. For relatively small survey volumes, the power spectrum - which relies on the orthogonality of plane waves using Cartesian coordinates in Fourier space - provides a convenient means of analysis, as it provides a natural basis for describing structure growth in an approximately homogeneous and isotropic universe. However, for surveys that cover a large fraction of the sky, the Fourier basis becomes inadequate due to the curvature of the sky and the significant extent of the observed volume. Instead, the spherical Fourier-Bessel (SFB) power spectrum and the angular power spectrum provide alternative formalisms that are better suited for full-sky analyses.

The SFB power spectrum decomposes the observed galaxy density field into spherical harmonics on the sky and spherical Bessel functions along the radial direction, naturally accounting for the survey’s geometry and redshift evolution. This formalism has been extensively explored as a means of studying LSS while incorporating RSD and GR corrections \cite{Heavens::1995sfb, Lanusse::2015sfb, Pratten::2013sfb, Wen::2024sfb, Semenzato::2025sfb}. Extensive work has been done in providing numerical implementations of the SFB formalism, such as \textsc{SuperFaB} \cite{Gebhardt::2021sfb_superfab}; they use a careful treatment of Bessel function transforms and mode coupling, particularly when incorporating relativistic effects. Furthermore, surveys often have limited redshift resolution, meaning that discrete tomographic bins remain a practical and widely used alternative to fully continuous radial decompositions.

Compared to the SFB power spectrum, its Fourier-Bessel basis dual, the angular power spectrum, is a much more widely used statistic that describes the clustering of galaxies tomographically. In this work, we will derive the relativistic model for the angular power spectrum using the total angular momentum (TAM) formalism \cite{Dai:2012bc}, which is an extension of the spherical Fourier-Bessel basis for general tensorial quantities. Given the upcoming era of full-sky surveys, it is crucial to develop efficient and accurate methods for computing the angular power spectrum with relativistic corrections. The angular power spectrum offers a computationally efficient approach that retains key relativistic effects while allowing for direct comparison with observational data, particularly in tomographic analyses. This paper focuses on improving the calculation of $C_\ell(r,r')$ in a fully relativistic framework, systematically incorporating wide-angle and relativistic effects.

 Although numerical calculations of the full relativistic angular power spectrum have been incorporated into linear Boltzmann codes, such as \texttt{CLASS} \cite{DiDio::2013class}, we wish to provide a robust alternative. In particular, we aim to expand upon \textsc{2-FAST} \cite{Gebhardt::2018twofast} -- a package that quickly and accurately computes integrals over one and two spherical Bessel functions, which is critical for computation of the two-point correlation function and angular power spectrum, respectively -- by using it to quickly and effectively extend the Newtonian and GR formalisms to the angular power spectrum. By refining the modeling of these corrections, we aim to enable more accurate constraints on fundamental cosmological parameters, including the effects of PNG, dark energy, and modifications to gravity, in current-generation surveys.

In this paper, we begin with a re-introduction of total angular momentum formalism in \refsec{TAM}, then apply it to model the impact of redshift-space distortions, relativistic effects, and primordial non-Gaussianity on the observed galaxy field in \refsec{gal_dens}. After this, we provide a brief review of the 2-FAST algorithm and overview of its relativistic overhaul, \textsc{PowerFull}, in \refsec{power_full}. Finally, in \refsec{spherex}, we apply this new code to make a forecast of the impact of these relativistic effects on the precision of primordial non-Gaussianity measurements in a SPHEREx-like survey and make some concluding remarks in \refsec{conclusion}.

For the main part of the paper, we use the fiducial cosmological parameters from the Planck 2018 results \cite{Planck::2020}: $\mathcal A_s = 2.105 \times 10^{-9}$, $f_\text{NL}=0$, $n_s = 0.9665$, $\alpha_s = 0$, $\Omega_b h^2 = 0.02242$, $\Omega_c h^2 = 0.11933$, $h = 0.6766$, and a flat $\Lambda$CDM background. The only exception is in the appendix \refapp{powerfull_details} where we use the same cosmological parameters as in Ref.~\citep{Gebhardt::2018twofast} to facilitate direct numerical comparisons.

\section{Total angular momentum formalism}\label{sec:TAM}
A set of total angular momentum (TAM) bases has been defined in Ref.~\cite{Dai:2012bc}, forming complete orthonormal sets of basis functions for scalar, vector, and tensor fields in three-dimensional Euclidean space. The TAM bases are eigenfunctions of the three-dimensional Laplacian operator as well as the total angular momentum operator, which makes them suitable to describe the observables beyond the flat-sky approximation. Specifically, unlike Fourier modes, which are plane-waves that reflect the translational symmetry of Euclidean space, TAM waves naturally incorporate the rotational symmetry of the curved sky by separating out the radial and angular structure.

In the context of the observed density contrast, we only have to deal with scalar and vector fields; although the relativistic effects are significantly more complex than Newtonian redshift-space distortion (RSD), the principal players are still the density field and the peculiar velocity field. In an effort to motivate and derive the mathematical tools within the total angular momentum formalism required to model the observed density field, we first outline the TAM wave expansion for scalar and vector fields, following the derivations laid out in Ref.~\cite{Dai:2012bc}.

\subsection{Scalar fields}\label{subsec:scalarTAM}

For scalar fields, TAM waves reduce to the familiar spherical Fourier-Bessel basis, for $\bfx = r\nhat$, 
\be
\Psi^k_{\lm}(\bfx) \equiv j_\ell(kr) Y_{\lm}(\nhat)\,,
\ee
defined with a spherical Bessel function of the first kind $j_\ell(z)$, and spherical harmonics $Y_{\lm}(\nhat)$.\footnote{In \cite{Dai:2012bc}, $J$ and $M$ were used for the total-angular-momentum eigenvalues.  Given that the total angular momentum coincides with the orbital angular momentum for scalar TAM waves, we use here the more common $\ell$ and $m$ for the TAM eigenvalues.} The TAM waves $\Psi^k_{\lm}(\bfx)$ are eigenfunctions of the three-dimensional Laplacian
\be
\nabla^2 \Psi^k_{\ell m}(\bfx) = -k^2 \Psi^k_{\ell m}(\bfx)\,,
\ee
as well as the angular momentum operator and its azimuthal component
{\allowdisplaybreaks
\ba
\nabla^2_{\Omega} \Psi^k_{\ell m}(\bfx) =& -\ell(\ell+1) \Psi^k_{\ell m}(\bfx)
\\
\widehat{L_z} \Psi^k_{\ell m}(\bfx) =& \,m \Psi^k_{\ell m}(\bfx)\,.
\ea
}
The TAM waves satisfy the completeness relation,
\be
     \sum_{\ell m} \int \frac{k^2\, \dd k}{(2\pi)^3} \left[ 4 \pi i^\ell \Psi^k_{\lm}(\bfx) \right]^*  \left[ 4 \pi i^\ell   \Psi^k_{\lm}(\bfx') \right] = \delta_D(\bfx-\bfx')\,,
\ee
where $\delta_D$ is the Dirac delta, as well as the orthogonality relation:
\be
\int \dd^3 x \, \left[4\pi i^{\ell}\Psi_{\ell m}^k(\bfx)\right]^* 
\left[4\pi i^{\ell'}\Psi_{\ell'm'}^{k'}(\bfx)\right] 
= \delta_{\ell\ell'}\delta_{mm'}\frac{(2\pi)^3}{k^2} \delta_D(k-k')\,,
\ee
where $\delta_{ij}$ is the Kronecker delta.

This implies that an arbitrary scalar field $\phi(\bfx)$ can be expanded as 
\be
\label{eq:scalar_TAM}
     \phi(\bfx) = \sum_{\ell m} \int \frac{k^2\, \dd k}{(2\pi)^3} \phi_{\lm}^k \, 4\pi i^\ell \Psi^k_{\lm}(\bfx),
\ee
in terms of TAM waves and expansion coefficients,
\be
\label{eq:phi_lm}
    \phi_{\lm}^k  = \int \, \dd^3x\, \phi(\bfx)  \left[ 4 \pi i^\ell \Psi^k_{\lm}(\bfx) \right]^*\,.
\ee
By using the plane-wave expansion,
\be
     e^{i \bfk \cdot \bfx} = \sum_{\ell m} 4 \pi i^\ell 
     Y_{\lm}^*(\khat) 
     \Psi^k_{\lm}(\bfx)\,,
	\label{eq:scalarplanewave} 
\ee
we can also write the TAM coefficients $\phi_{\lm}^k$ in terms of the Fourier coefficient $\tilde{\phi}(\bfk)$ as 
\be
     \phi^k_{\lm} = \int \dd^2\khat \, \tilde \phi(\bfk) Y_{\lm}^*(\khat) \label{eq:scalarspherical}.
\ee

\subsection{Vector fields}\label{subsec:vectorTAM}

For vector fields, $\bm V$, there are three corresponding TAM wave modes:
\ba
\label{eq:psi_vec_sol1}
    \Psi^{L,k}_{\JMp a}(\bfx) &\,= \frac{i}{k} \nabla_a \Psi^k_{\JM}(\bfx)\,, \\
\label{eq:psi_vec_sol2}
    \Psi^{B,k}_{\JMp a}(\bfx) &\,= - \frac{ i \widehat{L_a} }{\sqrt{J(J+1)}}  \Psi^k_{\JM}(\bfx) = -\frac{i}{k} \epsilon_{abc} \nabla^b \Psi^{E,k\,\,c}_{\JMp}(\bfx) \,, \\
\label{eq:psi_vec_sol3}
    \Psi^{E,k}_{\JMp a}(\mathbf x) &\,=  -\frac{ 1  }{k\sqrt{J(J+1)}} \epsilon_{abc}  \nabla^b \widehat{L^c} \Psi^k_{\JM}(\mathbf x) = \frac{i}{k} \epsilon_{abc} \nabla^b \Psi^{B,k\,\,c}_{\JMp}(\mathbf x)\,,
\ea
where $\widehat{L_a} \equiv -i\epsilon_{abc} x^b \nabla^c$ is an angular momentum operator and $\epsilon_{abc}$ are the Levi-Civita symbols (i.e., the antisymmetric tensor). Here, $\Psi^{L,k}_{\JMp a}(\mathbf x)$ are the components of a \textit{longitudinal} vector field, and $\Psi^{B,k}_{\JMp a}(\mathbf x)$ and $\Psi^{E,k}_{\JMp a}(\mathbf x)$ are the two polarization components (i.e., the B and E modes, with definite parity) of a {\it transverse} vector field.

Returning to the context of the RSDs, the peculiar velocity field can be treated as \textit{curl-free} in the linear regime, so we can ignore the transverse modes. Therefore, we can expand an arbitrary curl-free field in terms of the longitudinal component exclusively
\be
\label{eq:general_vector_field_TAM}
    \bm V(\bm x) = \sum_{\ell m} \int \frac{k^2 \dd k}{(2\pi)^3} V^{L,k}_{\lm}  4\pi i^\ell \bm \Psi^{L,k}_{\lm} (\bm x)\,,
\ee
where we define the basis functions of the longitudinal vector field component via
\be
\label{eq:longitudinal_basis_funcs}
    \bm \Psi^{L,k}_{\JM}(\bm x) = -i\sqrt{J(J+1)} \frac{j_J(kr)}{kr} \bm Y^E_{\JM}(\nhat) - i \frac{\dd}{\dd(kr)} \left[ j_J(kr) \right] \bm Y^L_{\JM}(\nhat)\,.
\ee
Here, $\bm Y^E_{\JM}$ and $\bm Y^L_{\JM}$ are the E-mode and longitudinal components of vector spherical harmonics, respectively. Returning to the context of $\ell$ and $m$, the line-of-sight projection (i.e., along $\nhat$) of the basis vectors becomes
\be
\label{eq:longitudinal_basis_funcs2}
    \nhat \cdot \bm \Psi^{L,k}_{\lm}(\bm x)  = i  j_\ell'(kr)  Y_{\lm}(\bm{\hat n})\,.
\ee
where the primed quantity, $j'_\ell(kr)$, of course refers to a derivative with respect to the argument $kr$. Here, we have utilized the fact that $\bm Y^L_{\lm}(\nhat) = -\nhat Y_{\lm}(\nhat)$ and the orthogonal component vanishes, $\nhat \cdot \bm Y^E_{\lm}(\nhat) = 0$. Therefore, along the line-of-sight direction, this expansion becomes
\be
\label{eq:general_LOS_vector_field_TAM}
     \nhat \cdot \bm V(\bm x) = i \sum_{\ell m} \int \frac{k^2 \dd k}{(2\pi)^3} V^{L,k}_{\lm}  4\pi i^\ell j_\ell'(kr) Y_{\lm}(\nhat)\,.
\ee
Finally, it may be convenient to write the vector field coefficients $V^{L,k}_{\lm}$ in terms of a scalar field. One natural choice is to exploit the divergence of the vector field, $\bm \nabla \cdot \bm V(\bfx)$. We know that
\ba
\label{eq:PSI_to_psi}
    \sum_{\ell m} 4 \pi i^\ell \left[ \bm \nabla \cdot \bm \Psi^{L,k}_{\lm}(\bm x) \right] = -i \sum_{\ell m} 4\pi i^\ell k \Psi^k_{\lm}(\bm x)\,,
\ea
so the divergence clearly becomes
\ba
\label{eq:vector_divergence}
    \psi(\bfx) \equiv \bm \nabla \cdot \bm V(\bm x) &\, =  \sum_{\ell m} \int \frac{k^2 \dd k}{(2\pi)^3} V^{L,k}_{\lm}  4\pi i^\ell \left[\bm \nabla \cdot \bm \Psi^{L,k}_{\lm}(\bfx) \right] \,, \vs
    &\,= -i \sum_{\ell m} \int \frac{k^2 \dd k}{(2\pi)^3} k\,V^{L,k}_{\lm}  4\pi i^\ell \Psi^{k}_{\lm}(\bfx)\,.
\ea
And, since $\psi(\bfx)$ itself can be expanded as a scalar field using \refeq{scalar_TAM}, this implies that 
\be
\label{eq:scalar_field_vector_field_coeffs}
    V^{L,k}_{\lm} = \frac{i}{k}\psi^k_{\lm}\,,
\ee
where we have successfully recast the vector field coefficients in terms of scalar field coefficients. With all of the tools we require to model the observed galaxy clustering statistics in a relativistic context, we move on to the density contrast.

\section{Observed galaxy density contrast}\label{sec:gal_dens}

In this section, we briefly review the derivation of the observed density contrast in both the Newtonian and general relativistic frameworks. We should begin with a definition of the density contrast,
\be
\label{eq:delta_g}
    \delta_g(\bfx) = \frac{n_g(\bfx) - \bar n_g(r)}{\bar n_g(r)}\,,
\ee
where $n_g(\bfx)$ and $\bar n_g(r)$ are the local and mean physical galaxy number density, respectively. Here, $\bfx \equiv r \nhat$, where $r$ is the comoving radial distance and $\nhat$ is the line-of-sight direction. That is, we consider galaxies in the light-cone coordinate system, which is defined by the redshift-distance relationship $r(z)$ and sky location $\nhat$. In this coordinate system, we can generically recast redshift evolution as radial evolution; in other words, we are free to use both $r$ and $z$ to handle distance-dependent quantities.

Galaxies are biased tracers of the underlying matter distribution, and the galaxy bias at the largest scales is well approximated by the linear relationship
\be
\label{eq:linear_bias}
    \delta_g(\bfx) = b_g(r) \delta(\bfx)\,,
\ee
with linear galaxy bias parameter $b_g$. The linear bias $b_g$ depends on the galaxy population, but it is expected to be spatially homogeneous and only dependent on time. Explicitly, it is given as the linear response of the background galaxy number density to any change in the average matter density $\bar{\rho}_m$,
\be
\label{eq:def_bg}
    b_g =\frac{{\rm d} \ln \bar{n}_g}{{\rm d}\ln \bar{\rho}_m}\,.
\ee

In the presence of the primordial non-Gaussianity (PNG), the linear bias relationship needs to be modified to include the extra modulations of galaxy number density due to the primordial three-point correlation function. For example, local-type PNG yields an extra linear bias in the primordial potential field $\Phi$, adding an additional scale-dependent bias term, as we show in \refsec{PNG}.

\subsection{Newtonian treatment}\label{sec:newtonian}

We begin with Newtonian RSD treatment, only considering the effect of the peculiar velocity, $\bfv$, on the observed density contrast. The radial peculiar velocity $v_r = \bfv \cdot \nhat$ moves a galaxy at $\bfx=r\nhat$ to a redshift-space coordinate
\be
    \bfs = \left(r + \frac{v_r}{aH}\right)\nhat \equiv s\nhat\,,
\label{eq:RSD}
\ee
where $a$ and $H$ are the scale factor and Hubble parameter at the location of the galaxy, respectively.

Since the redshift-space distortion is a mere coordinate remapping and must conserve the total number of galaxies, the relation between $\delta_g(\bfx)$, the galaxy density contrast in real space and redshift-space, $\delta_s(\bfs)$, is given as
\be
   a^3(r) \bar n_g(r)\left[1+\delta_g(\bfx)\right]\dd^3 \bfx = a^3(s)\bar n_g(s)\left[1+\delta_s(\bfs)\right]\dd^3 \bfs,
\label{eq:RSD2}
\ee
where $a^3(r) \bar n_g(r)$ is the comoving mean density of galaxies at a distance $r$. To linear order in the perturbation variables $\delta_g$, $\delta_s$, and $v_r$, this relation is
\ba
    \delta_s(\bfs) =&\,\delta_g(\bfx) - \frac{\alpha_1(r)}{r} \frac{v_r}{aH} - 
    \frac{\dd}{\dd r}\left(\frac{v_r}{aH}\right)\,,
\label{eq:linearRSD_Newtonian}
\ea
Here, the new quantity is present in proportion to the velocity, 
\ba
    \alpha_1(r) &\,\equiv \frac{\dd\ln \left[r^2 a^3(r) \bar n_g(r)\right]}{\dd\ln r}  =  2 - aHr b_e\,,
\label{eq:alpha1}
\ea
parameterizes the radial evolution of the mean number of galaxies within a unit solid angle. The new term can also be related to the evolution bias, $b_e$, defined in Ref.~\cite{Jeong::2012gr}: 
\be
\label{eq:def_be}
    b_e = \frac{\dd\ln (a^3 \bar{n}_g)}{\dd\ln a }\,.
\ee
If we assume the universality of the mass function -- that is, the galaxy number density is determined only by $\nu\equiv \delta_c/\sigma_R(M)$, with $\delta_c=1.686$ -- the effective bias, $b_e$, is related to the linear bias of galaxies via 
\be
\label{eq:b_e_halo}
    b_e = \delta_cf (b_g-1)\,.
\ee

When implementing \refeq{linearRSD_Newtonian}, the derivative with respect to $r$ is usually assumed to be a purely radial derivative. Under this assumption, however, the change in the linear velocity field due to time evolution is missed. In the light-cone coordinate system, we must treat the radial derivative as a \textit{total} derivative with $\dd/\dd r \equiv \partial/\partial r + (\dd t/\dd r)\,\partial/\partial t$. We also know that to linear order, the velocity divergence $\theta(t,\bfx) \equiv \bm \nabla \cdot \bfv(t,\bfx)$ is related to the matter density contrast $\delta(t,\bfx)$ by the continuity equation
\be
    \theta(t,\bfx) = -a(t)H(t)f(t) \delta(t,\bfx)\,,
\label{eq:theta_delta}
\ee
where
\be
\label{eq:linear_growth_rate}
    f(t) \equiv  \frac{\dd\ln D}{\dd\ln a}
\ee
is the linear growth rate and $D(a)$ is the linear growth factor. As a rather important aside, \refeq{theta_delta} is the entire reason why the peculiar velocities of galaxies cause redshift-space distortions. With these facts in mind, the radial derivative in \refeq{linearRSD_Newtonian} should be written as
\ba
\frac{\dd}{\dd r}
   \left(\frac{v_r}{aH}\right) 
   = & \,\frac{\dd}{\dd r}\left( \frac{v_r}{aHfD}fD \right) \vs 
    \equiv\,& \frac{\alpha_2(r)}{r} \frac{v_r}{aH} + fD \frac{\dd}{\dd r}\left(\frac{v_r}{aHfD}\right)
\ea
with 
\ba
    \alpha_2(r) &\,\equiv \frac{\dd\ln (fD)}{\dd\ln r} = -aHr \left(f + \frac{\dd\ln f}{\dd\ln a}\right)\,.
\label{eq:alpha2}
\ea

Combining all of the above, our final expression for the observed linear redshift-space density contrast becomes
\be
    \delta_s(\bfx) = b_g\delta(\bfx) - \frac{\alpha(r)}{r}\frac{v_r}{aH} - fD \frac{\dd}{\dd r}\left(\frac{v_r}{aHfD}\right)\,,
\label{eq:linearRSD_Newtonian_final}
\ee
where the coefficient becomes
\be
\label{eq:alpha}
    \alpha(r) \equiv \alpha_1(r) + \alpha_2(r) = \frac{\dd\ln[ r^2fD\, a^3\bar n_g(r)]}{\dd\ln r}\,.
\ee

We see that on large scales, the line-of-sight peculiar velocity induces an anisotropy in the galaxy density contrast. In fact, when one ignores radial evolution (i.e., $\alpha = 0$), the Newtonian formula in \refeq{linearRSD_Newtonian_final} reduces to the well-established Kaiser formula \cite{Kaiser:1987qv}.

Note that this is the case for the plane-parallel (or flat-sky) approximation when $\dd/\dd r\simeq k_r\gg 1/r$, but we must retain the contribution beyond the flat-sky approximation. That is, in previous generation surveys, which only covered a small fraction of the sky and the radial distance to the survey volume was far greater than the distances covered in the transverse direction, the $v_r/r$ term was much smaller than the derivative term, $\partial v_r/\partial r \simeq k_r v_r$, as these surveys probed length scales $1/k_r \ll r$. For the wide-angle surveys that we are interested in, we must restore the $v_r/r$ term.

Now that we have an expression for the observed density contrast in a Newtonian context, we can utilize the total angular momentum formalism outlined in \refsec{TAM}. First of all, the observed density contrast is a scalar field, so it can be expanded via \refeq{scalar_TAM} to obtain
\be
\label{eq:delta_s_tam}
    \delta_s(\bfx) = \sum_{\ell m} \int \frac{k^2 \dd k}{(2\pi)^3}\delta^k_{s,\lm}  4\pi i^\ell  \Psi^k_{\lm}(\bfx) \,,
\ee
with coefficients $\delta^k_{s,\lm}$. We can perform a similar procedure for the matter density contrast, $\delta(\bfx)$:
\be
\label{eq:delta_tam}
    \delta(\bfx) = D(r) \sum_{\ell m} \int \frac{k^2 \dd k}{(2\pi)^3}\delta^k_{\lm}  4\pi i^\ell  j_\ell(kr) Y_{\lm}(\nhat) \,,
\ee
which has coefficients $\delta^k_{\lm}$ and has been rescaled by the linear growth factor, $D(r)$, normalized to be unity at present time. Next, we have to handle the peculiar velocity field along the line-of-sight, $v_r \equiv \nhat \cdot \bfv$. We recall from \refeq{scalar_field_vector_field_coeffs} that we can relate the TAM coefficients of a curl-free vector field to a scalar field by utilizing the divergence. Also, the linear peculiar velocity divergence, $\theta(\bfx)$, is related to the underlying density field, $\delta(\bfx)$, via \refeq{theta_delta}, which yields that 
\be
\label{eq:v_coeffs_related_to_theta}
    v^{L,k}_{\lm} = -i\frac{aHfD(r)}{k} \delta_{\lm}^k\,.
\ee
The line-of-sight component of the peculiar velocity field follows directly from \refeq{general_LOS_vector_field_TAM}, and the velocity term in \refeq{linearRSD_Newtonian_final} can be written as
\be
\label{eq:v_r_tam}
    \frac{v_r}{aH} = f(r) D(r) \sum_{\ell m} \int  \frac{k^2 \dd k}{(2\pi)^3}\delta^{k}_{\lm} 4\pi i^\ell \frac{j_\ell '(kr)}{k} Y_{\lm} (\nhat)\,.
\ee
Finally, we require the velocity derivative term in the observed density contrast. This step is relatively simple; we utilize the following:
\be
\label{eq:j_ell_der}
    \frac{\dd}{\dd r} j_\ell'(kr) =  k j_\ell''(kr)\,,
\ee
where, again, a prime denotes a derivative with respect to $kr$. Combining this relation with \refeq{v_r_tam}, it is trivial to show that
\be
\label{eq:v_r_tam_der}
    \frac{\dd}{\dd r}\left(\frac{v_r}{aHfD}\right) = \sum_{\ell m} \int  \frac{k^2 \dd k}{(2\pi)^3}\delta^{k}_{\lm} 4\pi i^\ell j_\ell ''(kr) Y_{\lm} (\nhat)\,.
\ee
Putting everything together, we have
\ba
\label{eq:delta_s_tam_explicit}
    \delta_{s,\lm}^k &\,= \frac{2}{\pi}\int \dd k' k'^2  \,\delta^{k'}_{\lm}  \int \dd r\,r^2 \, j_\ell(kr) D(r) F_\ell(k',r)\,,
\ea
where we have defined                                                 
\be
\label{eq:redshift_space_kernel}
    F_\ell(k,r) = b_g(r) j_\ell(kr) - f(r) \frac{\alpha(r)}{r}\frac{j'_\ell(kr)}{k} - f(r) j_\ell''(kr)\,.
\ee
We refer to $F_\ell(k,r)$ as the \textit{redshift-space kernel}, which closely resembles \refeq{linearRSD_Newtonian_final}. We include a full step-by-step derivation of \refeq{delta_s_tam_explicit} in \refapp{delta_s_derivation}.

\subsection{Relativistic formalism}

In this section, we briefly summarize the full relativistic expression of the observed density contrast of galaxies \cite{Yoo::2009gr,Yoo::2010gr,Yoo::2014gr,Jeong::2012gr,Jeong_Schmidt::2015review,Alonso2015UltraLargeScale,Paul2026Visualising}. In a nutshell, the relativistic effects use the null geodesic equation for light propagation to connect the intrinsic galaxy clustering, where the usual galaxy bias relationship must be defined at the galaxies' constant-proper-time slicing, to the observed one, which is defined in the constant-observed-redshift slicing. As a result, the relativistic effects contain contributions proportional to the gravitational potential and line-of-sight velocity, both of which have extra $\propto k^{-\gamma}$ scaling (with $\gamma > 0$) compared to the density field; thus, the effect is more prominent as we move to larger scales.

The treatment of the relativistic effects follows that of Ref.~\cite{Jeong::2012gr}, and we adopt the synchronous-comoving gauge on a flat Friedmann-Lemaître-Robertson-Walker background. The perturbed metric is given as
\be
    \dd s^2 = a(\tau)^2 \left\{ -\dd\tau^2 +  \left[(1+2D(\tau,\bfx))\delta_{ij} + 2E_{ij}(\tau,\bfx) \right] \dd x^i\dd x^j \right\}\,, 
\ee
where $D(\tau, \bfx)$ is a scalar perturbation and $E_{ij}(\tau,\bfx)$ is a longitudinal and traceless tensor perturbation related to a scalar function $E(\tau,\bfx)$ via
\be
    E_{ij}(\tau,\bfx) = \left(\partial_i\partial_j - \frac13 \delta_{ij}\nabla^2 \right) E(\tau,\bfx)\,.
\ee
Here, $\tau$ is the conformal time, where $\dd t = \dd\tau\,a(t)$, and $\delta_{ij}$ is the Kronecker delta (with $i,j$ as spatial indices). Under the condition of this being a \textit{comoving} gauge, we know that the observer must follow the cosmic fluid's velocity; $T_0^j \propto v^j = 0$.

In this setting, the full expression for the observed galaxy density contrast is given by
\ba
\nonumber
    \delta_s(\bfx) =\:& b_g\delta_m + \bt\delta z - \frac{1}{aH} \partial_\parallel^2E'  +2(1-\mathcal{Q})\phi -\frac{\phi'}{aH} - \frac{2}{r} (1-\mathcal{Q}) \left[E'-E'_o\right] \vs 
    &  - \bigg[1 +2\mathcal{Q} -\frac{1}{aH}\frac{\dd H}{\dd z} + \frac{2}{aHr}(1-\mathcal{Q}) \bigg] \delta z + 2(1-\mathcal{Q})\partial_\parallel E'_o  \vs
    & -\frac{2}{r} (1-\mathcal{Q}) \int_0^{r}\dd r' \left(\phi + E''\right) - 2(1-\mathcal{Q})\kappa.
\label{eq:deltas_GR}
\ea
Here, all quantities are evaluated at the \textit{observed} position, $\bfx$ - which is given by the comoving distance, $r$ (or redshift, $z$) and line-of-sight, $\nhat$ - and time $\tau$, except those with the subscript $o$, which are evaluated at the observer's location (e.g., $E'_o$). Primed terms, such as $E'$, refer to the conformal time derivative (i.e., $E' = \partial E/\partial \tau$). Additionally, we have
\ba
    r(z) =&\, \int_0^{z} \frac{\dd z'}{H(z')}\,, \\[4pt]
    \tau(z) =&\, \tau_0 - r(z)\,,
\ea
for the observed comoving distance and conformal time for galaxies at redshift $z$, where $\tau_0$ is the conformal time at observation.

In \refeq{deltas_GR}, there are three parameters, $b_g$, $b_e$, $\mathcal{Q}$ which quantify the response of the \textit{local} mean galaxy number density to changes in the local matter density, cosmic time (or age), and flux cut (for a galaxy sample driven by some apparent flux criteria), respectively. We have defined $b_g$, the linear galaxy bias, in \refeq{def_bg}, and $b_e$, the effective galaxy bias in \refeq{def_be}. The magnification bias, $\mathcal Q$, is defined as
\ba
\mathcal{Q}(z) =&\, -\frac{\dd\ln \bar{n}_g( > L_\text{min},z)}{\dd \ln L_\text{min}(z)}\,,
\label{eq:def_Q}
\ea
where $L_\text{min}$ is the limiting luminosity of a flux-limited survey \cite{Moessner::1998Qmag,Matsubara::2000Qmag,Hui::2007Qmag,Yoo::2014gr,Montanari::2015Qmag}. Note that this parameter, as well as the biases $b_g$ and $b_e$ defined in Eqs.~\eqref{eq:def_bg} and \eqref{eq:def_be}, can be measured from surveys.

The other variables in \refeq{deltas_GR} can be written in terms of the scalar metric perturbations. First, combining the two scalar perturbations, a gauge-invariant curvature perturbation arises 
\be
\label{eq:phi_gauge}
    \phi(\tau,\bfx) = D(\tau,\bfx) - \frac13\nabla^2 E(\tau,\bfx)\,.
\ee
Then, we define $\delta z$ as the fractional difference between the observed redshift ($z$) and the background redshift we would observe in the unperturbed universe ($\bar{z}$):
\be
    1+z = (1+\bar{z})(1+\delta z)\,,
\ee
which can be recast as 
\be
 \delta z =\, \partial_\parallel E' + E'' - \partial_\parallel E'_o - E''_o + \int_0^{r} \dd r' (\phi+E'')'\,, \label{eq:deltaz_GR}
\ee
and, finally, the weak gravitational lensing convergence, $\kappa$, is given as 
\be
    \kappa =\, -\frac12 \int_0^{r} \dd r' \frac{(r'-r)r'}{r}\nabla_\perp^2(\phi + E'').
\label{eq:kappa}
\ee

The formalism up to this point is completely general for any metric theory of gravity. For definiteness, we further simplify \refeq{deltas_GR} using general relativity to relate the metric perturbations to the linear density contrast in synchronous comoving gauge $\delta^{\rm sc}(\tau,\bfx)\equiv \delta(\tau,\bfx)$. Following along with Ref.~\cite{Jeong::2012gr}, we find
\ba
\label{eq:nabla^2_phi_E'_E''}
    \nabla^2 \phi(\tau,\bfx) =&\, -a^2(\tau) H^2(\tau) \left[f(\tau) + \frac{3}{2}\Omega_m(\tau) \right] \delta(\tau,\bfx) \\
    \nabla^2 E'(\tau,\bfx) =&\, - a(\tau) H(\tau) f(\tau) \delta(\tau,\bfx)\,, \\
    \nabla^2 E''(\tau,\bfx) =&\, -a^2(\tau) H^2(\tau) \left[  \frac{3}{2}\Omega_m(\tau) - f(\tau) \right] \delta(\tau,\bfx)\,,
\ea
where $\bar \rho_m(\tau)$ is the mean matter density. We further rewrite them in terms of the density contrast coefficients, $\delta^k_\lm$, of the total angular momentum wave expansion as
\ba
\label{eq:tam_phi}
    \phi(\tau,\bfx) =&\, a^2(\tau) H^2(\tau) D(\tau) \left[f(\tau) + \frac{3}{2}\Omega_m(\tau) \right] \sum_{\lm} \int \frac{k^2 \dd k}{(2\pi)^3} \delta^k_\lm 4\pi i^\ell \frac{j_\ell(kr)}{k^2} Y_\lm(\nhat)\,, \\
    \label{eq:tam_E'}
    E'(\tau,\bfx) =&\, a(\tau) H(\tau) f(\tau) D(\tau) \sum_{\lm} \int \frac{k^2 \dd k}{(2\pi)^3} \delta^k_\lm 4\pi i^\ell \frac{j_\ell(kr)}{k^2} Y_\lm(\nhat)\,,\\
    \label{eq:tam_E''}
    E''(\tau,\bfx) =&\, a^2(\tau) H^2(\tau) D(\tau) \left[ \frac{3}{2}\Omega_m(\tau) - f(\tau) \right] \sum_{\lm} \int \frac{k^2 \dd k}{(2\pi)^3} \delta^k_\lm  4\pi i^\ell\frac{j_\ell(kr)}{k^2} Y_\lm(\nhat)\,,
\ea

The radial derivatives of $E'$ can be computed using \refeq{longitudinal_basis_funcs2} 
\ba
\label{eq:radial_der_GR}
    \partial_\parallel \Psi^k_\lm(\bfx) &\, \equiv \nhat \cdot \bm \nabla \Psi^k_\lm(\bfx) = k j_\ell'(kr) Y_\lm(\nhat)\,,\\
    \partial_\parallel^2 \Psi^k_\lm(\bfx) &\, \equiv (\nhat \cdot \bm \nabla)^2 \Psi^k_\lm(\bfx) = \nhat^i\nhat^j \nabla_i \nabla_j \Psi^k_\lm(\bfx) = k^2 j_\ell^{\prime\prime}(kr) Y_\lm(\nhat)\,,
\ea
for $\Psi_{\ell m}^k(\bfx=r\nhat)=j_\ell(kr)Y_{\ell m}(\nhat)$.
Applying this to the TAM wave expansion in \refeq{tam_E'} yields
\ba
\label{eq:radial_der_GR_TAM}
    \partial_\parallel E'(\tau,\bfx)  =&\, a(\tau) H(\tau) f(\tau) D(\tau) \sum_{\lm} \int \frac{k^2 \dd k}{(2\pi)^3} \delta^k_\lm 4\pi i^\ell \frac{j_\ell'(kr)}{k} Y_\lm(\nhat)\,,\\  
    \partial_\parallel^2 E'(\tau,\bfx)  =&\, a(\tau) H(\tau) f(\tau) D(\tau) \sum_{\lm} \int \frac{k^2 \dd k}{(2\pi)^3} \delta^k_\lm 4\pi i^\ell j_\ell''(kr) Y_\lm(\nhat)\,.
\ea
The transverse Laplacian, $\nabla^2_\perp$, which arises in the lensing convergence \refeq{kappa}, acts on the TAM wave as
\be
\label{eq:perp_der_TAM}
    \nabla^2_\perp \Psi^k_\lm(\bfx) = -\frac{\ell(\ell+1)}{r^2}\Psi^k_\lm(\bfx)\,.
\ee
The convergence then becomes
\ba
\label{eq:kappa_TAM}
    \kappa &\,= \sum_\lm 6\pi i^\ell\ell(\ell+1)Y_\lm(\nhat) \int \frac{k^2 \dd k}{(2\pi)^3} \delta^k_\lm    
    \int_0^r \dd r' \frac{r-r'}{rr'} [a^2H^2 \Omega_m D](r') \frac{j_\ell(kr')}{k^2}\,,
\ea

We are ready to recast the observed density contrast with the relativistic corrections from \refeq{deltas_GR} in terms of the TAM wave expansion. For quantities $E_o'$ and $\partial_\parallel E_o'$ evaluated at the observer's location, we use the fact that $j_\ell(kr = 0)$ is only non-zero for $
    j_0(0) = 1$, $\qquad j'_1(0) = \frac{1}{3}$.
Combining all of the results above, \refeq{deltas_GR}, we obtain
\ba
    \delta_s(\bfx) = & \sum_{\ell m}   \int \frac{k^2 \dd k}{(2\pi)^3}\delta_{\ell m}^k 4\pi i^\ell Y_{\ell m}(\nhat)  D(r)  \vs
    &\times \Bigg\{ b_g j_\ell(kr) - fj_\ell''(kr) + \bigg[ b_e  - (1+2\mathcal{Q}) - \frac{\dd \ln H}{\dd\ln a} - \frac2{aHr}(1-\mathcal{Q}) \bigg] \vs
    &\qquad  \times \bigg[ aHf\frac{j_\ell'(kr)}{k} + a^2H^2\left(\frac32\Omega_m-f\right)\frac{j_\ell(kr)}{k^2} -a_oH_of_o\frac{D_o}{D(r)}\frac{\delta^K_{\ell 1}}{3k} \vs
    &\qquad \qquad \,-a_o^2 H_o^2\left(\frac32\Omega_{mo}-f_o\right)\frac{D_o}{D(r)}\frac{\delta^K_{\ell 0}}{k^2} + \mathcal{I}_{\mathrm{ISW}}(k,r) \bigg] \vs
    & \qquad -2(1-\mathcal{Q})\mathcal{I}_{\kappa} + 2(1-\mathcal{Q})\, a^2 H^2 \left(f+\frac32\Omega_m\right)\frac{j_\ell(kr)}{k^2} +2(1-\mathcal{Q})a_oH_of_o\frac{D_o}{D(r)}\frac{\delta^K_{\ell1}}{3k}\vs
    & \qquad - \frac2r(1-\mathcal{Q})\left[aHf\frac{j_\ell(kr)}{k^2}-a_oH_of_o\frac{D_o}{D(r)}\frac{\delta^K_{\ell 0}}{k^2}\right]-2\frac{(1-\mathcal{Q})}{r} \mathcal{I}_\mathrm{TD}(k,r)  \Bigg\}\,.
\ea
We recall that $a_o$, $H_o$, $f_o$, $D_o$, and $\Omega_{mo}$ are all evaluated at the time of observation. Here, the Kronecker delta $\delta^K_{\ell 0}$ and $\delta^K_{\ell 1}$ signify that observer terms vanish outside of the monopole and dipole, respectively. Since the monopole observer terms cannot, in practice, be distinguished from the changes in the mean number density, we can safely ignore them. Additionally, the dipole observer components -- from which we can measure our motion with respect to large-scale structure -- are not stochastic, so we neglect their contribution as well.

In \refeq{deltas_GR}, we also defined three new quantities involving line-of-sight integration,
\ba
\label{eq:I_isw}
    \mathcal{I}_\mathrm{ISW}(k,r) \equiv &\, 3\int _0^r \dd r' \frac{a^3(r')H^3(r')\Omega_m(r')[f(r')-1]}{k^2} \frac{D(r')}{D(r)} j_{\ell}(kr')\,,
\\
\label{eq:I_td}
    \mathcal{I}_\mathrm{TD}(k,r) \equiv &\,3\int _0^r \dd r' \frac{a^2(r')H^2(r')\Omega_m(r')}{k^2} \frac{D(r')}{D(r)} j_{\ell}(kr')\,,
\\
\label{eq:I_kappa} 
    \mathcal{I}_\mathrm{\kappa}(k,r) \equiv &\, \frac{3}{2}\ell(\ell+1)\int _0^r \dd r'\frac{r-r'}{rr'} \frac{a^2(r')H^2(r')\Omega_m(r')}{k^2}\frac{D(r')}{D(r)} j_{\ell}(kr')\,,
\ea
which are the integral representations of the integrated Sachs-Wolfe effect, Shapiro time delay, and gravitational lensing components present in the relativistic expansion. With these definitions and safely ignoring the observer terms that would only affect monopole and dipole, the observed density contrast simplifies to
\ba
\label{eq:deltas_GR_TAM}
    \delta_s(\bfx) = \sum_{\ell m} &  \int \frac{k^2 \dd k}{(2\pi)^3}\delta_{\ell m}^k 4\pi i^\ell Y_{\ell m}(\nhat)  D(r) \Bigg[ b_g j_\ell(kr) - fj_\ell''(kr) + a^2 H^2 \mathcal A \frac{j_\ell(kr)}{k^2} \vs
    &+ aH \mathcal B \frac{j_\ell'(kr)}{k} + \frac{\mathcal B}{f} \,\mathcal I_\text{ISW}(k,r) -2\frac{(1-\mathcal{Q})}{r} \mathcal{I}_\mathrm{TD}(k,r) - 2(1-\mathcal Q) \mathcal I_\kappa(k,r) \Bigg]\,,
\ea
where we have defined the following coefficients:
\ba
\label{eq:A}
    \mathcal A  =&\, \frac{3}{2}\Omega_m \bigg[ \frac{\mathcal B}{f} \, \left(1-\frac{2f}{3\Omega_m}\right) + 2(1-\mathcal{Q}) \left(1 + \frac{2f}{3\Omega_m} \right) - \frac{4}{3}\frac{(1-\mathcal Q)}{aHr} \frac{f}{\Omega_m}  \vs
    &\, \qquad \quad - \frac{2f}{3\Omega_m} \bigg(\frac{3}{2}\Omega_m + \frac{\dd \ln H}{\dd \ln a} \bigg)\bigg]\,,\\[5pt]
\label{eq:B}
    \mathcal B  =&\, f(b_e + \mathcal C - 1 )\,,\\[5pt]
\label{eq:C}
    \mathcal C  =&\, -\frac{\dd \ln H}{\dd\ln a} - \frac2{aHr}(1-\mathcal{Q}) - 2\mathcal Q \,.
\ea
\begin{figure}[t]
    \centering
    \includegraphics[width=0.99\linewidth]{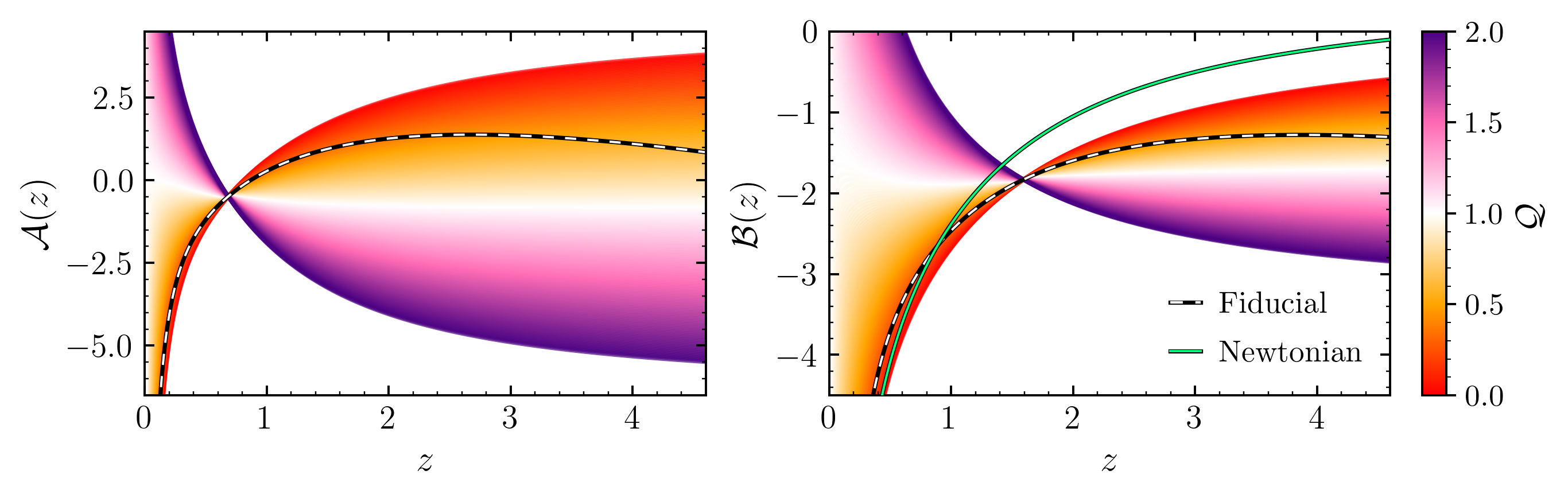}
    \caption{The redshift evolution of the relativistic coefficients $\mathcal A$ and $\mathcal B$ derived in Eqs.~\eqref{eq:A} and \eqref{eq:B}, respectively, assuming $\Lambda$CDM and for different constant values of magnification bias $\mathcal Q$, given in \refeq{def_Q}. For both panels, we use the fiducial evolution bias  $b_e(z)$ -- see Eq.~\eqref{eq:def_be} -- for the SPHEREx-like survey outlined in Sec.~\ref{sec:Q_choice}. The dashed lines show the coefficients fixing the magnification bias ${\cal Q}$ to the values from SPHEREx-like mock luminosity function. The green line in the right panel represents the Newtonian term $-f\alpha(r)/(aHr)$ with $\alpha(r) = \alpha_1(r) + \alpha_2(r)$, as presented in \refeq{alpha}.}
    \label{fig:AB_fig}
\end{figure}
As two of these coefficients ($\mathcal A$ and $\mathcal B$) appear in the full relativistic density contrast, we plot their redshift evolution for different magnification bias values in \reffig{AB_fig}. The figure shows the similarity of $\mathcal B(\mathcal Q = 0 )$ to the $\alpha(r)$ term present in the Newtonian treatment of the density contrast in \refeq{redshift_space_kernel}; here we assume $\Lambda$CDM, so $\dd \ln H/\dd \ln a = -3\Omega_m/2$. A detailed comparison of the two terms is presented in \refapp{newt_gr_comp}.

Now that we have presented the observed density contrast using the total angular momentum formalism in the relativistic context, we can clearly see that the redshift-space kernel in this case is
\ba
\label{eq:F_ell_GR}
    F_\ell(k,r) =&\, b_g j_\ell(kr) - fj_\ell''(kr) + {\cal H}^2 \mathcal A \frac{j_\ell(kr)}{k^2} + {\cal H} \mathcal B \frac{j_\ell'(kr)}{k} + \frac{\mathcal B}{f} \,\mathcal I_\text{ISW}(k,r) \vs
    &\,-2\frac{(1-\mathcal{Q})}{r} \mathcal{I}_\mathrm{TD}(k,r) - 2(1-\mathcal Q) \mathcal I_\kappa(k,r)\,,
\ea
where ${\cal H}\equiv aH$ is the conformal Hubble expansion rate.

\subsection{Local primordial non-Gaussianity}\label{sec:PNG}

Primordial non-Gaussianity refers to the statistics of seed cosmological perturbations deviating from Gaussian initial conditions. As single-field inflation models firmly predict nearly Gaussian initial conditions, detection of primordial non-Gaussianity will be a ``smoking gun" signature for new physics, such as interaction between inflaton fields or a non-standard vacuum state at the beginning of the Universe \cite{Maldacena::2003png,Creminelli::2004png,Bartolo_2004::png,Komatsu_2010::png,Chen_2010::png}. 

The local-type primordial non-Gaussianity is the most well-studied model. In the local model, the non-Gaussian deviation of Bardeen potential $\Phi$ (the large-scale gravitational potential) can be modeled with the Gaussian field, $\Phi_G$, as
\be
\label{eq:png}
    \Phi(\bfx) = \Phi_G(\bfx) + f_\text{NL} \left[\Phi_G^2(\bfx) - \left<\Phi_G^2\right> \right] + \mathcal O(\Phi_G^3)\,,
\ee
where $f_\text{NL}$ is the principal observable that quantifies the amplitude of the non-Gaussianity. 

The Bardeen potential is related to the underlying matter density field in comoving gauge via Poisson's equation, 
\be
\label{eq:poisson_png}
    \delta(\bfk,\tau) = \mathcal T_\delta(k,\tau) \Phi(\bfk)\,,
\ee
where $\mathcal T_\delta(k,\tau)$ is the transfer function between the two quantities at epoch $\tau$:
\be
\label{eq:T_delta}
    \mathcal T_\delta(k,z) = \frac{2}{3}\frac{g(0) D(z)}{\Omega_{m,0}\,H_0^2} k^2 T(k)\,.
\ee
Here, $T(k)$ is the usual matter transfer function, $D(z)$ is the linear growth factor normalized to unity at present time. The $g(0) = D_\text{md}(0)\sim 0.79$ factor further normalizes growth deep into the matter-domination epoch. Both the matter density parameter $\Omega_{m,0}$ and the Hubble expansion rate $H_0$ are evaluated at present time. Finally, the linear-order galaxy density contrast can be written as  \cite{Dalal::2008png,Matarrese::2008png,Giannantonio::2010png,McDonald::2008png,Baldauf::2011png,Assassi::2015png}
\be
\label{eq:delta_g_png}
    \delta_g(\bfk,z) = b_g(z) \delta(\bfk,z) + b_\Phi(z) f_\text{NL} \Phi(\bfk) + \varepsilon(\bfk,z)\,,
\ee
with the usual linear galaxy bias $b_g(z)$ defined in \refeq{def_bg} and the stochastic bias $\varepsilon(\bfk,z)$, and the linear non-Gaussian bias $b_\Phi$ (see Ref.~\cite{Desjacques::2018galbias} for a review). In the peak-background split (PBS) picture, the linear non-Gaussian bias parameter is given as the linear response of the galaxy number density to the change of local gravitational potential as
\be
\label{eq:b_phi}
    b_\Phi = \frac{\dd\ln \bar{n}_g}{\dd(f_\text{NL} \Phi)} \,.
\ee
Assuming universality of the halo mass function, similar to the evolution bias defined in \refeq{def_be}, the linear non-Gaussian bias can be simplified as
\be
\label{eq:b_phi_universal}
    b_\Phi = 2 (b_g - 1) \delta_c\,,
\ee
where $\delta_c = 1.686$ is once again the threshold linear matter density contrast from the spherical-collapse model. Using \refeq{delta_g_png} in combination with Eqs.~\eqref{eq:poisson_png} and \eqref{eq:T_delta}, we arrive at
\ba
\label{eq:delta_g_png_2}
    \delta_g(\bfk,z) &\,= b_g(z) \delta(\bfk,z) + 2 \left[b_g(z) - 1 \right] \delta_c\, f_\text{NL}  \mathcal T_\delta^{-1}(k,z) \delta(\bfk,z) + \varepsilon(\bfk,z) \vs
    &\, = \left \{ b_g(z)  + 3 \left[b_g(z) - 1 \right] f_\text{NL} \frac{\Omega_{m,0}\,H_0^2 g(0)^{-1} \delta_c}{k^2 T(k) D(z)}  \right\} \delta(\bfk,z) + \varepsilon(\bfk,z)\,,
\ea
where we see the addition of a scale-dependent bias ($\propto k^{-2}$) as a result of the local primordial non-Gaussianity. As a result of this bias relation, the redshift-space kernels defined in Eqs.~\eqref{eq:redshift_space_kernel} and \eqref{eq:F_ell_GR} for the Newtonian and general relativistic treatments of RSDs, respectively, get an additional term
\ba
\label{eq:F_ell_modification}
    F_\ell(k,r) &\,= F_\ell^{G}(k,r) + b_\Phi(r) f_\text{NL}  \mathcal T_\delta^{-1}(k,r) j_\ell(kr)\,, \vs
    &\, = F_\ell^{G}(k,r) + 3 \left[b_g(r) - 1 \right] f_\text{NL} \frac{\Omega_{m,0}\,H_0^2 \delta_c}{g(0) D(r)}  \frac{j_\ell(kr)}{k^2 T(k)}
\ea
where $F_\ell^G$ is the Gaussian component of the kernel (i.e., $f_\text{NL} = 0$) and the second line, of course, is under the assumption of mass function universality. We can see the effect of this additional component of the redshift-space kernel in \reffig{fell_fig}.

One thing important to note is that while we assume the functional form of $b_\Phi$, in general, such an expression may not be exact \cite{Reid_2010_bphi1,Biagetti_2017_bphi2,Barreira_2020_bphi3,Voivodic_2021_bphi4,Barreira_2022_bphi5,Barreira_2022_bphi6,Lazeyras_2023_bphi7,barreira2023_bphi8}. Of course, without knowing (or assuming) the form of this bias parameter, the only form of primordial non-Gaussianity detectable from two-point statistics would be the degenerate combination $b_\Phi f_\text{NL}$. In the following sections, we use \refeq{b_phi_universal} in an effort to make a forecast for $f_\text{NL}$, as opposed to the degenerate alternative.

\subsection{The shape of density kernel {$F_\ell(k,z)$}}\label{sec:fell_relativistic}
\begin{figure}[h]
    \centering
    \includegraphics[width=0.93\linewidth]{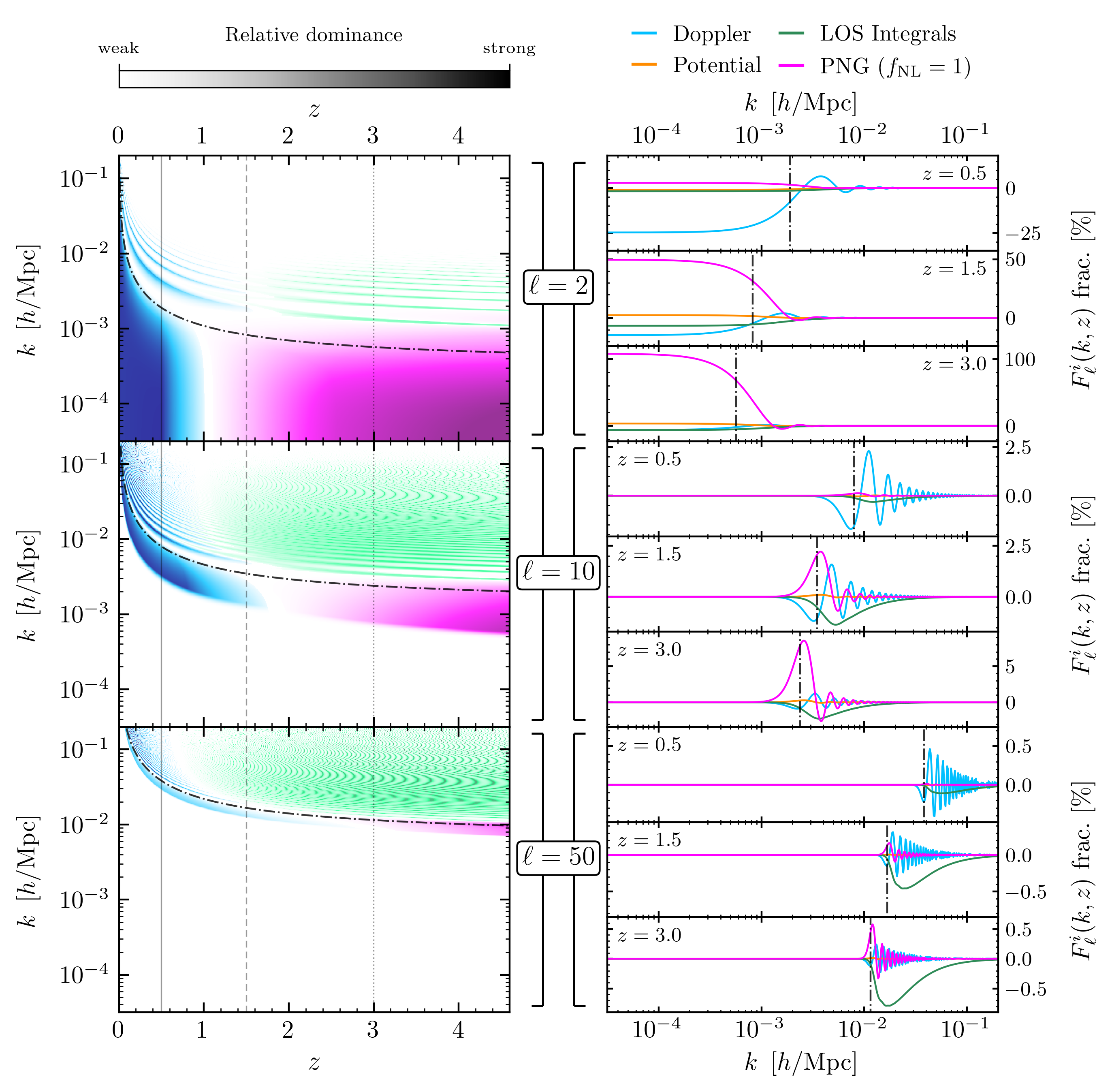}
    \caption{
    Comparison of the relativistic contributions to $F_\ell(k,z)$
    in Eq.~\eqref{eq:F_ell_GR}, including primordial non-Gaussianity (Eq.~\eqref{eq:F_ell_modification}) with
    $f_{\rm NL}=1$. Left: dominant contribution in the $(k,z)$
    plane, excluding density and Kaiser RSD terms. Right: relative amplitudes
    of the Doppler (scaling with $\mathcal B$, blue), local potential (goes as $\mathcal A$, orange),
    integrated photon geodesic (e.g., lensing; green), and PNG (magenta) contributions at $z=\{0.5,1.5,3\}$, indicated by the grey lines in the left panels.
    Rows show $\ell=2$, $10$, and $50$. Color shading indicates the degree of
    dominance. For both panels, the black dot-dashed lines mark the Limber scale $k=(\ell+1/2)/r(z)$.
    }
    \label{fig:fell_fig}
\end{figure}

The relativistic redshift-space kernels of \refeq{F_ell_GR} (Gaussian) and \refeq{F_ell_modification} (local PNG) extend the Newtonian form of \refeq{redshift_space_kernel} in two structural ways: they promote the velocity prefactor from $\alpha$ to $\mathcal{B}$, and they add a potential term together with the photon-geodesic line-of-sight terms (ISW, Shapiro time delay, and lensing) that the Newtonian derivation omits. Excluding the density and Kaiser RSD terms, \reffig{fell_fig} shows the remaining contributions to $F_\ell(k,z)$: the velocity term ($\propto\mathcal{B}$), the potential term ($\propto\mathcal{A}$), the line-of-sight integrals, and the PNG term. For the numerical parameters, we use the SPHEREx-like mock cases defined in Sec.~\ref{sec:Q_choice}.

The left panels of \reffig{fell_fig} show which term dominates across the $(k,z)$ plane at $\ell=2$, $10$, and $50$ (from top to bottom); the right panels give the fractional amplitudes at $z=0.5$, $1.5$, and $3$. The potential and PNG terms scale as $1/k^2$ and the velocity term as $1/k$, so all three rise toward small $k$, with the $1/k^2$ pair steepest. The spherical Bessel function $j_\ell(kr)\sim(kr)^\ell$ suppresses each one for $kr\lesssim\ell$; the resulting large-scale cutoff sits at $k\sim\ell/r$, coincident with the Limber scale $k=(\ell+1/2)/r$ \cite{LoVerde::2008limber} (dot-dashed), and moves to larger $k$ at higher $\ell$ and lower $z$.

Near the Limber scale, large-scale dominance shifts from the velocity term to PNG with increasing redshift at every multipole. At $\ell=2$ the right panels place the crossover between the $z=0.5$ and $z=1.5$ slices: the dominant color runs from blue at $z=0.5$ to magenta at $z=1.5$ and $z=3$. The crossover is not fixed; it marks where the velocity and PNG amplitudes balance, so it moves to lower redshift as $f_{\rm NL}$ or $(b_g-1)$ increases, and tracks the redshift dependence of $\mathcal{B}$ through $b_e$ and $\mathcal{Q}$.

The potential term carries the dominant color at no multipole. The velocity term overtakes it through the kernel ratio $(\mathcal{B}/\mathcal{A})(k/\mathcal{H})(j_\ell'/j_\ell)$, which carries $j_\ell'/j_\ell\simeq\ell/(kr)$ at small argument and grows as $\ell/(\mathcal{H}r)$ toward small $r$, while the shared $1/(\mathcal{H}r)$ in both prefactors cancels in $\mathcal{B}/\mathcal{A}$. PNG overtakes it as well: at large scales $T(k)\to1$, the two share the kernel $j_\ell(kr)/k^2$, and their ratio reduces to $3(b_g-1)f_{\rm NL}\delta_c/(g(0)\mathcal{A})$, the Poisson factor $\Omega_{m,0}H_0^2/\mathcal{H}^2\propto a$ canceling the $1/D(r)$ growth in matter domination so that only $\mathcal{A}(z)$ carries the redshift dependence. Since we know that $3\delta_c/g(0)\simeq 6$, the ratio only exceeds unity once $(b_g-1)\gtrsim g(0)\mathcal{A}/(3\delta_c f_{\rm NL})\sim0.15/f_{\rm NL}$. The potential is not parametrically small ($\mathcal{A}\sim\mathcal{O}(1)$, \reffig{AB_fig}); both ratios hold at every multipole, and the dominance map shows their effect where the line-of-sight terms are weak, at low $\ell$. In other words, while it remains important, it is simply exceeded by the PNG signal.

The line-of-sight integrals occupy the opposite corner. The transverse Laplacian acts on the angular modes as $-\ell(\ell+1)/r^2$, so these terms scale as $\ell(\ell+1)$: negligible at $\ell=2$, comparable to the velocity term at $\ell=10$, and dominant at $\ell=50$ (green). Unlike the local terms, which carry a single $j_\ell(kr)$ and cut off for $kr\lesssim\ell$, each integral runs over $j_\ell(kr')$ along the line-of-sight and peaks where its first oscillation overlaps the projection kernel, at the Limber scale $k\sim\ell/r(z)$. They strengthen toward high redshift, where the path lengthens and the lensing efficiency spans a wider range.

The shape dependence of $F_\ell$ examined so far gives a glimpse of how the relativistic contributions bear on a local $f_{\rm NL}$ measurement. In the case of auto-correlation, the potential shares the PNG kernel $j_\ell(kr)/k^2$ at large scales, so neglecting it biases the inferred $f_{\rm NL}$. Cross-correlation, however, weights the product of two redshift sections centered at $z_i$ and $z_j$. Therefore, terms localized in $k$ contribute only where the two $j_\ell(kr)$ share phase, on scales above both Bessel turnovers, $k\lesssim\ell/\max(r_i,r_j)$; at smaller scales, the two sections oscillate out of step and their product cancels. At $\ell=2$, this aligned range is wide and the large-scale PNG of a $z_i\neq z_j$ pair persists; at higher $\ell$ it narrows, the dominant terms of the two sections differ, and the PNG product cancels. The line-of-sight integrals, set by the path rather than a single shell, persist across redshift regardless, so the cross-correlated signal carries a galaxy-lensing contribution alongside PNG. Their relative size and sign do not follow from the single-redshift kernel; we compute them, and the bias from ignoring the relativistic contributions, in Sec.~\ref{sec:bias_fnl}.

\section{PowerFull: a relativistic extension of the 2-FAST algorithm}\label{sec:power_full}

The 2-FAST algorithm \cite{Gebhardt::2018twofast} achieves remarkable speed and accuracy in evaluating overlap integrals involving one or two spherical Bessel functions. It employs the FFTLog transformation \cite{Hamilton::2000FFTlog}, which reformulates the highly oscillatory integrals into analytic expressions, using Gamma functions for single-Bessel cases and hypergeometric functions ${}_2F_1$ for double-Bessel cases, whose backward recursion relations ensure numerical stability. The algorithm efficiently transforms the galaxy power spectrum $P(k)$ into configuration space via the two-point correlation function $\xi(r)$ and into harmonic space via the angular power spectrum $C_\ell(r,r')$.

Earlier, we showed that relativistic effects contribute proportionally to ${\cal A}(k/aH)^{-2}$ and ${\cal B}(k/aH)^{-1}$, both of which are significant on scales near the horizon ($k_H \simeq aH$). A faithful statistical characterization of cosmological observables on these large scales requires incorporating the underlying spherical geometry, for which the harmonic-space angular power spectrum $C_{\ell}(r,r')$ provides the most natural framework.

In this section, we first define the angular power spectrum and identify the family of integrals it requires (\refsec{Cl}), then present the explicit expression for the full relativistic angular power spectrum (\refsec{full_GR_derivation}). A review of the 2-FAST algorithm and our method for evaluating integrals involving odd-order derivatives of spherical Bessel functions is given in \refapp{twofast_review}. We name the code package {\textsc{PowerFull}}, as it computes the full relativistic angular power spectrum.

\subsection{The angular power spectrum}\label{sec:Cl}

In \refsec{gal_dens}, we presented the relativistic effects at linear order -- which arise from the generalization of redshift-space distortion effects by following the light geodesic from the source to the observer -- on the observed density contrast across the full sky. The final expression \refeq{deltas_GR_TAM} for the observed galaxy density contrast may be written as
\be
\delta_s(\bfx) =  \sum_{\ell m} \int \frac{k^2 \dd k}{(2\pi)^3}\delta_{\ell m}^k 4\pi i^\ell Y_{\ell m}(\nhat)  D(r)  F_\ell(k,r)\,,
\label{eq:deltas}
\ee
with the relativistic kernel $F_{\ell}(k,r)$, where \refeq{F_ell_modification} shows its full form including the scale-dependent bias from the local-type primordial non-Gaussianities.
Here, we use the TAM or Fourier-Bessel basis because they reflect the correct underlying geometry. The summary statistics we focus on in this work are the angular power spectra of the observed density field. Specifically, we consider a galaxy survey where galaxies can be divided into tomographic redshift slices, and compute all auto and cross angular power spectra in \refsec{window}.

As a first step, however, we begin with expanding the observed density contrast $\delta_s$ in \refeq{deltas} into the spherical-harmonics coefficients,
\be
\label{eq:alm}
    a_\lm(r) = \int \dd^2 \nhat \,\delta_s(\bfx) Y^*_\lm(\nhat)\,.
\ee
The angular power spectrum, $C_\ell(r,r')$, is then defined as
\ba
\label{eq:cl_def}
    C_\ell(r,r') &\,\equiv \left<a_\lm(r) a_\lm^*(r') \right> = D(r)D(r')\frac{2}{\pi} \int \dd k \, k^2 P(k) F_\ell(k,r) F_\ell(k,r')\,.
\ea
In our final expression for the kernels \refeq{F_ell_modification} and \refeq{F_ell_GR}, the relativistic kernels $F_\ell(k,r)$ consist of various combinations of $j_\ell$, $j_\ell'$, and $j_\ell''$, multiplied by powers of the Fourier wavenumber, $k^p$, with $p=(0,-1,-2)$. The computation of the angular power spectrum, therefore, requires evaluating the following family of integrals:
\be
\label{eq:wljj'}
    w_{\ell,jj'}^p(r,r') = \frac{2}{\pi}\int \dd k\,k^{2+p} P(k) j_\ell^{(j)}(kr) j_\ell^{(j')}(kr')\,.
\ee
The methods for evaluating these integrals are presented in \refapp{twofast_review}; here, we focus on the physical content of the angular power spectrum.

\subsection{Newtonian angular power spectrum}\label{sec:ClN}
Using the Newtonian redshift-space kernel in \refeq{redshift_space_kernel} with $\alpha = 0$, the Kaiser angular power spectrum takes the form
\ba
\label{eq:kaiser_Cl}
    C_\ell^\text{K}(r_1,r_2) = b_{g,1} b_{g,2} D_1 D_2  &\,\Big( w_{\ell,00}^0 - \beta_1 w_{\ell,20}^{0} -\beta_2 w_{\ell,02}^0 + \beta_1 \beta_2 w_{\ell,22}^0\Big)\,,
\ea
where $\beta_i \equiv f(r_i)/b_g(r_i)$ is the usual RSD amplitude, $D_i = D(r_i)$, and $b_{g,i} = b_g(r_i)$.

\begin{figure}[t!]
    \centering
    \includegraphics[width=0.93\linewidth]{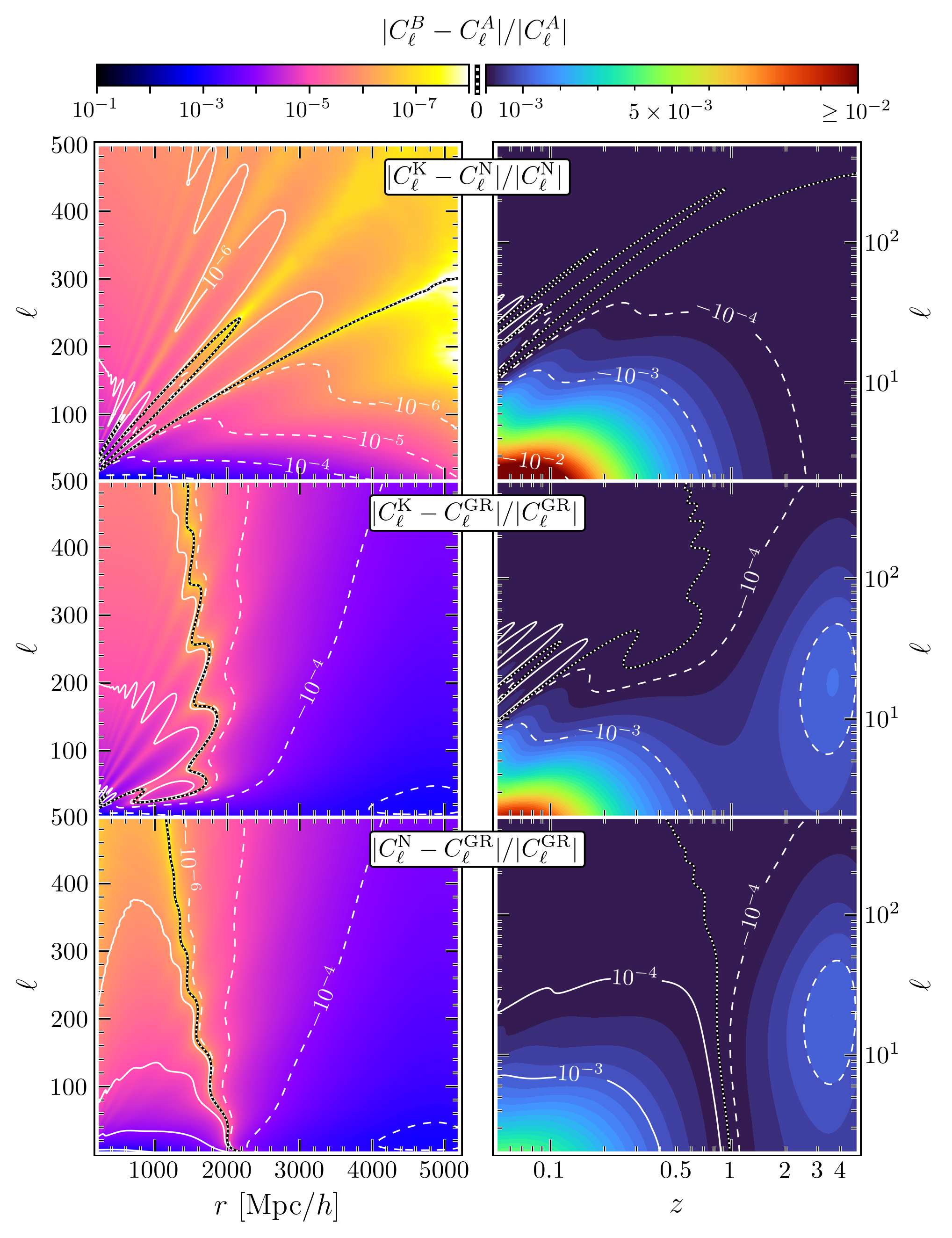}
    \caption{The relative differences between the Kaiser, Newtonian, and full relativistic -- given in Eq.~\eqref{eq:kaiser_Cl}, \eqref{eq:newtonian_Cl}, and \eqref{eq:Cl_GR_expanded}, respectively -- auto-correlated angular power spectra evaluated with $f_\text{NL} = 0$. \textit{Left}: The log-scaled differences plotted as $\ell$ versus $r$, the comoving distance. \textit{Right}: Linear-scaled relative differences, but this time plotted as $\log \ell$ versus $\log z$. Signed difference contours (i.e., no absolute value) are plotted, as well as the zero contour (the dotted black and white line). }
    \label{fig:gr_newt_fig}
\end{figure}

Including the full Newtonian kernel from \refeq{redshift_space_kernel}, the angular power spectrum becomes
\ba
\label{eq:newtonian_Cl}
    C_\ell^\text{N}(r_1,r_2) = b_{g,1} b_{g,2} D_1 D_2 \Big ( &\, w^0_{\ell,00} - \beta_1 w^0_{\ell,20} - \beta_2 w^0_{\ell,02}   + \beta_1 \beta_2 w^{0}_{\ell,22} \vs
    &   - \beta_1 \frac{\alpha_1}{r_1} w^{-1}_{\ell,10}
     - \beta_2 \frac{\alpha_2}{r_2} w^{-1}_{\ell,01}   + \beta_1 \beta_2 \frac{\alpha_1}{r_1} w^{-1}_{\ell,12}     \vs
    &    + \beta_1 \beta_2 \frac{\alpha_2}{r_2} w^{-1}_{\ell,21}  +  \beta_1 \beta_2 \frac{\alpha_1}{r_1}\frac{\alpha_2}{r_2} w^{-2}_{\ell,11}\Big)\,,
\ea
where we have 5 new terms proportional to $\alpha_i = \alpha(r_i)$, the quantity present in the Newtonian derivation that contains information on the radial evolution of cosmological parameters across the survey volume. If we also factor in the impact of local primordial non-Gaussianity on the redshift-space kernel in \refeq{F_ell_modification}, we see that the Newtonian angular power spectrum in \refeq{newtonian_Cl} becomes
\ba
\label{eq:newtonian_Cl_fnl}
    C_\ell^{\text{N,PNG}}(r_1,r_2) =  C^{\text{N}}_\ell(r_1,r_2)  & \,+ \frac{3}{2} H_0^2 \Omega_{m,0} D_2 f_\text{NL} b_{\Phi,1} b_{g,2}   \Big( u_{\ell,00}^{-2} - \beta_2 u_{\ell,02}^{-2} - \beta_2 \frac{\alpha_2}{r_2} u_{\ell,01}^{-3}\Big) \vs
    & \,+ \frac{3}{2} H_0^2 \Omega_{m,0} D_1 f_\text{NL}  b_{\Phi,2} b_{g,1}  \Big(  u_{\ell,00}^{-2} - \beta_1 u_{\ell,20}^{-2} - \beta_1 \frac{\alpha_1}{r_1} u_{\ell,10}^{-3}\Big) \vs
    & \,+ \frac{9}{4} H_0^4 \Omega_{m,0}^2 f_\text{NL}^2 b_{\Phi,1}b_{\Phi,2}  v_{\ell,00}^{-4} \,,
\ea
where we use $b_{\Phi,i} = b_\Phi(r_i)$ and have defined new integral terms necessary for integrating a combination of the transfer function, $T(k)$, and the power spectrum, $P(k)$:
\ba
\label{eq:uljj'}
    u_{\ell,jj'}^p(r,r') &\,= \frac{2}{\pi}\int \dd k\,k^{2+p} \,T^{-1}(k) P(k) j_\ell^{(j)}(kr) j_\ell^{(j')}(kr')\,, \\
\label{eq:vljj'}
    v_{\ell,jj'}^p(r,r') &\,= \frac{2}{\pi}\int \dd k\,k^{2+p} \,T^{-2}(k) P(k) j_\ell^{(j)}(kr) j_\ell^{(j')}(kr')\,.
\ea
The use of $p=-2$ with $u_{\ell,jj'}^p$ in \refeq{newtonian_Cl_fnl} reflects the scale-dependence $\propto k^{-2}$ present in the local primordial non-Gaussianity component of the redshift-space kernel defined in \refeq{F_ell_modification}, while the $v_{\ell,jj'}^p$ term with $p=-4$ handles the auto-correlation of the aforementioned non-Gaussianity component.

\subsection{Full relativistic angular power spectrum}\label{sec:full_GR_derivation}

The full relativistic angular power spectrum, $C_\ell^{\text{GR}}$, involves all pairwise cross-correlations among the contributions listed in \refeq{F_ell_GR} and \refeq{F_ell_modification}. The number of terms scales as the square of the number of kernel contributions: real space has one, the Kaiser treatment has two (yielding four), the Newtonian treatment has three (yielding nine; cf.\ \refeq{newtonian_Cl}), and the relativistic kernel contains seven, producing 49 terms. Including primordial non-Gaussianity adds an eighth contribution, bringing the total to 64. We present the complete expression below without abbreviation; doing so requires several new integral quantities, which we define first.

The singly-integrated cross terms follow from \refeqs{I_isw}{I_kappa}:
\ba
\label{eq:s_jj'}
    s_{\ell,jj';r}^{p}(r,r') &\,\equiv \frac{3}{D(r)}\int_0^r \dd r^{\prime\prime} a^3(r^{\prime\prime})H^3(r^{\prime\prime})\Omega_m(r^{\prime\prime})[f(r^{\prime\prime})-1] D(r^{\prime\prime}) \,w_{\ell,jj'}^{p}(r^{\prime\prime},r')\,,
\\
\label{eq:t_jj'}
    t_{\ell,jj';r}^{p}(r,r') &\,\equiv \frac{3}{D(r)}\int _0^r \dd r^{\prime\prime} a^2(r^{\prime\prime})H^2(r^{\prime\prime})\Omega_m(r^{\prime\prime}) D(r^{\prime\prime}) \,w_{\ell,jj'}^{p}(r^{\prime\prime},r')\,,
\\
\label{eq:l_jj'}
    l_{\ell,jj';r}^{p}(r,r') &\,\equiv \frac{3}{2}\frac{\ell(\ell+1)}{D(r)} \int _0^r \dd r^{\prime\prime} \left(\frac{r-r^{\prime\prime}}{rr^{\prime\prime}}\right) a^2(r^{\prime\prime})H^2(r^{\prime\prime})\Omega_m(r^{\prime\prime})D(r^{\prime\prime}) \,w_{\ell,jj'}^{p}(r^{\prime\prime},r')\,,
\ea
where the addition of $r$ in the subscript indicates which comoving distance coordinate is integrated. To demonstrate the case of $r'$ as the integrated coordinate, we have
\ba
\label{eq:s_jj'_example}
    s_{\ell,jj';r'}^{p}(r,r') &\,\equiv \frac{3}{D(r')}\int_0^{r'} \dd r^{\prime\prime} a^3(r^{\prime\prime})H^3(r^{\prime\prime})\Omega_m(r^{\prime\prime})[f(r^{\prime\prime})-1] D(r^{\prime\prime}) \,w_{\ell,jj'}^{p}(r,r^{\prime\prime})
\ea
for the case of \refeq{s_jj'}. Clearly, we use the letter ``s" to denote quantities involving the integrated Sachs-Wolfe effect, the letter ``t" to handle Shapiro time delay, and ``l" for gravitational lensing.

We also define similar integrated quantities that arise from cross-correlation with the scale-dependent bias of the primordial non-Gaussianity term. Specifically, they are
\ba
\label{eq:s_jj'_png}
    \mathfrak s_{\ell,jj';r}^{p}(r,r') &\,\equiv \frac{3}{D(r)}\int_0^r \dd r^{\prime\prime} a^3(r^{\prime\prime})H^3(r^{\prime\prime})\Omega_m(r^{\prime\prime})[f(r^{\prime\prime})-1] D(r^{\prime\prime}) \,u_{\ell,jj'}^{p}(r^{\prime\prime},r')\,,
\\
\label{eq:t_jj'_png}
    \mathfrak t_{\ell,jj';r}^{p}(r,r') &\,\equiv \frac{3}{D(r)}\int _0^r \dd r^{\prime\prime} a^2(r^{\prime\prime})H^2(r^{\prime\prime})\Omega_m(r^{\prime\prime}) D(r^{\prime\prime}) \,u_{\ell,jj'}^{p}(r^{\prime\prime},r')\,,
\\
\label{eq:l_jj'_png}
    \mathfrak l_{\ell,jj';r}^{p}(r,r') &\,\equiv \frac{3}{2}\frac{\ell(\ell+1)}{D(r)} \int _0^r \dd r^{\prime\prime} \left(\frac{r-r^{\prime\prime}}{rr^{\prime\prime}}\right) a^2(r^{\prime\prime})H^2(r^{\prime\prime})\Omega_m(r^{\prime\prime})D(r^{\prime\prime}) \,u_{\ell,jj'}^{p}(r^{\prime\prime},r')\,,
\ea
where we recall the difference between $u^p_{\ell,jj'}$ and $w^p_{\ell,jj'}$ given in Eqs.~\eqref{eq:uljj'}, and $\eqref{eq:vljj'}$, respectively, is the presence of the transfer function $T(k)$ in the integral of the former.

Next, we have the relativistic auto-correlation terms
\ba
\label{eq:S_jj'}
    \mathcal S_{\ell,jj';r,r'}^{p}(r,r') &\,\equiv \frac{9}{D(r)D(r')}\int_0^r \dd r^{\prime\prime} \int_0^{r'}  \dd r^{\prime\prime\prime} \bigg\{  \mathcal H^3(r^{\prime\prime})\mathcal H^3(r^{\prime\prime\prime}) \Omega_m(r^{\prime\prime})\Omega_m(r^{\prime\prime\prime}) D(r^{\prime\prime}) D(r^{\prime\prime\prime}) \vs
    & \qquad \qquad \qquad \qquad \qquad \qquad \times [f(r^{\prime\prime})-1][f(r^{\prime\prime\prime})-1]   \,w_{\ell,jj'}^{p}(r^{\prime\prime},r^{\prime\prime\prime}) \bigg\}\,,
\\
\label{eq:T_jj'}
    \mathcal T_{\ell,jj';r,r'}^{p}(r,r') &\,\equiv \frac{9}{D(r)D(r')}\int_0^r \dd r^{\prime\prime} \int_0^{r'}  \dd r^{\prime\prime\prime} \bigg\{  \mathcal H^2(r^{\prime\prime})\mathcal H^2(r^{\prime\prime\prime}) \Omega_m(r^{\prime\prime})\Omega_m(r^{\prime\prime\prime}) D(r^{\prime\prime}) D(r^{\prime\prime\prime}) \vs
    & \qquad \qquad \qquad \qquad \qquad \qquad \qquad \times    w_{\ell,jj'}^{p}(r^{\prime\prime},r^{\prime\prime\prime}) \bigg\}\,,
\\
\label{eq:L_jj'}
    \mathcal L_{\ell,jj';r,r'}^{p}(r,r') &\,\equiv \frac{9}{4}\frac{\ell^2(\ell+1)^2}{D(r)D(r')} \int _0^r \dd r^{\prime\prime}\int_0^{r'}  \dd r^{\prime\prime\prime} \bigg\{ \left(\frac{r-r^{\prime\prime}}{rr^{\prime\prime}}\right) \left( \frac{r'-r^{\prime\prime\prime}}{r'r^{\prime\prime\prime}}\right)  \mathcal H^2(r^{\prime\prime}) \mathcal H^2(r^{\prime\prime\prime}) \vs
    & \qquad \qquad \qquad \qquad \qquad \times    \Omega_m(r^{\prime\prime}) \Omega_m(r^{\prime\prime\prime}) D(r^{\prime\prime}) D(r^{\prime\prime\prime}) \,w_{\ell,jj'}^{p}(r^{\prime\prime},r^{\prime\prime\prime}) \bigg\}\,,
\ea
where the notation $r,r'$ in the subscript now indicates which integrand each coordinate is assigned to, and we use the conformal Hubble parameter $\mathcal H = aH$. We see that reversing the integration order for arbitrary $j,j'$ yields the following symmetry,
\ba
\label{eq:symmetry_of_STL}
    \{\mathcal S,\mathcal T,\mathcal L\}_{\ell,jj';r',r}^p(r,r') = \{\mathcal S,\mathcal T,\mathcal L\}_{\ell,j'j;r,r'}^p(r,r')\,.
\ea
The version of these integrals that is used in the full relativistic expression in Eq.~\eqref{eq:Cl_GR_expanded} utilizes $j=j' = 0$, so the integration order does not matter.

Lastly, for the cross-correlation among the integrated terms, we have
\ba
\label{eq:X_jj'}
    \mathcal X_{\ell,jj';r,r'}^{p}(r,r') &\,\equiv  \frac{9}{D(r)D(r')}\int_0^r  \dd r^{\prime\prime}\int_0^{r'} \dd r^{\prime\prime\prime} \bigg\{ \mathcal H^3(r^{\prime\prime}) \mathcal H^2(r^{\prime\prime\prime}) \Omega_m(r^{\prime\prime}) \Omega_m(r^{\prime\prime\prime}) D(r^{\prime\prime}) D(r^{\prime\prime\prime}) \vs
    & \qquad \qquad \qquad \qquad  \qquad \qquad \qquad \times   [f(r^{\prime\prime})-1]  \,w_{\ell,jj'}^p(r^{\prime\prime},r^{\prime\prime\prime})\bigg\}\,,
\\
\label{eq:Y_jj'}
    \mathcal Y_{\ell,jj';r,r'}^{p}(r,r') &\,\equiv  \frac{9}{2}\frac{\ell(\ell+1)}{D(r)D(r')}\int_0^r \dd r^{\prime\prime} \int_0^{r'}  \dd r^{\prime\prime\prime} \bigg\{ \left(\frac{r'-r^{\prime\prime\prime}}{r'r^{\prime\prime\prime}}\right) \mathcal H^3(r^{\prime\prime}) \mathcal H^2(r^{\prime\prime\prime}) \Omega_m(r^{\prime\prime}) \Omega_m(r^{\prime\prime\prime}) \vs
    & \qquad \qquad \qquad \qquad  \qquad \qquad \times  D(r^{\prime\prime}) D(r^{\prime\prime\prime}) [f(r^{\prime\prime})-1]  \,w_{\ell,jj'}^p(r^{\prime\prime},r^{\prime\prime\prime})\bigg\}\,,
\\
\label{eq:Z_jj'}
    \mathcal Z_{\ell,jj';r,r'}^{p}(r,r') &\,\equiv \frac{9}{2}\frac{\ell(\ell+1)}{D(r)D(r')}\int_0^r \dd r^{\prime\prime} \int_0^{r'}  \dd r^{\prime\prime\prime} \bigg\{ \left(\frac{r'-r^{\prime\prime\prime}}{r' r^{\prime\prime\prime}}\right) \mathcal H^2(r^{\prime\prime}) \mathcal H^2(r^{\prime\prime\prime}) \Omega_m(r^{\prime\prime}) \Omega_m(r^{\prime\prime\prime}) \vs
    & \qquad \qquad \qquad \qquad \qquad \qquad \qquad \times   D(r^{\prime\prime}) D(r^{\prime\prime\prime}) \,w_{\ell,jj'}^p(r^{\prime\prime},r^{\prime\prime\prime})\bigg\}\,.
\ea
Here, it is clear that $\mathcal X$ involves the ISW effect and time delay, $\mathcal Y$ involves the ISW effect and lensing, and $\mathcal Z$ incorporates time delay and lensing. In this case, the integration order is important, as it dictates which effect is applied to each coordinate in $(r,r')$. For instance, since $\mathcal X_{\ell,00;r,r'}^{-4}(r,r')$ encodes ISW $\times$ time delay, its counterpart $\mathcal X_{\ell,00;r',r}^{-4}(r,r')$ gives time delay $\times$ ISW.

\begin{figure}[p!]
    \centering
    \includegraphics[width=0.99\linewidth]{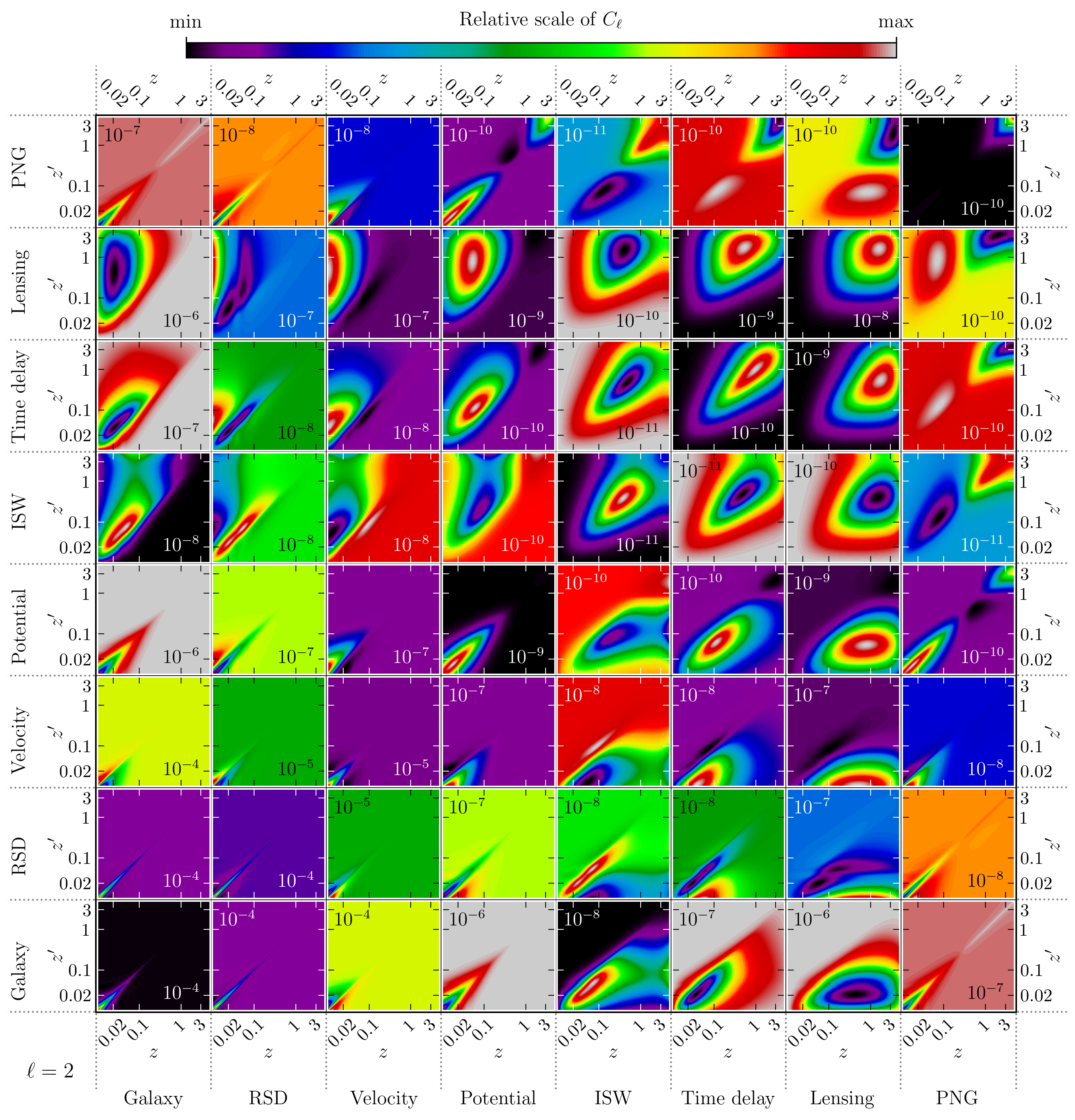}
    \caption{Each term in the full relativistic power spectrum defined in \refeq{Cl_GR_expanded} at $\ell = 2$ using 1000 Dirac delta bins from $z=0$ to 5, computed with $f_\text{NL} = 1$ so that the terms carrying the primordial non-Gaussianity kernel do not vanish. In order to fully illustrate the features of each term, the colors range from the minimum to maximum of each plot. Then, to provide a sense of absolute scale, the average order of magnitude is shown for each term. As expected, the terms are symmetric across the diagonal (auto-correlation) terms. The eight different effects present in the observed galaxy density contrast are plotted. In order as they appear in \refeq{F_ell_GR} and on the plot: the real space galaxy density contrast, the first-order redshift-space distortion, the additional peculiar velocity term from the Newtonian expression, the potential term, the ISW effect, Shapiro time delay, and gravitational lensing. Furthermore, the PNG term added in \refeq{F_ell_modification} comprises the final quantity plotted. Obviously, the galaxy-galaxy (bottom left) component of the angular power spectrum dominates, followed by the RSD-galaxy and RSD-RSD terms. While the auto-correlation of the relativistic terms has some diagonal signal, the relativistic effects dominate in the cross-correlations. Additionally, the ISW, time delay, and lensing terms are larger at farther redshifts; this makes sense, as they are integrated effects. Lastly, we see that the PNG signal dominates in the galaxy-PNG and RSD-PNG terms, particularly at low redshifts. Due to this, galaxy surveys aiming at probing the PNG, such as SPHEREx, will gain most of their signal on $f_\text{NL}$ at lower redshifts.}
    \label{fig:cl_2_grid}
\end{figure}

Collecting all terms, the full relativistic angular power spectrum is
\ba
\label{eq:Cl_GR_expanded}
    C_\ell^{\text{GR}}(r_1,r_2) = D_1 D_2 \Bigg\{ &\, b_{g,1} b_{g,2} \big(w_{\ell,00}^0 - \beta_1 w_{\ell,20}^{0} - \beta_2 w_{\ell,02}^0 + \beta_1 \beta_2 w_{\ell,22}^0 \big) \vs
    &\,  + b_{g,1}\, \mathcal H_2 \mathcal B_2 \big( w_{\ell, 01}^{-1}  -  \beta_1 w_{\ell,21}^{-1} \big) + b_{g,2}\, \mathcal H_1 \mathcal B_1 \big( w_{\ell, 10}^{-1}  -  \beta_2 w_{\ell,12}^{-1} \big) \vs
    &\,  + b_{g,1} \,\mathcal H_2^2 \mathcal A_2 \bigg( w_{\ell,00}^{-2}  - \beta_1 w_{\ell,20}^{-2} +  \frac{\mathcal H_1}{b_{g,1}} \mathcal B_1 w_{\ell,10}^{-3} \bigg) \vs
    &\,  + b_{g,2} \,\mathcal H_1^2 \mathcal A_1 \bigg( w_{\ell,00}^{-2} - \beta_2 w_{\ell,02}^{-2} +  \frac{\mathcal H_2}{b_{g,2}} \mathcal B_2 w_{\ell,01}^{-3} \bigg) \vs
    &\,  + b_{g,1} \frac{\mathcal B_2}{\beta_2} \bigg( s_{\ell,00;r_2}^{-2}  - \beta_1  s_{\ell,20;r_2}^{-2}   +  \frac{\mathcal H_1}{b_{g,1}} \mathcal B_1  s_{\ell,10;r_2}^{-3} +  \frac{\mathcal H_1^2}{b_{g,1}}   \mathcal A_1 s_{\ell,00;r_2}^{-4} \bigg)  \vs %
    &\,  + b_{g,2}\frac{\mathcal B_1}{\beta_1} \bigg( s_{\ell,00;r_1}^{-2}   - \beta_2 s_{\ell,02;r_1}^{-2}  + \frac{\mathcal H_2}{b_{g,2}} \mathcal B_2  s^{-3}_{\ell,01;r_1}  +   \frac{\mathcal H_2^2}{b_{g,2}}  \mathcal A_2 s_{\ell,00;r_1}^{-4}  \bigg)  \vs %
    &\,  - 2b_{g,1}\frac{(1-\mathcal Q_2)}{r_2} \left(  t_{\ell,00;r_2}^{-2}  - \beta_1 t_{\ell,20;r_2}^{-2} + \frac{\mathcal H_1}{b_{g,1}}\mathcal B_1 t_{\ell,10;r_2}^{-3}  + \frac{\mathcal H_1^2}{b_{g,1}}  \mathcal A_1 t_{\ell,00;r_2}^{-4}\right) \vs
    &\,  - 2b_{g,2}\frac{(1-\mathcal Q_1)}{r_1} \bigg(t_{\ell,00;r_1}^{-2}  - \beta_2 t_{\ell,02;r_1}^{-2} + \frac{\mathcal H_2}{b_{g,2}} \mathcal B_2 t_{\ell,01;r_1}^{-3} + \frac{\mathcal H_2^2}{b_{g,2}} \mathcal A_2 t_{\ell,00;r_1}^{-4} \bigg) \vs
    &\,  - 2b_{g,1}(1-\mathcal Q_2) \left(l_{\ell,00;r_2}^{-2}  - \beta_1 l_{\ell,20;r_2}^{-2}  + \frac{\mathcal H_1}{b_{g,1}} \mathcal B_1 l_{\ell,10;r_2}^{-3} + \frac{\mathcal H_1^2}{b_{g,1}} \mathcal A_1 l_{\ell,00;r_2}^{-4}\right) \vs
    &\,  - 2b_{g,2} (1-\mathcal Q_1) \left(l_{\ell,00;r_1}^{-2} - \beta_2 l_{\ell,02;r_1}^{-2}  + \frac{\mathcal H_2}{b_{g,2}}\mathcal B_2 l_{\ell,01;r_1}^{-3}   + \frac{\mathcal H_2^2}{b_{g,2}} \mathcal A_2 l^{-4}_{\ell,00;r_1}\right) \vs   
    &\,  - 2(1-\mathcal Q_2) \bigg[\frac{\mathcal B_1}{f_1 r_2} \mathcal X_{\ell,00;r_1,r_2}^{-4} + \frac{\mathcal B_1}{f_1} \mathcal Y_{\ell,00;r_1,r_2}^{-4} - 2\frac{(1-\mathcal Q_1)}{r_2} \mathcal Z_{\ell,00;r_2,r_1}^{-4}\bigg] \vs
    &\,  - 2(1-\mathcal Q_1) \bigg[\frac{\mathcal B_2}{f_2 r_1} \mathcal X_{\ell,00;r_2,r_1}^{-4} + \frac{\mathcal B_2}{f_2} \mathcal Y_{\ell,00;r_2,r_1}^{-4} - 2\frac{(1-\mathcal Q_2)}{r_1} \mathcal Z_{\ell,00;r_1,r_2}^{-4}\bigg] \vs
    &\,  + \frac{3}{2} \frac{H_0^2 \Omega_{m,0}}{g(0)}  b_{g,1} \frac{b_{\Phi,2}}{D_2} f_\text{NL} \bigg(u_{\ell,00}^{-2} - \beta_1 u_{\ell,20}^{-2} + \frac{\mathcal H_1}{b_{g,1}} \mathcal B_1 u_{\ell,10}^{-3} + \frac{\mathcal H_1^2}{b_{g,1}} \mathcal A_1 u_{\ell,00}^{-4} \bigg) \vs
    &\,  + \frac{3}{2} \frac{H_0^2 \Omega_{m,0}}{g(0)} b_{g,2} \frac{b_{\Phi,1}}{D_1} f_\text{NL}\bigg(u_{\ell,00}^{-2} - \beta_2 u_{\ell,02}^{-2} + \frac{\mathcal H_2}{b_{g,2}} \mathcal B_2 u_{\ell,01}^{-3} + \frac{\mathcal H_2^2}{b_{g,2}} \mathcal A_2 u_{\ell,00}^{-4} \bigg) \vs
    &\,  + \frac{3}{2} \frac{H_0^2 \Omega_{m,0}}{g(0)} \frac{b_{\Phi,1}}{D_1} f_\text{NL} \bigg[\frac{\mathcal B_2}{f_2} \mathfrak s_{\ell,00;r_2}^{-4} - 2(1-\mathcal Q_2)\bigg(\frac{1}{r_2} \mathfrak t_{\ell,00;r_2}^{-4} +  \mathfrak l_{\ell,00;r_2}^{-4}\bigg) \bigg] \vs
    &\,  + \frac{3}{2} \frac{H_0^2 \Omega_{m,0}}{g(0)} \frac{b_{\Phi,2}}{D_2} f_\text{NL} \bigg[\frac{\mathcal B_1}{f_1} \mathfrak s_{\ell,00;r_1}^{-4} - 2(1-\mathcal Q_1)\bigg(\frac{1}{r_1} \mathfrak t_{\ell,00;r_1}^{-4} +  \mathfrak l_{\ell,00;r_1}^{-4}\bigg) \bigg] \vs
    &\,  + \mathcal H_1 \mathcal H_2 \mathcal B_1 \mathcal B_2 w_{\ell,11}^{-2} + \mathcal H_1^2 \mathcal H_2^2 \mathcal A_1 \mathcal A_2 w_{\ell,00}^{-4} + \frac{\mathcal B_1}{f_1} \frac{\mathcal B_2}{f_2} \mathcal S_{\ell,00;r_1,r_2}^{-4} \vs
    &\,+ 4\frac{(1-\mathcal Q_1)}{r_1}\frac{(1-\mathcal Q_2)}{r_2} \mathcal T_{\ell,00;r_1,r_2}^{-4} + 4(1-\mathcal Q_1) (1-\mathcal Q_2) \mathcal L_{\ell,00;r_1,r_2}^{-4} \vs
    &\,+ \frac{9}{4} \frac{H_0^4 \Omega_{m,0}^2}{g(0)^2} \frac{b_{\Phi,1}}{D_1}\frac{b_{\Phi,2}}{D_2} f_\text{NL}^2   v_{\ell,00}^{-4} \Bigg \} \,,
\ea
where subscripts $1$ and $2$ denote evaluation at $r_1$ and $r_2$, respectively, and $\mathcal H_i \equiv a_i H_i$. We implement the full expression in \refeq{Cl_GR_expanded} within {\textsc{PowerFull}}. Each integral class ($w$, $u$, $v$, and the singly- and doubly-integrated quantities $s$, $t$, $l$, $\mathcal S$, $\mathcal T$, $\mathcal L$, $\mathcal X$, $\mathcal Y$, $\mathcal Z$) requires a separate choice of the biasing parameter $q$ defined in \refeq{biased_pk}; the values we adopt are collected in Table~\ref{tab:q_val_table}, and their selection is discussed in \refapp{q_optimization}. For each $(p,n)$ combination, we validated the $q$ value by comparing the 2-FAST output against the Lucas algorithm \cite{LUCAS1995269}. As a final check, we compare the output of {\textsc{PowerFull}} with \texttt{CLASSgal} \cite{DiDio::2013class} and find good agreement. \reffig{gr_newt_fig} compares the three treatments directly: it shows the relative differences between the Kaiser spectrum of \refeq{kaiser_Cl}, the Newtonian spectrum of \refeq{newtonian_Cl}, and the full relativistic spectrum of \refeq{Cl_GR_expanded}, for the auto-correlations at $f_\text{NL} = 0$, as functions of multipole and comoving distance.

\subsection{Implementation of window functions and selection effects}\label{sec:window}
\begin{figure}[t!]
    \centering
    \includegraphics[width=0.93\linewidth]{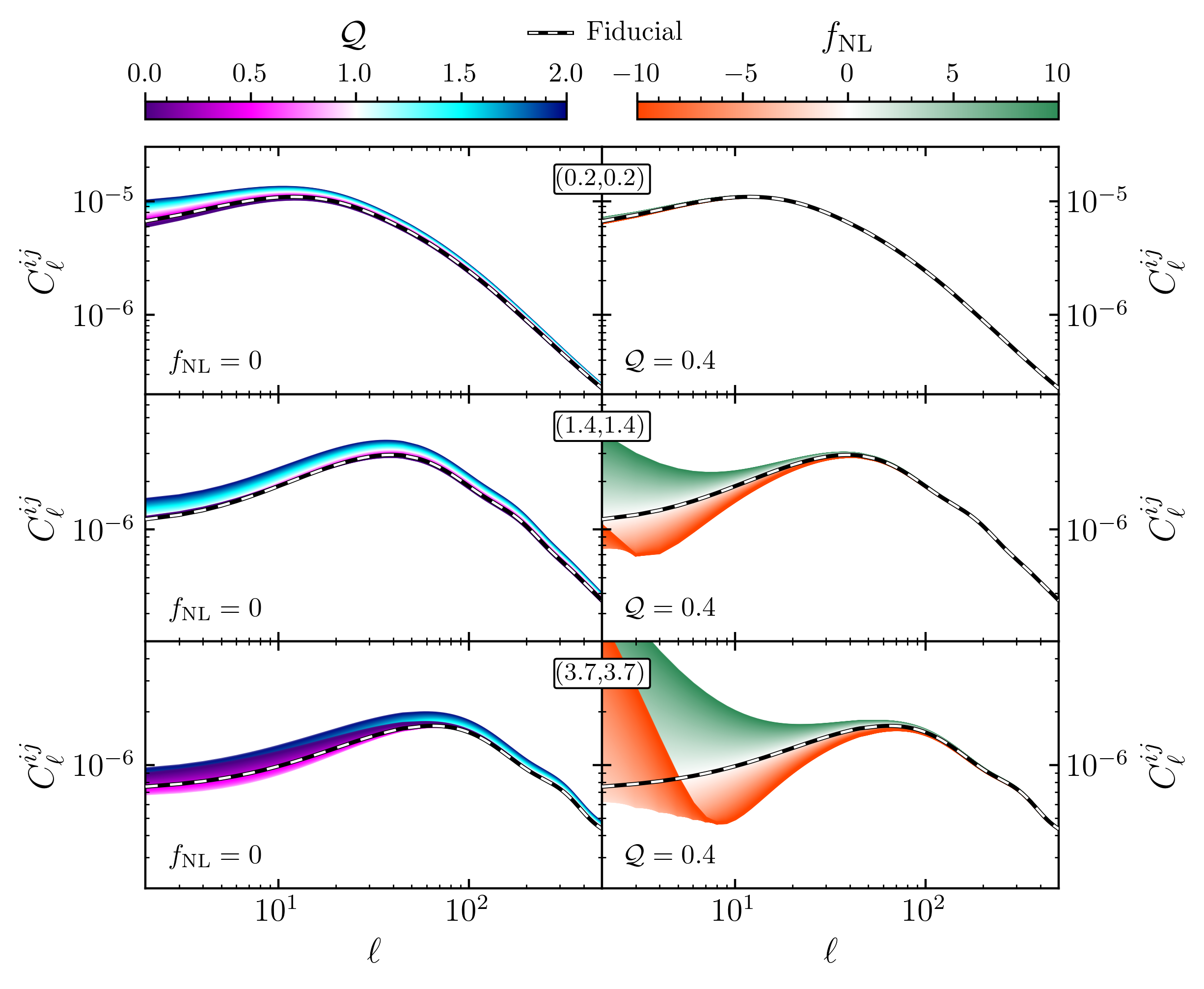}
    \caption{A comparison of three different diagonal components of the angular power spectrum, using top-hat window functions convolved with Gaussian redshift scatter as described in Eq.~\eqref{eq:top-hat_erf_windows}. As explained in \refsec{spherex}, the characteristic width of each of the five redshift bins based on SPHEREx sample 5, where $\sigma_z/(1+z) = 0.2$. Importantly, the left panels show the angular power spectrum, $C_\ell^{ii}$, at $z_i \sim0.2$, $1$, and $3.4$ (top to bottom) for varying values of \textit{constant} magnification bias, $\mathcal Q$, defined in \refeq{def_Q}, and a fixed $f_\text{NL} = 0$. The right panels show the same angular power spectra but using a fixed magnification bias $\mathcal Q = 0.4$ and a variable $f_\text{NL}$. We overplot the fiducial power spectrum (the dashed line), which uses \textit{both} $\mathcal Q = 0.4$ and $f_\text{NL} = 0$, as a point of comparison. These plots serve to illustrate the loose similarity of $\mathcal Q$ and $f_\text{NL}$ in modifying the angular power spectrum, particularly at low multipole.}
    \label{fig:Q_fnl_comp_plot}
\end{figure}

Analyzing galaxy clustering within tomographic redshift bins is still common practice, so it makes sense to add that capability to \textsc{PowerFull}. Such an implementation requires integrating the angular power spectrum $C_\ell(z,z')$ over the window function of each bin. Since 2-FAST evaluates $C_\ell(z,z')$ at fixed redshifts (i.e., infinitely thin bins), we perform this integration as a post-processing step: the 2-FAST output serves as the integrand, and the window and selection functions supply the weights. In principle, one could instead integrate over the hypergeometric function in $M_{\ell,\ell'}$ directly \cite{Gebhardt::2018twofast}; we do not pursue that route here. Below we describe how the window functions, radial selection functions, and shot noise enter the observed angular power spectrum.

For the $i$th redshift bin, the tomographic density contrast on the sky is
\be
\label{eq:delta_tomo}
    \delta_i(\nhat) = \int \dd z \, \phi(z) \, \mathcal W_i(z) \, \delta_s(z, \nhat)\,,
\ee
where $\mathcal W_i(z)$ is the window function of the bin and $\phi(z) \equiv \dd N/\dd z$ is the radial selection function. Expanding $\delta_s({\bm x})$ in the TAM basis as in \refeq{deltas}, the spherical-harmonic coefficients of $\delta_i$ inherit the windowed redshift-space kernel
\be
\label{eq:window_func}
    \mathcal F_\ell^{\,i}(k) = \int \dd z \,\phi(z) \mathcal W_i(z) D(z) F_\ell(k,z)\,,
\ee
with which we compute the binned angular power spectrum,
\ba
\label{eq:Cl_with_window_func}
    C_\ell^{\,ij} &\,= \frac{2}{\pi} \int \dd k\, k^2 P(k) \mathcal F^{\,i}_\ell(k) \mathcal F^{\,j}_\ell(k) \vs
    &\, = \frac{2}{\pi} \int \dd k \, k^2 P(k) \int \dd z \, \phi(z) \mathcal W_i(z) D(z) F_\ell(k,z) \int \dd z'\, \phi(z')\mathcal W_j(z') D(z') F_\ell(k,z') \vs
    &\, = \int \dd z \, \phi(z) \mathcal W_i(z) \int \dd z' \, \phi(z') \mathcal W_j(z') \, C_\ell(z,z')\,,
\ea
where $\mathcal W_i(z)\,\phi(z)$ is normalized to unity for each bin. The observed power spectrum also includes the Poisson shot noise,
\ba
\label{eq:observed_Cell}
    \bar C_\ell^{\,ij} &\, \equiv C_\ell^{\,ij} + N_\ell^{\,ij} 
\ea
where $N_\ell^{ij} = \delta^{ij}/\bar n_{g,i}$; here, $\bar n_{g,i}$ the mean number density of galaxies in the $i$th bin (per steradian), and the Kronecker delta restricts the noise to diagonal bins. \reffig{Q_fnl_comp_plot} shows examples of the windowed angular power spectrum for several redshift bins.

\subsection{Connection to other two-point statistics}\label{sec:SFB_xi}
The angular power spectrum $C_\ell(r_1,r_2)$ serves as the building block for two related two-point statistics. The spherical Fourier-Bessel (SFB) power spectrum further decomposes the radial direction into Fourier-Bessel modes:
\ba
\label{eq:SFB}
    C_\ell(k_1,k_2) &\,\equiv \left<\delta_{s,\lm}^k \delta_{s,\lm}^{k'}\right> \vs
        &\,= \frac{4}{\pi^2}\int \dd r_1 \, r_1^2 \,j_\ell(k_1r_1) D(r_1) \int \dd r_2 \, r_2^2 \,j_\ell(k_2r_2)D(r_2) \vs
        &\, \qquad \qquad  \qquad \qquad \qquad \qquad \times  \int \dd k \,  k^2 P(k) F_\ell(k,r_1) F_\ell(k,r_2) \vs
        &\, = \frac{2}{\pi} \int \dd r_1 \, r_1^2 \,j_\ell(k_1r_1) \int \dd r_2 \, r_2^2 \,j_\ell(k_2r_2) C_\ell(r_1,r_2)\,.
\ea
In a perfectly homogeneous and isotropic universe, $ C_\ell(k_1,k_2)$ is diagonal in $k$, reducing to a three-dimensional power spectrum. Once radial evolution, redshift-space distortions, and relativistic effects are included, this symmetry is broken and the off-diagonal structure carries no computational advantage over $C_\ell(r_1,r_2)$ itself. The SFB basis does, however, retain a useful correspondence: its radial wavenumber $k$ maps directly onto the Fourier modes that parameterize primordial fluctuations, providing a natural connection to inflationary predictions and to the flat-sky limit.

Numerical implementation of the SFB power spectrum is fairly difficult in practice though, as codes like \textsc{SuperFaB} illustrate \cite{Gebhardt::2021sfb_superfab,Khek_etal}. As we showed in Eq.~\eqref{eq:SFB}, in order for the $r$-integration to be accurate and $j_\ell(kr)$ to act like an orthogonal basis function for each mode $k$, the integration bounds must go from $r \in [0,\infty]$. In practice, of course, surveys do not have infinite depth. To mitigate this, \textsc{SuperFaB} discretizes the SFB power spectrum calculation, which requires mapping the nicely behaved $j_\ell$ radial basis functions to a linear combination of $j_\ell$ and $y_\ell$ (spherical Bessel functions of the second kind); these hybrid Bessel functions act as a new basis for $r \in [r_\text{min}, r_\text{max}]$ with eigenvalues $k_{n\ell}$, which are more applicable for a real survey. However, since $C_\ell(r,r')$ does not use any additional integration to map from configuration space to $k$ space, it does not require this extra discretization step.

Another common two-point statistic, the angular correlation function, collapses the harmonic expansion into configuration space:
\ba
\label{eq:xi_theta}
    \xi(\theta; r_1,r_2) &\,\equiv \left<\delta_s(\bfx_1) \delta_s(\bfx_2) \right> \vs
    &\, = \frac{2}{\pi} \sum_\lm \int \dd k_1 \, k_1^2 \int \dd k_2 \, k_2^2 \Psi^{k_1}_\lm(\bfx_1) \Psi^{k_2}_\lm(\bfx_2) C_\ell(k_1,k_2)\vs
    &\, = \sum_\ell \frac{2\ell + 1}{4\pi} C_\ell(r_1,r_2) \mathcal P_\ell(\cos \theta)\,,
\ea
where $\mathcal P_\ell(\cos \theta)$ are Legendre polynomials and $\cos \theta = \nhat_1 \cdot \nhat_2$. While $C_\ell(r_1,r_2)$ is diagonal in $(\ell,m)$ by statistical isotropy, $\xi(\theta)$ mixes all multipoles into a single angular variable; the resulting covariance matrix couples every angular bin, making $\xi(\theta)$ substantially harder to use for parameter inference than $C_\ell$ itself.

\section{Application to SPHEREx}\label{sec:spherex}
\begin{figure}[t!]
    \centering
    \includegraphics[width=0.9\linewidth]{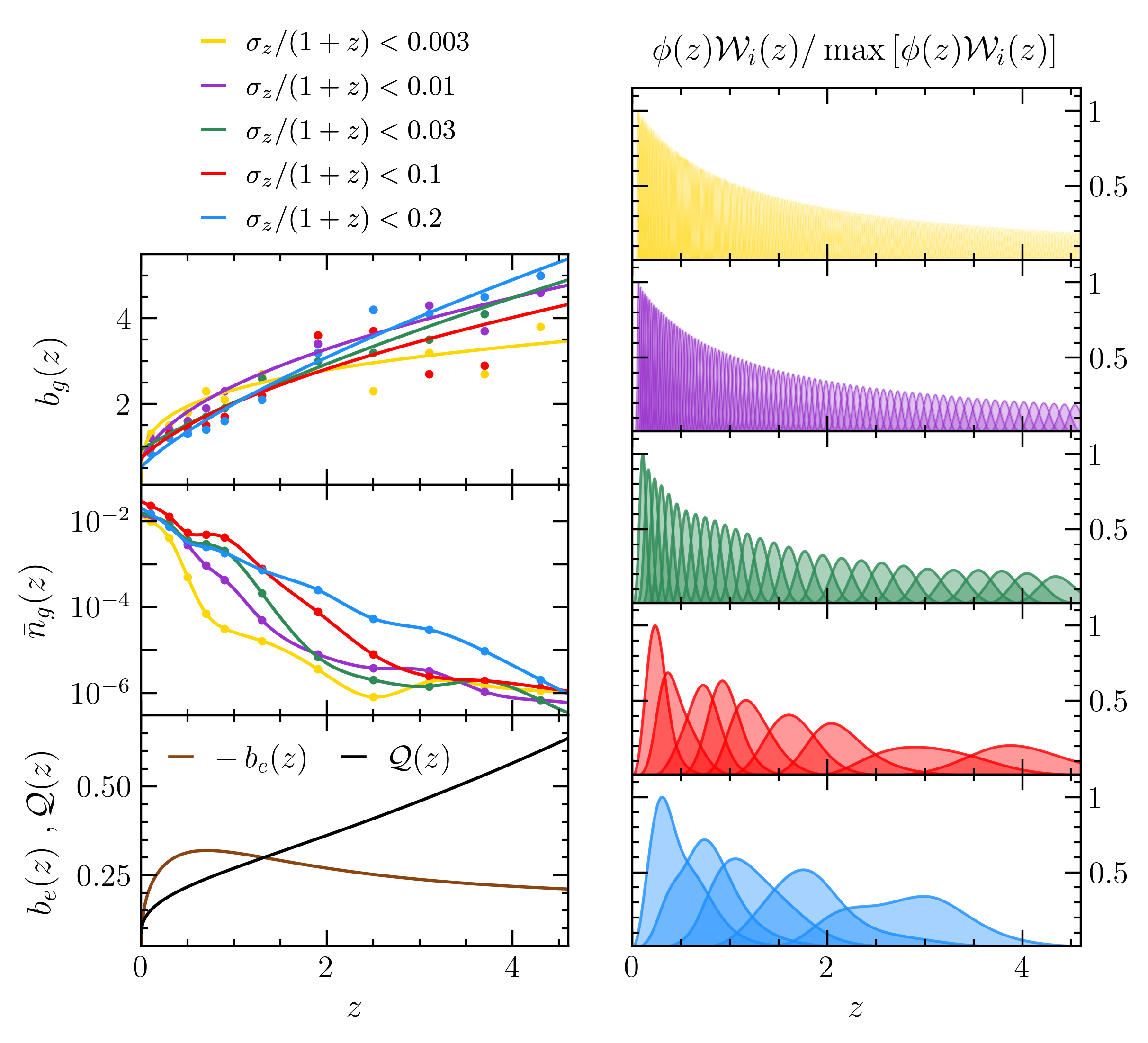}
    \caption{\textit{Top left}: the linear galaxy bias, $b_g$, for each SPHEREx sample and fitted using \refeq{linear_bias_fit}. \textit{Middle left}: the average galaxy number density, $\bar n_g$, in each redshift bin. \textit{Bottom left}: The evolution bias, $b_e$, and magnification bias, $\mathcal Q$, computed using Eqs.~\eqref{eq:def_be} and \eqref{eq:def_Q}, respectively, while assuming a Schechter luminosity function (see \refsec{Q_choice} for more details). \textit{Right column}: the window functions used for each sample, as defined by Eq.~\eqref{eq:top-hat_erf_windows}.}
    \label{fig:spherex_param_fig}
\end{figure}

Here, we apply {\textsc{PowerFull}} to a mock version of the full-sky SPHEREx mission and compute the angular power spectrum $C_\ell(z_1,z_2)$ retaining all linear-order relativistic contributions. We then assess the systematic effects introduced by ignoring these relativistic terms in the two-point statistics. Specifically, we perform a Fisher information matrix analysis for three treatments (Kaiser, Newtonian, and fully relativistic, in order of completeness) with fiducial $f_{\rm NL}=0$ for local primordial non-Gaussianity. We find that the recovered $f_{\rm NL}$ shifts by an amount non-trivially exceeding the statistical uncertainty when the relativistic corrections are omitted.

The ${\cal A}$ term in \refeq{F_ell_GR} carries the same $k^{-2}$ scaling as the local-type primordial non-Gaussianity contribution in \refeq{F_ell_modification}; Ref.~\citep{Jeong::2012gr} introduces the effective nonlinearity parameter $f_{\rm NL}^{\rm eff}$ on this basis, finding it of order unity. Ref.~\citep{Yoo::2012gr} later shows that the degeneracy can be broken in principle by combining the multi-tracer method with mass-weighted shot-noise reduction to isolate the line-of-sight dependent (via $\mu$) redshift-space distortion on the large scales where the non-Gaussianity signal resides. These results, however, rely upon Fourier-based analyses that do not accommodate the wide redshift bins and light-cone projection effects relevant on those scales, and omit the integrated contributions altogether. Because the degeneracy can be broken in principle, we do not recompute $f_{\rm NL}^{\rm eff}$; instead, we quantify the shift in the recovered $f_{\rm NL}$ that arises when the theoretical template omits the relativistic contributions.

While we are not the first to use this implied shift in $f_\text{NL}$ (e.g., see Ref.~\cite{Semenzato::2025sfb}) to motivate the necessity of the relativistic corrections at linear order in $C_\ell$, to our knowledge we are the first to rigorously handle SPHEREx survey geometry (e.g., redshift uncertainty and binning) and other observational effects (i.e., the luminosity function and magnification bias). Our analysis here is therefore an even stronger argument for the use of the full relativistic angular power spectrum when estimating primordial non-Gaussianity in full-sky surveys like SPHEREx.

\subsection{Survey parameters}\label{sec:spherex_params}

The mock survey scheme we use here is inspired by the original SPHEREx forecast \cite{Dore::2014spherex}, which employs a large number of photometric channels to approximate a spectroscopic survey, with the potential to measure redshifts for a far larger number of galaxies. The survey comprises five subsamples, each with differing selection functions and galaxy bias parameters from $z \sim 0$ to 4.6. The samples are divided by the upper limit on their redshift uncertainties,
\ba
\label{eq:sigma_z0}
    \sigma_{0} \equiv \frac{\sigma_z}{1+z} = \{0.003\,,0.01\,,0.03\,,0.1\,,0.2\}\,,
\ea
for samples 1--5, respectively.

We extracted the fiducial values for the linear galaxy biases and galaxy number densities (in units of galaxies per $({\rm Mpc}/h)^{3}$) for each sample from the SPHEREx public GitHub repository\footnote{The SPHEREx galaxy bias and number-density data for each redshift bin and sample are available in the \href{https://github.com/SPHEREx/Public-products/blob/master/galaxy_density_v28_base_cbe.txt}{SPHEREx public GitHub repository}.}, and then converted the latter into a surface density of galaxies per steradian. The linear galaxy bias was fit using nonlinear least squares regression while the galaxy densities were fit with cubic splines for a smoother radial selection function, similar to previous mocks \cite{Khek_etal}. We fit the biases with the functional form
\be
\label{eq:linear_bias_fit}
    b_g(z) = A_{b} (1+\beta_b z)^{\gamma_b}\,,
\ee
where, for a given sample, $A_{b}$ is the amplitude of the fit, $\gamma_b$ is the index of the bias fit, and $\beta_b$ dictates the redshift scaling of the bias. The specific values of each fitting parameter for the ``best-fit" curves  in each sample are reported in Table~\ref{tab:tab1}. We also plot the original values and fits in \reffig{spherex_param_fig}.

To retain the radial cosmological information accessible given the photometric-redshift uncertainty while avoiding excessively redundant neighboring bins, we use top-hat bins logarithmically spaced in observed redshift with the edges of the top-hats defined at
\ba
\label{eq:bin_edges}
    1 + z_{s,i}^{\text{obs}} = (1+z_\text{min}) e^{\Delta z_s (i-1)}\,, \quad i \in [1,N_{z,s}+1]\,,
\ea
where we use a bin width $\Delta z_s \sim 2 \sigma_{0,s}$ for sample $s$, with $\sigma_{0,s}$ the photo-$z$ error of Eq.~\eqref{eq:sigma_z0}. The number of bins in a given sample is given as
\ba
\label{eq:N_z_s}
    N_{z,s} = \frac{1}{\Delta z_s}\ln \bigg(\frac{1 + z_\text{max}}{1 + z_\text{min}} \bigg)\,,
\ea
where $z_{\min}$ and $z_{\max}$ are the minimum and maximum observed redshifts in the survey; since this number is usually not an integer, we take the ceiling of its result and then rescale the bin width to $\Delta z_s = \ln[(1+z_\text{max})/(1+z_\text{min})]/N_{z,s}$, so that the $N_{z,s}$ bins tile $[z_\text{min}, z_\text{max}]$ exactly; the resulting widths are $\Delta z_s = 0.0060$, $0.0199$, $0.0598$, $0.1860$, and $0.3348$ for samples 1 to 5. For SPHEREx, $z_\text{max} = 4.6$, while we chose $z_\text{min} = 0.05$, as it corresponds to a comoving distance $r_\text{min} \sim 150$ Mpc$/h$ using our fiducial cosmology. We list the resulting number of bins $N_{z,s}$ in \reftab{tab1}.

The observed angular power spectrum is then
\ba
\label{eq:Cl_ij_obs_z}
    C_{\ell}^{ij} = \int \dd z_\text{obs}\, T_i(z_\text{obs}) \int \dd z_\text{obs}' T_j(z'_\text{obs}) \int \dd z \,\phi(z) p(z_\text{obs}| z) \int \dd z' \phi(z') p(z_\text{obs}'|z') C_\ell(z,z')
\ea
with the observed-redshift top-hat bins $T_i(z_\text{obs})$ and Gaussian redshift scatter kernel
\ba
\label{eq:pzobs}
    p\big(z_{\rm obs}\mid z\big)
    = \frac{1}{\sqrt{2\pi}\,\sigma_{0,s} (1+z)}
      \exp\bigg[-\frac{\big(z_{\rm obs}-z\big)^2}{2\,(\sigma_{0,s})^2 (1+z)^2}\bigg]\,,
\ea
for each sample. Here, we conservatively assign every galaxy in a given sample the corresponding upper limit on the photometric redshift scatter, $\sigma_{0,s}$; see Eq.~\eqref{eq:sigma_z0}. The same conservative choice was made in Refs.~\cite{Dore::2014spherex,Khek_etal}. We can handle the integral over the observed-redshift coordinate and work in true-redshift by convolving the two kernels; this yields
\ba
\label{eq:top-hat_erf_windows}
    \mathcal W_i(z) &\,\equiv \int \dd z_\text{obs} \, T_i(z_\text{obs}) p(z_\text{obs}|z) \vs
    &\, = \frac{1}{2} \bigg[ \text{erf} \bigg(\frac{z_{s,i+1}^\text{obs} - z}{\sqrt{2} \sigma_{0,s} (1+z) } \bigg) - \text{erf} \bigg(\frac{z_{s,i}^\text{obs} - z}{\sqrt{2} \sigma_{0,s} (1+z) } \bigg) \bigg]\,,
\ea
which depends on the SPHEREx sample analyzed.

The choice of the top-hat bin width balances two failure modes: for $\Delta z_s \ll 2 \sigma_{0,s}$, the photo-$z$ scatter would strongly mix neighboring bins, leading to significant linear dependencies between them and a poorly conditioned covariance matrix; for $\Delta z \gg 2 \sigma_{0,s}$, we would discard radial information that photo-$z$ measurements could still resolve (though this effect is not particularly important for $f_\text{NL}$). Setting $\Delta z_s \simeq 2\sigma_{0,s}$ therefore balances the information retained on the primordial power spectrum parameters against the computational cost of a larger number of bins; $f_\text{NL}$ itself requires far fewer bins than $n_s$ and $\alpha_s$ do. We defer to a forthcoming paper the quantitative demonstration that this width retains nearly all of the available information. A similar analysis was conducted using the SFB power spectrum in Eq.~\eqref{eq:SFB} \cite{Khek_etal}; in reference to this study, the number of bins (or radial modes, in their case) $N_{z,s}$ plays the role that the choice of maximum wavenumber $k_{\rm max}$ played in their analysis.

The right-hand column of Fig.~\ref{fig:spherex_param_fig} shows substantial overlap between adjacent bins once the photo-$z$ scatter is included, most prominently for Samples 4 and 5, which suffer from significant redshift uncertainty. However, since the top-hat bins defined by Eq.~\eqref{eq:bin_edges} are disjoint in observed-redshift, a single galaxy in the survey can only be assigned to one bin $i$ within a given sample. As such, the shot noise of the survey remains diagonal and Poisson; i.e., $N_\ell^{ij} = \delta^{ij}/\bar n_{g,i}$ for the galaxy distribution $\bar n_{g}$ in the sample.

We plot the full mock survey configuration in \reffig{spherex_param_fig}. In the following sections, we provide forecasts for each SPHEREx sample individually, as well as a comprehensive prediction for all of the samples combined.

\begin{table}[t!]
    \centering
    \begin{tabular}{|c|c|c|c|c|c|}
    \hline
        &  \multicolumn{5}{c|}{\textbf{Sample Upper Bound} $\sigma_z/(1+z)$} \\[1.5pt]
        & 0.003 & 0.01 & 0.03 & 0.1 & 0.2 \\[1.5pt]
        \hline
        $A_b$      & 0.13 & 0.69 & 0.93 & 0.72 & 0.52 \\
        $\beta_b$  & $6.9\times 10^4$ & 14 & 2.0 & 5.6 & 5.6 \\
        $\gamma_b$ & 0.26 & 0.46 & 0.72 & 0.55 & 0.71 \\
        $N_z$ & 279 & 84 & 28 & 9 & 5 \\
        \hline
    \end{tabular}
    \caption{The values of the best-fit parameters ($A_b, \beta_b, \gamma_b$) for the linear galaxy bias given by the fitting function in Eqs.~\eqref{eq:linear_bias_fit} for each SPHEREx subsample. Additionally, $N_z$ represents the total number of observed-redshift bins in each SPHEREx subsample, with the width of each bin determined by the respective upper bounds of $\sigma_{0,s}$; we outline the computation of this quantity in Eq.~\eqref{eq:N_z_s}.} 
    \label{tab:tab1}
\end{table}

\subsection{Luminosity function}\label{sec:Q_choice}

To construct a more physically motivated forecast for a SPHEREx-like survey, we require a model for the shape and redshift dependence of the galaxy luminosity function, $\Phi_L(L,z)$, from which we determine the evolution and magnification biases. SPHEREx intends to utilize WISE \cite{WISE::2010} photometry in its effort to map the full sky, with wavelength coverage overlapping the WISE W1 and W2 bands \cite{Dore::2014spherex}. We therefore adopt the WISE luminosity function, focusing on the W1 band, as a physically motivated approximation for the galaxy population in our SPHEREx-like forecast. 

We assume a Schechter luminosity function of the form
\ba
\label{eq:phi_L}
    \Phi_L(L,z) = \frac{\phi_*}{L_*}\left( \frac{L}{L_*}\right)^\alpha e^{-L/L_*}\,,
\ea
where we have 
\ba
    \phi_*(z) = \phi_{0} e^{-R_\phi t_L(z)}\,,
\ea
and
\ba
    L_*(z) = L_{0} e^{-R_L t_L(z)} \left(1 - \frac{t_L(z)}{t_0} \right)^{n_0}\,,
\ea
where $t_L$ is the lookback time to a given redshift, and the value of each parameter can be computed from Ref.~\cite{Lake::2018_WISE}; see Table~\ref{tab:phi_tab} for the specifics.

We recall that for a luminosity function, the number of galaxies above some flux limit is given as
\be
\label{eq:n_g_L}
    \bar n_g( > L_\text{min}, z) = \int_{L_\text{min}}^\infty \dd L\,\Phi_L(L,z)\,.
\ee
We take SPHEREx's expected depth in the W1 band to correspond to a limiting flux $F_{2.4\,\mu\text{m}}\sim 2 \times 10^{-16}$ erg/cm$^2$/s at 2.4 microns \cite{Dore::2014spherex}. From this, we estimate the form of $L_\text{min}$ and directly compute the evolution bias, $b_e$, and magnification bias, $\mathcal Q$, as functions of redshift; here we use Eqs.~\eqref{eq:def_be} and \eqref{eq:def_Q}, respectively. The resulting curves can be found in the bottom left panel of Fig.~\ref{fig:spherex_param_fig}. These quantities derived from the luminosity function are the evolution and magnification biases used throughout the forecasts below. Our definitions of the evolution and magnification biases are the same as those of Ref.~\cite{Maartens2021Biases}, where both are defined as responses of the galaxy number density to redshift evolution and the luminosity threshold, respectively. Importantly, for a flux-limited galaxy sample these quantities are determined from the galaxy luminosity function rather than the simplified relation that assumes the universality of the halo mass function in Eq.~\eqref{eq:b_e_halo}.

One caveat is in order. The WISE luminosity function presented here is measured for lower-redshift galaxies \cite{Lake::2018_WISE}, while SPHEREx will have to observe different emission-line galaxies to probe large-scale structure out to redshifts $z \gtrsim 3$; these galaxies would have different luminosity functions. We nonetheless extrapolate the low-redshift luminosity function to higher redshifts to provide an approximate but physically motivated forecast for a SPHEREx-like survey.

\begin{table}[t!]
    \centering
    \renewcommand{\arraystretch}{1.4}
    \begin{tabular}{|c|c|c|c|c|c|c|}
        \hline
         $\alpha$ & $\phi_*(0)$ & $\phi_*(0.38)$ & $\phi_*(1.5)$ & $L_*(0)$ & $L_*(0.38)$ & $L_*(1.5)$\\
          & \multicolumn{3}{c|}{$h^3\,\,$Mpc$^{-3}$} & \multicolumn{3}{c|}{$10^{10} L^{2.4\,\mu \text{m}}_\odot h^{-2}$} \\
         \hline 
         $-1.05$ & 0.0169 & 0.016 & 0.015  & 3.12 & 5.6 & 9.9\\
         \hline
    \end{tabular}
    \caption{The values of each parameter in the Schechter luminosity function given in \refeq{phi_L} at $z=0$, $0.38$, and $1.5$. We took these values from Ref.~\cite{Lake::2018_WISE} for the WISE survey in the W1 band. Here, $\phi_*$ values are in units of $(h/$Mpc$)^3$ and $L_*$ is given in terms of $10^{10} L^{2.4\,\mu \text{m}}_\odot h^{-2}$, where $L^{2.4\,\mu \text{m}}_\odot = 3.344 \times 10^{-8} $ Jy Mpc$^2$ is the luminosity of the Sun at 2.4 microns.}
    \label{tab:phi_tab}
\end{table}

\subsection{Effect of redshift uncertainty on relativistic terms}\label{sec:bins}

\begin{figure}[t!]
    \centering
    \includegraphics[width=0.94\linewidth]{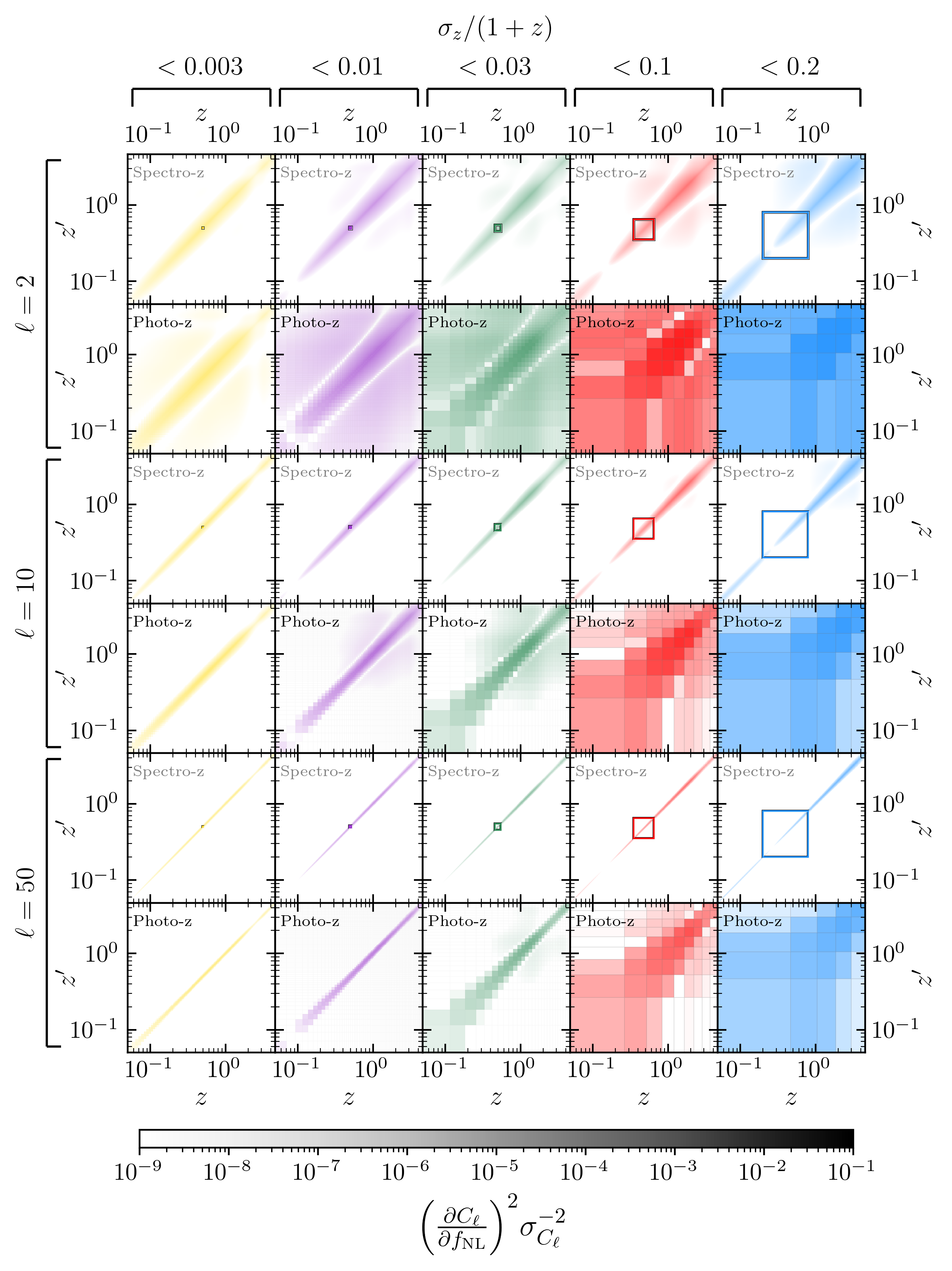}
    \vspace{-0.3cm}
    \caption{The sensitivity of the full relativistic angular power spectrum with respect to $f_\text{NL}$ for three different multipole values ($\ell = 2,10,50$) for all 5 subsamples. To quantify this sensitivity, we plot the square of the quantity $\partial C_\ell/\partial f_\text{NL}$ rescaled by the variance of $C_\ell$. Additionally, for each $\ell$ and subsample, we show the unbinned (purely spectroscopic) and binned versions; the former is labeled and is in the top row of each case. Lastly, we overlay the 1-$\sigma$ extent of the Gaussian photo-$z$ scatter at $z = 0.5$ for each sample on the unbinned plots to provide a sense of scale.}
    \vspace{-0.2cm}
    \label{fig:der_bin_fig}
\end{figure}

Before proceeding to the main results of our forecast, we examine how the photometric redshift scatter and the general survey setup condition the role of the relativistic effects in the two-point statistics. We have already shown in \refsec{Cl} that relativistic effects are most dominant in the cross-correlation, in which there is foreground structure that can be correlated with lensing and other perturbed photon geodesic signals from background sources. 

Furthermore, Fig.~\ref{fig:cl_2_grid} shows that the auto-correlated signal is dominated by galaxy-galaxy clustering and linear order redshift-space distortions. Therefore, in the Dirac-delta window function case (i.e., a spectroscopic survey with virtually no redshift uncertainty), that cross-correlated relativistic signal is effectively untouched by the shot noise contained in the auto-correlated bins. 

For non-trivial redshift uncertainty, however, the broader true-redshift window functions naturally mix the auto- and cross-correlated signal. We emphasize that the shot-noise contribution itself remains diagonal, $N_\ell^{ij}=\delta^{ij}/\bar n_{g,i}$, so the off-diagonal cross-correlations do not acquire an additive shot-noise contribution. Rather, the broader window functions mix some of the near-diagonal relativistic signal into a region where the covariance is increasingly dominated by the much larger galaxy-galaxy and RSD auto-correlated signal, together with its associated shot noise. As a result, the sensitivity of these near-diagonal modes to the relativistic contribution can be reduced. Fig.~\ref{fig:der_bin_fig} shows this effect clearly; there we plot the sensitivity of the angular power spectrum to changes in $f_\text{NL}$.

A substantial relativistic contribution also appears in the cross-correlations. At low redshift, the relativistic contribution is concentrated at larger separations from the auto-correlated diagonal, whereas at higher redshift appreciable relativistic contributions extend closer to the diagonal. The relativistic effects would therefore contaminate the diagonal at low to moderate bin width, particularly at $z \gtrsim 3$. Additionally, in the case of $f_\text{NL}$, Fig.~\ref{fig:der_bin_fig} shows that the sensitivity is not confined to the exact equal-redshift ($i=j$) correlations in the Dirac-delta limit, but extends substantially into cross-correlations between distinct radial shells.

\subsection{Mode selection and handling non-linearity}\label{sec:ellmax}

The lightcone treatment of the galaxy clustering, the bias relationship between the galaxy field and the underlying field, and the underlying matter power spectrum $P(k)$ used in our angular power spectra created with \textsc{PowerFull} rely solely upon \textit{linear} perturbation theory, which is only valid for $\delta_m \ll 1$. On scales of order $\lambda \lesssim 30\,\,\text{Mpc}/h$, however, the perturbations rise to order unity, and the linear theory used to describe their behavior is ill-defined \cite{jeong/komatsu:2006}. To avoid contaminating the model with non-linearity that we do not treat, we use a maximum wavenumber $k_\text{max} = 2\pi/\lambda_\text{min}$ corresponding to a minimum scale $\lambda_\text{min}$ where the non-linearity starts to become non-trivial; for our analysis, we use $k_\text{max} = 0.2 \,\,h/\text{Mpc}$.

Moving back to our spherical Fourier-Bessel basis, this simple cut in Fourier wavenumber becomes a much more complicated cut in $\ell$ and along redshift $z$. The effect on multipole $\ell$ is naturally analogous to a cut along the transverse direction, i.e.,
\ba
\label{eq:k_perp_cut}
    k_\perp(z) = \frac{\ell + \frac{1}{2}}{r(z)}\,.
\ea
Since the associated multipole values for the non-linearity cut are fairly large in general, we can neglect this factor of $1/2$. For the purpose of defining the nonlinear cut, we characterize the effective radial extent of each tracer using the moments of its galaxy-weighted radial selection function. Specifically, since we defined the photo-$z$-broadened window in Eq.~\eqref{eq:top-hat_erf_windows}, the effective mean redshift and width of the $i$th tracer are then
\ba
\label{eq:moments_of_number_dens}
    \bar z_i=\int \dd z\,z\,\phi_i(z)\,,
    \qquad
    \sigma_{z,i}^2=\int \dd z\,(z-\bar z_i)^2\phi_i(z)\,.
\ea
Thus, $\bar z_i$ and $\sigma_{z,i}$ are moments of the number-density-weighted true-redshift distribution of galaxies assigned to bin $i$; here $\phi_i(z) \propto \mathcal W_i(z)\,\phi(z)$ is the product of the bin window and the radial selection function of \refsec{window}, normalized to unity. We use these quantities to define the effective radial boundaries of the tracer window as
\ba
\label{eq:z_i_bounds}
z_i^-=\max(0,\bar z_i-2\sigma_{z,i}),
\qquad
z_i^+=\bar z_i+2\sigma_{z,i}\,.
\ea
We then define the corresponding comoving radial boundaries as
\ba
r_i^\pm = r(z_i^\pm)\,,
\ea
and the effective comoving distance to the $i$th bin is evaluated at its mean redshift,
\ba
\label{eq:r_i}
\bar r_i = r(\bar z_i)\,.
\ea
For the cross-correlation between bins $i$ and $j$, we similarly define the characteristic comoving distance of the pair as
\ba
\label{eq:r_ij}
\bar r_{ij} = \frac{\bar r_i+\bar r_j}{2}\,.
\ea
These quantities specify the characteristic transverse distance of each bin pair, while the radial extent $[r_i^-,r_i^+]$ is used below to determine the minimum line-of-sight separation between the two windows. Throughout this construction, $r(z)$ is the comoving distance of our fiducial cosmology, held fixed when the forecast parameters are varied. From this, we know 
\ba
\label{eq:lambda_perp}
    \lambda_\perp^{ij} = \frac{2\pi }{ \ell }\begin{cases} \bar r_i & i =j \\  \bar r_{ij} & i \neq j\end{cases}\,.
\ea

\begin{figure}[t!]
    \centering
    \includegraphics[width=0.83\linewidth]{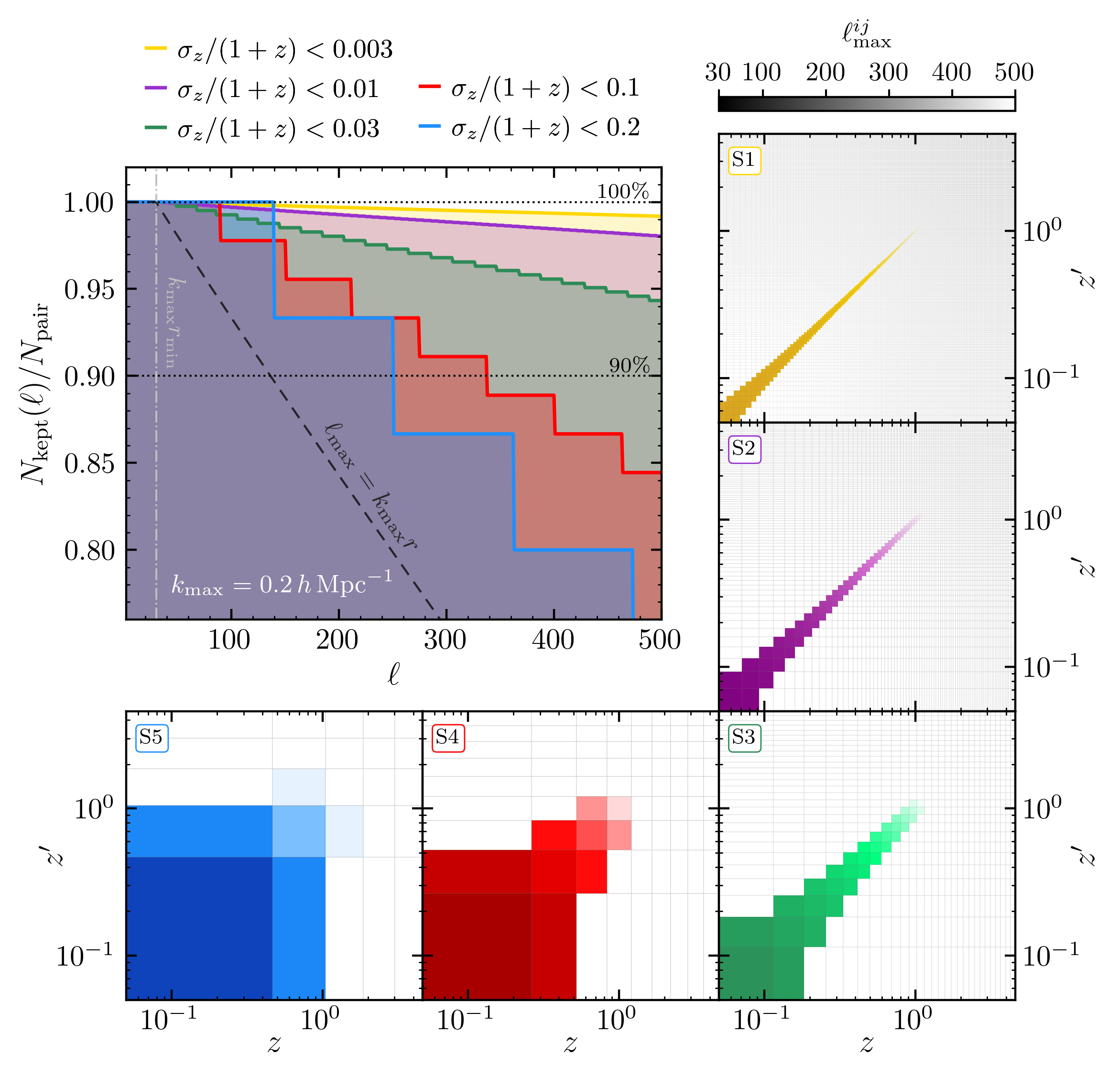}
    \vspace{-0.3cm}
    \caption{\textit{Top left}: the fraction of bin pairs $(i,j)$ within a given sample kept as a function of $\ell$; this $N_\text{kept}(\ell)$ follows directly from Eq.~\eqref{eq:ellmax_of_z}, where $\lambda_\parallel^{ij}$ is sample dependent. Additionally, we show the diagonal limit of the fraction of pairs kept at the Limber scale for $k_\text{max} = 0.2 \,\,h/\text{Mpc}$; this represents the contribution of auto-correlation pairs to the total fraction for each sample. Lastly, we denote the minimum $\ell_\text{max}$ value, which is set by the minimum redshift of the survey; we used $z_\text{min} = 0.05$, which yields $k_\text{max} r(z_\text{min}) \sim 30$. \textit{Surrounding plots}: For each sample, we show the corresponding $\ell_\text{max}$ up to $\ell = 500$ for a pair of bins centered at $(z,z') =(z_i,z_j)$; the axes here are logarithmic and the shading denotes the value of $\ell_\text{max}$ (white indicates that no nonlinear $\ell$ cut is applied to that bin pair within the plotted multipole range).}
    \vspace{-0.2cm}
    \label{fig:ellmax_fig}
\end{figure}

Along the radial direction, we instead consider the separation of the bins in our survey. Specifically, the minimum radial separation between the two bins is given as
\ba
\label{eq:lambda_par}
    \lambda_\parallel^{ij} = \max\big(0, r_i^- - r_j^+, r_j^- - r_i^+ \big)\,,
\ea
which follows from Ref.~\cite{DiDio::2014}. This definition handles bin overlaps and vanishes for auto-correlation when $i = j$. Taken together, the two quantities set the smallest scale a given bin pair can probe: at multipole $\ell$ the pair resolves transverse structure of size $\lambda_\perp^{ij}$, while its windows cannot resolve radial structure smaller than their separation $\lambda_\parallel^{ij}$. In a spatially flat background these combine in quadrature, so the pair is sensitive to the three-dimensional scale $[(\lambda_\perp^{ij})^2 + (\lambda_\parallel^{ij})^2]^{1/2}$, and requiring that scale to exceed $2\pi/k_\text{max}$ is the projected counterpart of the three-dimensional condition $k \leq k_\text{max}$. A pair whose windows are already separated by more than $2\pi/k_\text{max}$ therefore never probes the non-linear regime, however large $\ell$ becomes, which is the origin of the unbounded branch below. Since the cut for non-linearity is given as
\ba
\label{eq:lambda_of_ell}
    \sqrt{\big(\lambda_\perp^{ij}\big)^2 + \big(\lambda_\parallel^{ij}\big)^2} \geq \frac{2\pi}{k_\text{max}}\,, 
\ea
we know that the cut as a function of maximum multipole, $\ell_\text{max}$, for a given set of bins $(i,j)$ is then
\ba
\label{eq:ellmax_of_z}
    \ell_\text{max}^{ij} = \begin{cases}   k_\text{max} \bar r_{ij} \bigg[1 - \Big(\frac{k_\text{max} \lambda_\parallel^{ij}}{2\pi}\Big)^2 \bigg]^{-1/2} & 0 \leq \,\,\lambda_\parallel^{ij} < \frac{2\pi}{k_\text{max}}  \\ 
     \infty & \lambda_{\parallel}^{ij} \geq \frac{2\pi}{k_\text{max}} \end{cases} \,.
\ea

We illustrate the resulting $\ell_\text{max}$ for each SPHEREx sample in Fig.~\ref{fig:ellmax_fig}. The fraction of bin pairs retained to a given multipole is generally larger for the samples with better redshift precision (e.g., Samples 1 and 2). However, these samples also contain by far the largest number of radial bins, and therefore the largest total number of bin pairs. As a result, even though a smaller fraction of their pairs is removed by the nonlinear cut, the absolute number of affected pairs can still be larger than for the samples with poorer redshift precision (e.g., Samples 4 and 5).

Additionally, we see that much of the truncation in multipole occurs near or along the diagonal, where the auto-correlations and cross-correlations of neighboring radial bins have relatively small $\lambda_\parallel^{ij}$ and therefore a lower $\ell_\text{max}^{ij}$. This behavior occurs for every sample, but is most consequential for Samples 1 and 2 because their finer radial binning produces many more adjacent and near-adjacent pairs. We therefore discard some of the high-$\ell$ radial information that lies within the nonlinear regime. However, since the information on local primordial non-Gaussianity is concentrated primarily on large scales, the corresponding loss of constraining power on $f_\text{NL}$ from this cut is relatively small.

\subsection{Impact of relativistic effects on $f_\mathrm{NL}$ and other parameters}\label{sec:bias_fnl}

\begin{figure}[t!]
    \centering
    \includegraphics[width=0.89\linewidth]{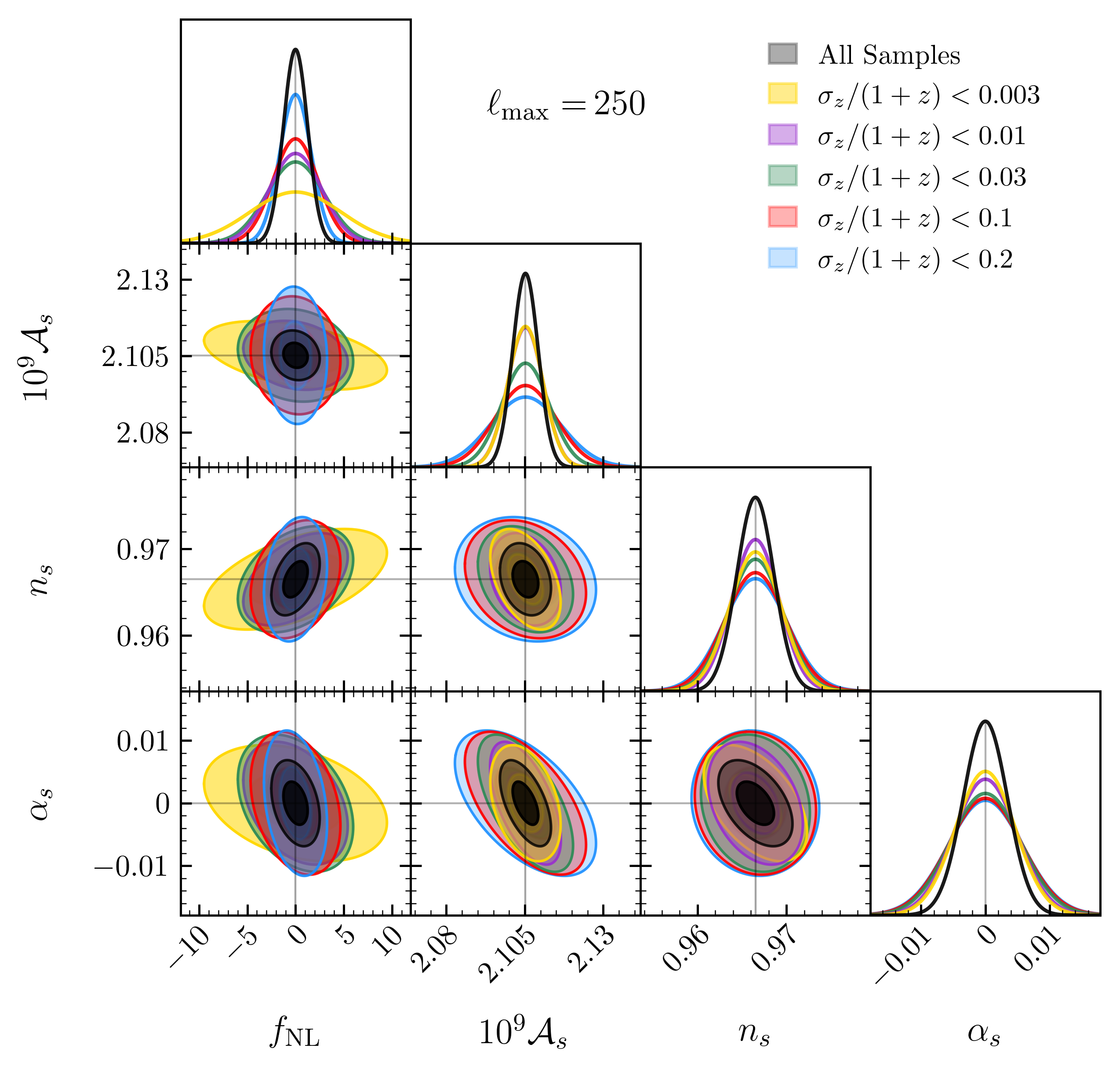}
    \caption{A corner plot of the marginalized parameters, $f_\text{NL}$, $\mathcal A_s$, $n_s$, and $\alpha_s$ computed via the Fisher information matrix outlined in \refeq{Fisher} up to a global $\ell_\mathrm{max} = 250$, applied in addition to the per-pair $\ell_\text{max}^{ij}$ cut of \refsec{ellmax}, for all of the SPHEREx samples and their combination (in black). Among the individual subsamples, the two samples with the poorest redshift precision (in red and blue) provide the tightest constraints on $f_\text{NL}$, while the higher-precision samples provide tighter individual constraints on the remaining primordial power spectrum parameters. We see that the expected level of uncertainty on $\sigma_{f_\text{NL}} \sim 1$ is achieved when using the multi-sample result.}
    \label{fig:corner_all_samps}
\end{figure}

To forecast SPHEREx constraints, we compute the Fisher information matrix for the angular power spectrum; the matrix quantifies the amount of information that an observable contains about a set of parameters and defines the best possible constraints achievable under the assumption of Gaussian uncertainty. We adopt a multivariate Gaussian likelihood function, which yields the following Fisher matrix \cite{Tegmark::1997fisher}
\be
\label{eq:Fisher}
    F_{ij} = f_\text{sky} \Delta \ell \sum_{\,\,\ell = \ell_\text{min}}^{\ell_\text{max}}  \frac{2 \ell + 1}{2}  \text{Tr} \bigg(\bm C_\ell^{-1} \frac{\partial  \bm C_\ell}{\partial \theta_i} \bm C_\ell^{-1} \frac{\partial \bm C_\ell}{\partial \theta_j} \bigg)\,,
\ee
where $\theta_i$ is the $i$th cosmological parameter, $f_\text{sky}$ is the fraction of the sky covered, and $\Delta \ell$ describes the multipole binning. We parameterize the amplitude by $10^9 A_s$, for which the derivative of each signal spectrum is analytic, $\partial C_\ell^{ij}/\partial (10^9 A_s) = C_\ell^{ij}/(10^9 A_s)$; the derivatives with respect to $n_s$ and $\alpha_s$ are central finite differences of the re-tilted linear power spectrum, with a total step of $5 \times 10^{-4}$ about the pivot scale $k_\text{piv} = 0.05\,\text{Mpc}^{-1}$, the running entering the tilt as $\frac{1}{2} \Delta\alpha_s \ln^2(k/k_\text{piv})$.

For the purposes of this study, we use $f_\text{sky} = 0.75$, to account for foreground galactic and stellar contamination in the survey, and $\Delta \ell = 1$. We also include Gaussian Planck priors on the primordial power spectrum parameters, taking $\sigma(n_s)=0.0042$, $\sigma(\alpha_s)=0.0067$, and  $\sigma[\ln(10^{10}A_s)]=0.014$; no external prior is imposed on $f_\text{NL}$. These priors are included to prevent the forecast from being driven by degeneracies in the primordial power-spectrum shape while leaving the constraint on $f_\text{NL}$ determined by the SPHEREx clustering information. The priors enter as a Gaussian prior matrix added to the Fisher matrix, $F \to F + F_\text{prior}$, in both Eq.~\eqref{eq:uncertainty} and Eq.~\eqref{eq:theta_bias} below. Additionally, as mentioned previously, the monopole and dipole are contaminated by the mean number density and by our motion with respect to the large-scale background, respectively; we therefore use a minimum multipole $\ell_\text{min} = 2$. We have also defined the data matrix, $\bm C_\ell$, as
\be
\label{eq:data_matrix}
    \bm C_\ell = \begin{pmatrix}
    \bar C_\ell^{\,11} & \cdots & \bar C_\ell^{\,1n}  \\
    \vdots & \ddots & \vdots \\
    \bar C_\ell^{\,n1} & \cdots & \bar C_\ell^{\,nn} 
    \end{pmatrix}
\ee
for a given multipole $\ell$ and $n$ redshift bins. Here, we recall the definition of the observed angular power spectrum, $\bar C_\ell^{\,ij}$, given in \refeq{observed_Cell}. For the multi-tracer analysis, we use the combined data matrix
\be
\label{eq:data_matrix_multi_tracer}
    \bm C_\ell^\text{all} = \begin{pmatrix}
    \bm C_\ell^{(\text{s}1 \times \text{s}1)} & \cdots & \bm C_\ell^{(\text{s}5 \times \text{s}1)}  \\
    \vdots & \ddots & \vdots \\
    \bm C_\ell^{(\text{s}1 \times \text{s}5)} & \cdots & \bm C_\ell^{(\text{s}5 \times \text{s}5)} 
    \end{pmatrix}\,,
\ee
where $\bm C_\ell^{\text{s}i\times \text{s}j}$ is the data matrix of Sample $i$ cross-correlated with Sample $j$. Since the samples trace the same underlying density field with different galaxy biases and shot noise, their cross-correlations can provide information beyond that obtained from treating the samples as statistically independent. Because the five samples are disjoint sets of galaxies, no shot-noise term arises between samples; the shot noise remains $N_\ell^{ij} = \delta^{ij}/\bar n_{g,i}$ within each sample.

At each multipole, the pair-dependent nonlinear cut defined in Sec.~\ref{sec:ellmax} is applied before constructing the covariance which enters in Eqs.~\eqref{eq:data_matrix} and \eqref{eq:data_matrix_multi_tracer}. Specifically, any spectra $\bar C_\ell^{ij}$ for which $\ell>\ell_{\max}^{ij}$ are removed from the data vector, and the Fisher matrix is evaluated using the covariance of the retained set of spectra. Equivalently, the covariance is restricted to the surviving information after the $\ell$ cuts but prior to inversion; this order of operations is crucial, as inverting \textit{before} doing the $\ell$ cuts would yield different (and improper) results.

The marginalized uncertainty of a given cosmological parameter, $\theta_i$, is then given by the Fisher matrix via
\be
\label{eq:uncertainty}
    \sigma_{\theta_i} = \sqrt{(F^{-1})_{ii}}\,.
\ee

\begin{figure}[t!]
    \centering
    \includegraphics[width=0.89\linewidth]{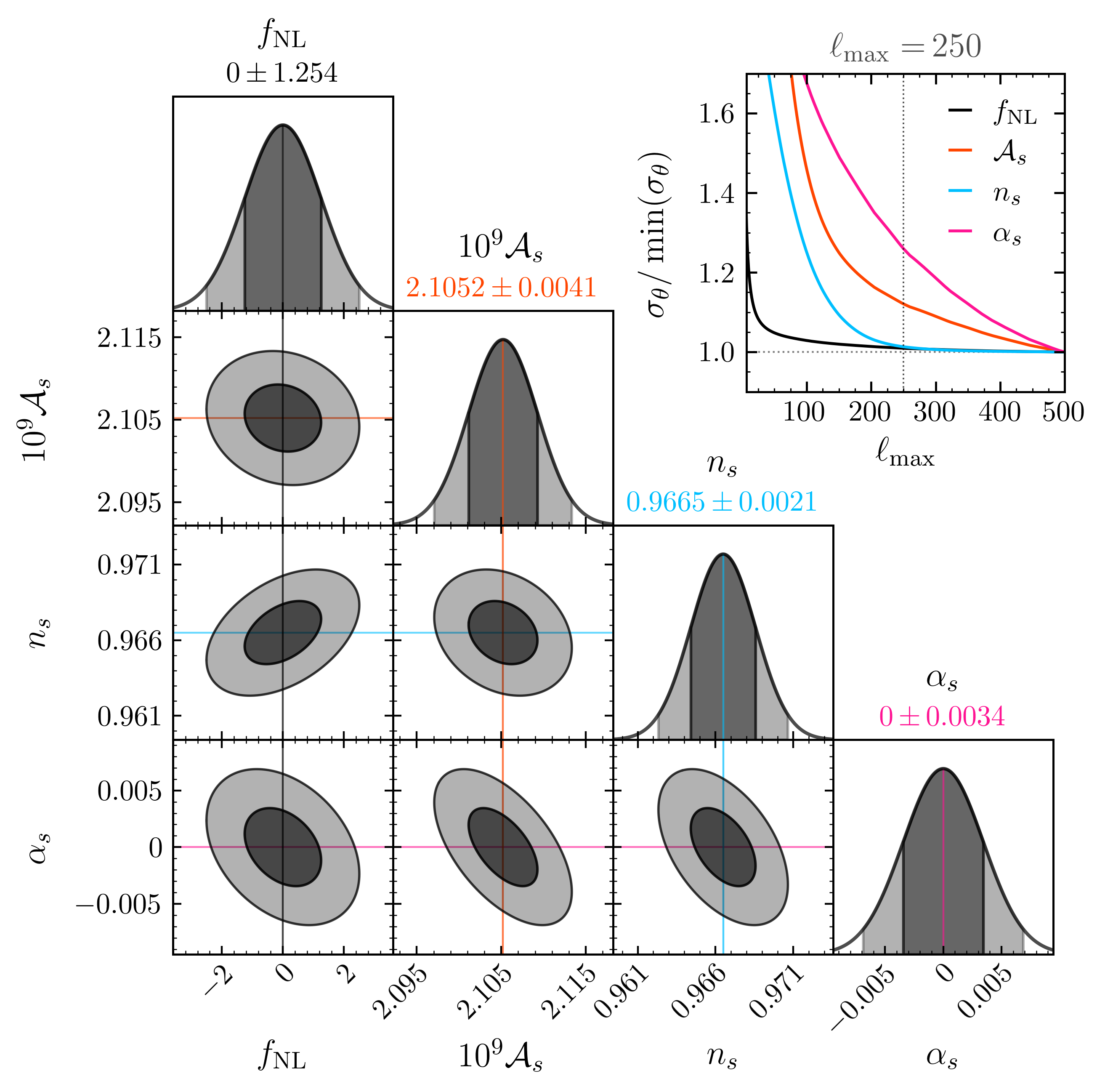}
    \caption{The corner plot for the multi-sample analysis, using the complete five-sample data vector including all auto- and cross-correlations, up to a global $\ell_\text{max} = 250$ applied in addition to the per-pair $\ell_\text{max}^{ij}$ cut of \refsec{ellmax}; these are the same as those presented in Fig.~\ref{fig:corner_all_samps} (the black contours). We see that $\sigma_{f_\text{NL}}\sim 1$ is indeed achieved, as well as $\sigma_{n_s}\,,\sigma_{\alpha_s}\,,\sigma_{10^9\mathcal A_s} \sim 10^{-3}$, as is expected for SPHEREx \cite{Dore::2014spherex}. Also, we plot the dependence of the marginalized uncertainty on each parameter as a function of $\ell_\text{max}$; as anticipated, there is very little additional information to be gained on $f_\text{NL}$ beyond $\ell \sim 100$.}
    \label{fig:corner_tot}
\end{figure}

We show the uncertainty on our set of cosmological parameters for all five SPHEREx samples in Fig.~\ref{fig:corner_all_samps}, as well as the multi-sample result in Fig.~\ref{fig:corner_tot}. Among the individual subsamples, Samples 4 and 5 provide the tightest constraints on $f_\text{NL}$, while the samples with better redshift precision provide stronger individual constraints on the other primordial power spectrum parameters. The strongest overall constraints are obtained from the complete multi-sample analysis.

The goal of this forecast is to quantify the systematic shift in the inferred cosmological parameters that arises when the theoretical model neglects wide-angle and/or relativistic contributions. Using the definition of the Fisher matrix in \refeq{Fisher}, we can estimate the expected systematic bias on a cosmological parameter, $\theta_i$, induced by a slight shift in our angular power spectrum from the fiducial model, $\Delta \bm C_\ell \equiv \bm C_\ell^{\text{test}} - \bm C_\ell^{\text{fid}}$, where $\bm C_\ell^{\text{test}}$ is the less accurate model. From Ref.~\cite{Bernal::2020paramshift}, we have
\be
\label{eq:theta_bias}
    \left< \Delta \theta_i \right> = f_\text{sky}  \Delta \ell\, F_{ij}^{-1} \sum_{\,\ell = \ell_\text{min}}^{\ell_\text{max}} \frac{2\ell + 1}{2} \text{Tr} \left(\frac{\partial \bm C_\ell}{\partial \theta_j} \bm C_\ell^{-1}\Delta \bm C_\ell \,\bm C_\ell ^{-1} \right)\,,
\ee
where $\left<\Delta \theta_i \right>$ is the expected shift on the cosmological parameter $\theta_i$ due to the use of the test model as compared to the fiducial. 

\begin{figure}[t!]
    \centering
    \includegraphics[width=0.89\linewidth]{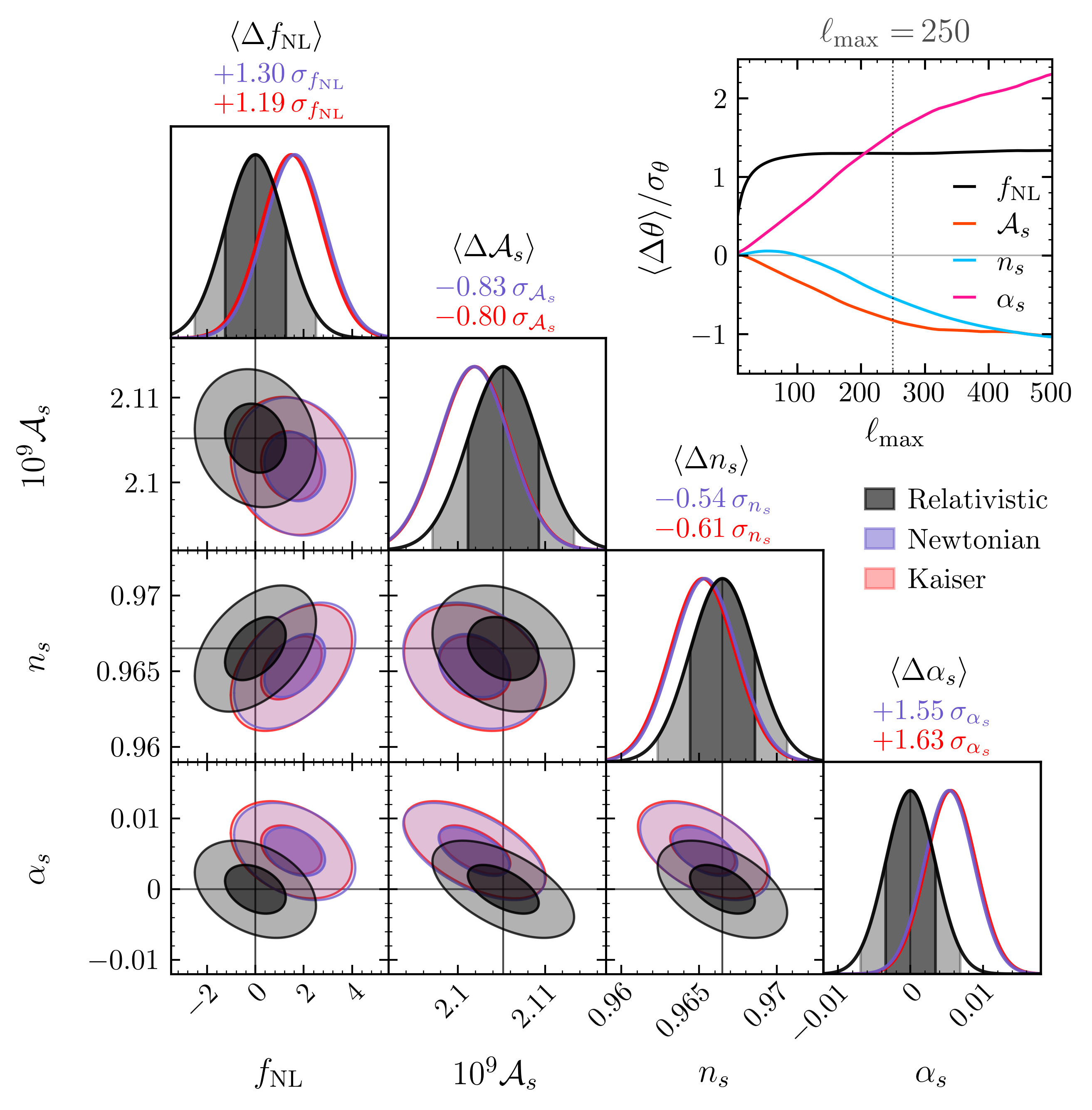}
    \caption{A corner plot for the complete five-sample analysis (all auto- and cross-correlations) up to a global $\ell_\text{max}=250$ applied in addition to the per-pair $\ell_\text{max}^{ij}$ cut of \refsec{ellmax}, showing the systematic shift in each cosmological parameter due to ignoring relativistic effects. The quoted shifts are expressed in units of the marginalized uncertainty from the fiducial relativistic model for the same survey and parameter configuration. The black contours are the fiducial model; all relativistic effects are taken into account. The blue contours show the Newtonian treatment of the angular power spectrum; this ignores relativistic corrections but keeps the Doppler velocity term. The red contours show the Kaiser treatment; this only includes galaxy clustering and first-order RSDs. The top right plot shows the dependence of this shift (in the Newtonian case) as a function of $\ell_\text{max}$ for each parameter.} 
    \label{fig:corner_shift}
\end{figure}

We investigate the shift of $f_\text{NL}$ marginalized over three other parameters: $n_s$, $\alpha_s$, and $\mathcal A_s$, the tilt and running of the spectral index, and the scalar amplitude, respectively. For this calculation, we use the complete multi-sample data vector introduced in Eq.~\eqref{eq:data_matrix_multi_tracer} including the auto- and cross-correlations among all five SPHEREx subsamples. In Figure \ref{fig:corner_shift}, we illustrate the shift of these cosmological parameters using the evolution and magnification biases we derived from the WISE luminosity function presented in \refsec{Q_choice} and an absolute maximum multipole of $\ell_\text{max} = 250$. We denote the fiducial relativistic model with the black contours, and investigate the shift on the parameters for two test cases: ignoring only the relativistic terms (i.e., using the Newtonian treatment, in blue), and then further ignoring the Newtonian peculiar velocity term (i.e., the Kaiser treatment, in red). To identify which relativistic contribution is primarily responsible for these shifts, we repeat the same forecast setup described above, but now add relativistic effects to the Newtonian angular power spectrum model one at a time and compute the corresponding bias on parameter inference; the results are summarized in Table~\ref{tab:tab2}.

We also investigated the shift of the parameters with respect to $\ell_\text{max}$; this can be found in the upper right panel of Fig.~\ref{fig:corner_shift}. We see that $\langle\Delta f_\text{NL}\rangle/\sigma_{f_\text{NL}}$ plateaus beyond $\ell\sim100$. This behavior is expected because the relativistic contributions most closely degenerate with local primordial non-Gaussianity are concentrated on large scales: the Doppler and potential contributions carry inverse powers of $k$, while the local-PNG contribution itself scales as $k^{-2}$. Integrated contributions such as gravitational lensing can remain important to substantially higher multipoles, but they have a different scale and radial dependence from the local-PNG signal. In addition, the nonlinear $\ell_\text{max}^{ij}$ cut introduced in Sec.~\ref{sec:ellmax} removes part of the high-$\ell$ information from closely separated radial bins. Consequently, increasing the absolute maximum multipole beyond $\ell\sim100$ adds relatively little to the systematic shift in $f_\text{NL}$, even though the shifts in parameters such as $n_s$ and $\alpha_s$ can continue to evolve at larger multipoles.

The shift in cosmological parameters, and in $f_\text{NL}$ in particular, is therefore non-trivial when relativistic effects are ignored in a full-sky survey like SPHEREx. Furthermore, a survey of this nature intending to constrain $f_\text{NL} \sim 0$ would shift the recovered primordial non-Gaussianity by $\langle\Delta f_\text{NL}\rangle\simeq1.3\,\sigma_{f_\text{NL}}$ at $\ell_\text{max}=250$, corresponding to an absolute shift in $f_\text{NL}$ of order unity. We also illustrated in passing that the shift from just ignoring the extra Doppler velocity term (i.e., the so-called Newtonian term) is negligible by comparison.

Table~\ref{tab:tab2} shows the corresponding parameter shifts when we include individual relativistic contributions to a Newtonian angular power spectrum model at $\ell_{\rm max}=250$. The dominant reduction in the cosmological parameter bias comes from weak gravitational lensing: including the full lensing contribution reduces the residual shift in $f_{\rm NL}$ from approximately $1.3\,\sigma_{f_{\rm NL}}$ for the purely Newtonian template to only $\sim-0.1\,\sigma_{f_{\rm NL}}$. By contrast, replacing the Newtonian Doppler velocity term by the corresponding relativistic term, or adding the potential, ISW, or time delay contributions individually, produces only modest changes to the residual parameter shift.

\begin{table}[t!]
    \centering
    \begin{tabular}{|c|c|c|c|c|c|c|}
    \hline
        & \multicolumn{6}{c|}{\textbf{Relativistic contribution added to Newtonian $C_\ell$}} \\
      Parameter
        & None 
        & Doppler
        & Potential
        & ISW
        & Time delay
        & Lensing \\
    \hline
      $f_{\rm NL}$  & $+1.30$ & $+1.33$ & $+1.41$ & $+1.37$ & $+1.19$ & $-0.12$ \\
      $\mathcal A_s$     & $-0.83$ & $-0.83$  & $-0.85$ & $-0.84$ & $-0.80$ & $+0.03$ \\
      $n_s$         & $-0.54$ & $-0.54$ & $-0.51$ & $-0.52$ & $-0.57$ & $-0.04$ \\ 
      $\alpha_s$    & $+1.55$ & $+1.56$ & $+1.54$ & $+1.54$ & $+1.57$ & $+0.02$  \\ 
    \hline
    \end{tabular}
    \caption{Systematic shifts in the recovered cosmological parameters at $\ell_{\rm max}=250$, expressed in units of the corresponding marginalized uncertainty from the fiducial relativistic model. The ``None'' column gives the shift obtained with a purely Newtonian angular power spectrum template (i.e. with all relativistic corrections neglected). The ``Doppler" case replaces the Newtonian peculiar velocity contribution with its relativistic counterpart. The other columns show the parameter shift when the Newtonian model is augmented with an additional relativistic contribution. For instance, the “Lensing” scenario includes the total lensing contribution together with the Newtonian model.} 
    \label{tab:tab2}
\end{table}

To understand this reduction, we further separate the lensing contribution according to the number and ordering of lensing fields entering the angular power spectrum. Denoting the Newtonian observed field by N, which accounts for the density, Kaiser RSD, and Newtonian velocity contributions, the singly integrated lensing terms are the cross-correlations $C_\ell^{N\kappa}$ and $C_\ell^{\kappa N}$. We compare these terms and their sum with the complete lensing model, which contains both cross terms as well as the lensing auto-correlation $C_\ell^{\kappa\kappa}$. Of course, for the auto-correlation case where $z_i = z_j$, the two cross-terms are identical, $C_\ell^{N\kappa}=C_\ell^{\kappa N}$. This symmetry does not hold for cross-correlations between distinct redshift bins or samples, however. Lensing of a background population by foreground structure is geometrically distinct from the reverse ordering, so in general $C_\ell^{N_i\kappa_j}\neq C_\ell^{\kappa_iN_j}$. 


\begin{figure}[t!]
    \centering
    \includegraphics[width=0.99\linewidth]{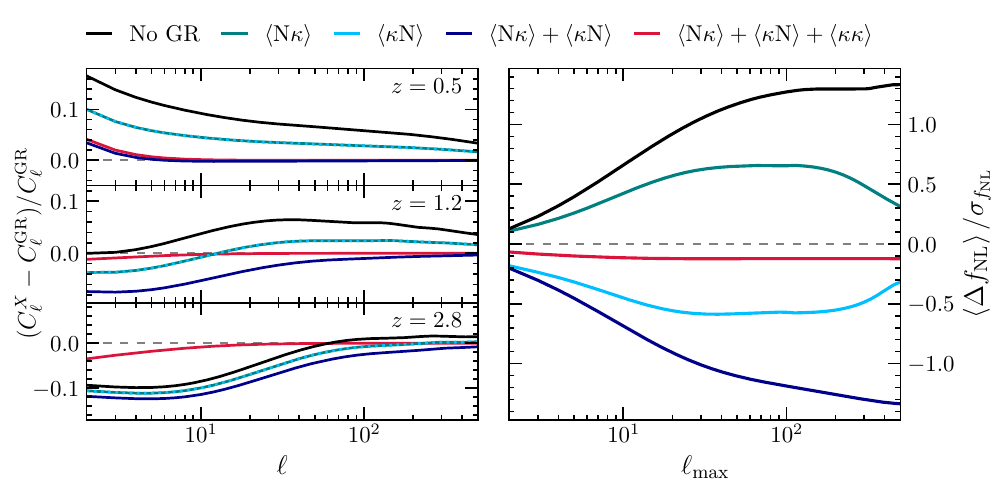}
    \caption{The role of weak gravitational lensing in the systematic shift of $f_{\rm NL}$. \textit{Left}: Fractional difference between angular power spectrum models and the full relativistic prediction for three representative auto-correlations of SPHEREx Sample 5, with mean redshifts at $z\sim 0.5,1.2,$ and $2.8$. The black curve (“No GR”) denotes the purely Newtonian template. The colored curves denote the Newtonian model augmented by different combinations of lensing contributions: teal corresponds to $\text{N}+\langle \text{N}\kappa\rangle$, light blue to $\text{N}+\langle \kappa \text{N}\rangle$, dark blue to $\text{N}+\langle \text{N}\kappa\rangle+\langle \kappa \text{N}\rangle$, and red to the full lensing template $\text{N}+\langle \text{N}\kappa\rangle+\langle \kappa \text{N}\rangle+\langle \kappa\kappa\rangle$, where N denotes the Newtonian density, Kaiser, and Newtonian velocity contributions. \textit{Right}: The systematic shift in the recovered $f_{\rm NL}$, in units of its marginalized uncertainty, for the complete five-sample analysis as a function of $\ell_{\rm max}$. Including only one set or both sets of lensing cross-correlation terms is not sufficient to reproduce the full relativistic result; only after including the lensing-lensing signal does the shift in $f_\text{NL}$ become small.}
    \label{fig:lensing_physical_shift}
\end{figure}

Figure~\ref{fig:lensing_physical_shift} illustrates the effect of the lensing terms on both the angular power spectrum and shift in $f_\text{NL}$. In the left panels, we show representative Sample 5 auto-correlations (we choose Sample 5 since it drives the signal on PNG), where the broad redshift windows make the lensing correction particularly visible. The Newtonian and lensing cross term model can lie on opposite sides of the full relativistic $C_\ell$, whereas adding $C_\ell^{\kappa\kappa}$ term brings the complete lensing model substantially closer to the relativistic template. The right panel shows how the same cancellation projects onto $f_{\rm NL}$. The Newtonian model produces a positive residual shift that saturates at order $\sim 1 \sigma$, while including either lensing cross term alone reduces the residual but does not eliminate it. Even still, the sum of the cross-correlations added to the Newtonian template reverses the sign without substantially reducing its magnitude. It is only when the lensing-lensing signal is included that this remaining shift is canceled, leaving only the small shift associated with the complete lensing model.

This distinction is important because it shows that an incomplete treatment of lensing can be misleading even when its constituent terms are individually large. In the present SPHEREx configuration, it is the full family of lensing contributions -- not any single lensing cross-correlation -- that accounts for most of the difference between the Newtonian and full relativistic angular power spectrum models. Once lensing is modeled consistently, the residual shift associated with neglecting the other relativistic contributions is small.

This lensing-dominated behavior is qualitatively consistent with the Fourier space results of Ref.~\cite{Addis2026Unbiased}, which emphasized the importance of the integrated relativistic contributions (i.e., lensing, ISW, and time delay) for unbiased inference of $f_{\rm NL}$. This also provides a useful point of comparison with Ref.~\cite{Guedezounme_2025}, where partial cancellation between the integrated and non-integrated relativistic contributions was found to suppress the net $f_{\rm NL}$ shift for some survey configurations. Our decomposition further shows that, in the present angular power spectrum analysis, an important cancellation occurs within the lensing contribution itself: the lensing cross terms generate a large parameter shift that is substantially canceled by the lensing-lensing signal, $C_\ell^{\kappa\kappa}$. In essence, our angular power spectrum results point to a similar physical picture, while further indicating that lensing is the dominant integrated contribution in our SPHEREx analysis; once the complete lensing contribution is included in the model, the residual parameter shift associated with neglecting the remaining relativistic terms is small.

A few minor caveats apply to our methods. First, the parameter space we probe is relatively small; we have fixed many of the cosmological parameters that SPHEREx can constrain. For instance, Ref.~\cite{Dore::2014spherex} included the parameter set $\{\Omega_c h^2, \Omega_b h^2, \Omega_\nu,\Omega_\text{de},w_0, w_a \}$ as well as marginalizing over the galaxy bias $b_g$ in each of their forecasted redshift bins. While we expect that the local primordial non-Gaussianity constraints are relatively insensitive to these extra parameters, studying dark energy constraints from SPHEREx, for example, would demand such an extension.  

On a related note, the level of impact of the individual relativistic contributions depends on the survey parameters, including the choices of $\mathcal Q(z)$ and $b_e(z)$. The dominance of the lensing contribution in the systematic shift demonstrated in Table~\ref{tab:tab2} and Fig.~\ref{fig:lensing_physical_shift} is therefore specific to the bias and selection functions adopted here. For other choices of these quantities, the local Doppler and potential terms can contribute more substantially to the recovered primordial non-Gaussianity. For example, Ref.~\cite{Jeong::2012gr} characterized these relativistic corrections in terms of an effective $f_{\rm NL}$, corresponding to the apparent shift in local PNG that would mimic their large-scale contribution, and found values of order unity for some choices of galaxy and magnification bias. This sensitivity provides further motivation for examining the treatment of the bias functions entering our forecast.  

We also hold the galaxy, evolution, and magnification biases fixed in this work. Ref.~\cite{Addis2026Unbiased} found that marginalizing over uncertainties in these bias functions degrades the resulting $f_\text{NL}$ constraints, and that the lensing contribution itself carries information about the magnification bias. We therefore also investigated the impact of marginalizing over $b_g$ and $\mathcal Q$ in our analysis. Although we do not explicitly present the results of this investigation, we found that marginalizing over these two nuisance parameters degrades $\sigma_{f_\text{NL}}$ by approximately $10$--$20\%$ and reduces the corresponding systematic shift by a commensurate amount. These bias functions, however, are not completely unconstrained nuisance quantities: the galaxy bias can be further constrained using higher-order clustering statistics such as the bispectrum, while the evolution and magnification biases can be informed by measurements of the luminosity function. Consequently, treating these quantities as freely marginalized parameters provides a conservative upper limit on the degradation of the cosmological constraints. A more realistic forecast would incorporate the statistical information from these complementary observables jointly with the angular power spectrum. Nevertheless, even under this conservative treatment, neglecting the relativistic contributions can induce a shift in the recovered $f_{\rm NL}$ of order its statistical uncertainty. A more complete treatment of bias marginalization and its effect on cosmological inference from SPHEREx will be included in a forthcoming analysis.

Additionally, as mentioned in \refsec{PNG}, we assumed the universality of the halo mass function to get our expression for $b_\Phi(z)$ in \refeq{b_phi}. Therefore, while we do provide the independent measure of $f_\text{NL}$, future forecasts and actual measurements of primordial non-Gaussianity will likely include constraints on this bias parameter; these constraints will naturally degrade the accuracy of any measurement of $f_\text{NL}$.

The absolute parameter uncertainties should be interpreted in light of the simplified parameter space adopted here, while the systematic shifts quantify the bias induced by neglecting relativistic effects within our fiducial analysis. A broader parameter space may redistribute this bias among correlated parameters, but there is no reason to expect the underlying systematic to disappear. What we have shown is that the survey setup (e.g., redshift bins, choice of $\ell_\text{max}$) can impact the amount of information that can be extracted on parameters like $f_\text{NL}$. Nonetheless, the expected accuracy of primordial non-Gaussianity was achieved in the multi-sample analysis. Furthermore, ignoring relativistic effects can have a statistically significant impact on the estimate of cosmological parameters; therefore, they must be taken into account.


\subsection{Uncertainty and bias on $f_\text{NL}$ as a function of magnification bias}\label{sec:Q_bias_fnl_dep}

\begin{figure}[t!]
    \centering
    \includegraphics[width=0.99\linewidth]{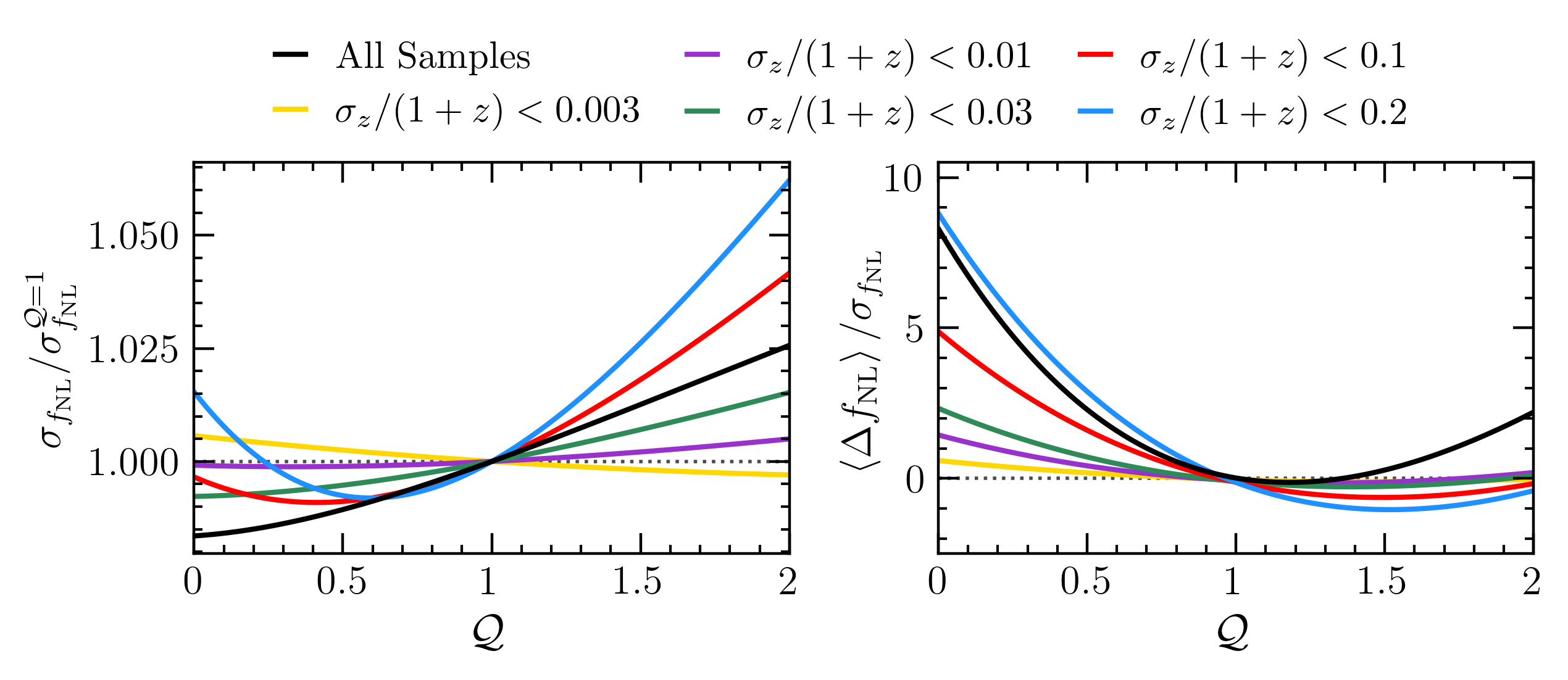}
    \caption{\textit{Left}: The uncertainty on $f_\text{NL}$, unmarginalized and with the other cosmological parameters held fixed, using the same survey setup of the five subsamples, but with constant magnification bias  $\mathcal Q$, and neglecting evolution bias $b_e(z) = 0$. The uncertainty is compared to the $\mathcal Q = 1$ case, where the relativistic effects are near their minimum effect. Among the individual subsamples, those with poorer redshift precision show the strongest variation in $\sigma_{f_\text{NL}}$ as $\mathcal Q$ is varied. \textit{Right}: The systematic shift in the recovered $f_\text{NL}$ from using the Newtonian $C_\ell$ expressed in units of its uncertainty, as a function of the assumed constant magnification bias $\mathcal Q$. The induced $f_\text{NL}$ shift reaches a minimum near $\mathcal Q \simeq 1$ in this calculation. At $\mathcal Q = 1$, the terms proportional explicitly to $1-\mathcal Q$, including the lensing and time delay contributions in our kernel, vanish; the remaining relativistic terms, however, do not. Additionally, at lower magnification bias values, the shift becomes less trivial.} 
    \label{fig:constQbias}
\end{figure}

To examine the robustness of our result, we computed the uncertainty and shift in $f_\text{NL}$ using the same radial selection functions and redshift distributions for each SPHEREx subsample, as presented in \reffig{spherex_param_fig}, but with $b_e(z) = 0$ and $\mathcal Q = \text{constant}$. Additionally, we compared the results to previous studies of the effect of relativistic corrections in the two-point function on estimates of $f_\text{NL}$. We plot the results of this analysis in \reffig{constQbias}. The left panel illustrates that an unmarginalized uncertainty of $\sigma_{f_\text{NL}} \sim 1$ is achieved regardless of magnification bias in the multi-sample result, while in contrast, the shift in $f_\text{NL}$ from neglecting relativistic effects is dependent on magnification bias. Additionally, we show that the strength of this dependence is largely related to the number of redshift bins in the sample; the more spectroscopic the survey (i.e., more bins, finer bins), the less the magnification bias matters.

The results presented here show a shift $\left<\Delta f_{\text{NL}}\right>$ comparable to previous work. For instance, a bias of order $4 \sigma_{f_\text{NL}}$ was estimated using the angular power spectrum \cite{Camera2015Einstein}. A separate analysis with the SFB power spectrum $ C_\ell(k,k')$ -- outlined in Eq.~\eqref{eq:SFB} -- showed a dependence of this shift on the magnification bias $\mathcal Q$ and maximum multipole $\ell_\text{max}$, but most of their results showed a shift of order $\sigma_{f_\text{NL}}$ \cite{Semenzato::2025sfb}. Another study of the usual power spectrum $P(k)$ extended to a smaller fraction of the sky showed a similar shift, of the same order as our result \cite{Guedezounme_2025}. Our setup here best mimics the methods of the former study, with one major difference: they investigate the effect of redshift binning, but do not vary the number of redshift bins in their analysis, as we do. Additionally, they do not explore the difference in the shift in $f_\text{NL}$ from a mock survey with closer to spectroscopic redshift precision (like SPHEREx Samples 1 and 2) versus those with large photo-$z$ errors (like Samples 4 and 5).

The right-hand side of Fig.~\ref{fig:constQbias} resolves some of the minor apparent differences between our $f_\text{NL}$ shift and that of previous studies like Ref.~\cite{Semenzato::2025sfb}. The difference is most pronounced at low magnification bias. For example, near $\mathcal Q=0$, the nearly spectroscopic Samples 1 and 2 yield shifts of order the marginalized uncertainty, $\langle\Delta f_\text{NL}\rangle/\sigma_{f_\text{NL}}\sim\mathcal O(1)$, whereas the samples with poorer redshift precision can yield shifts of several $\sigma_{f_\text{NL}}$, with the complete multi-sample result exceeding $5\sigma_{f_\text{NL}}$. All of these shifts are strongly reduced as $\mathcal Q$ approaches unity. Therefore, the shift in $f_\text{NL}$ from ignoring relativistic effects we see in \reffig{corner_shift} seems to be driven in part by the choice of flux limit, luminosity function, and, consequently, magnification bias, as well as redshift uncertainty. This effect can be seen in \reffig{Q_fnl_comp_plot}: generally speaking, the impacts of varying $f_\text{NL}$ and $\mathcal Q$ are somewhat degenerate. The qualitative similarity between varying $f_\text{NL}$ and varying $\mathcal Q$ is most apparent at low multipole. The shifts shown here should not be compared directly with those of \refsec{bias_fnl} as a pure measure of the effect of marginalization, because the two calculations also adopt different evolution and magnification bias prescriptions. In Sec.~\ref{sec:bias_fnl}, we use the fiducial WISE-derived $b_e(z)$ and $\mathcal Q(z)$, whereas the present test fixes $b_e = 0$ and assumes constant magnification bias.

These results illustrate that the gain in $f_\text{NL}$ information at low magnification bias (i.e., $\mathcal Q \rightarrow 0$) comes at a cost: one \textit{must} take into account the relativistic corrections to the angular power spectrum (or an equivalent two-point statistic), especially in this regime of $\mathcal Q$, lest the recovered $f_{\rm NL}$ differ from the true value by order unity or more. Additionally, we have shown that while surveys with less photometric redshift error are less sensitive to this shift in cosmological parameters from ignoring relativistic effects, they will still misestimate cosmological parameters like $f_\text{NL}$ by $\mathcal O(1)$ or more.

\section{Conclusion}\label{sec:conclusion}

In this work, we motivated the need to include the general relativistic terms in the observed galaxy density contrast that previous generations of galaxy surveys have ignored; the need is particularly acute for current surveys that probe significant fractions of the sky (e.g., SPHEREx). To account for these additional terms, we introduced \textsc{PowerFull}, an extension to \textsc{2-FAST}, a \texttt{Julia}-based code that efficiently evaluates integrals involving one or two spherical Bessel functions. We then use this new code to compute the full relativistic galaxy angular power spectrum and to estimate the sensitivity of a SPHEREx-like survey to parameters like $f_\text{NL}$, which parameterizes local primordial non-Gaussianity at leading order.

Next, we investigated the uncertainty and bias on these parameters assuming an approximate but physically motivated luminosity function and redshift binning scheme. We also computed the shift in the parameters as a result of ignoring relativistic effects as a function of maximum multipole $\ell_\text{max}$. We showed that the expected uncertainty on cosmological parameters for SPHEREx is obtained from the multi-sample analysis, and that, even though we do not include non-linearity in our power spectrum, the results remain representative of a full-sky galaxy survey on the retained linear scales. More broadly, ignoring relativistic effects induces a non-trivial amount of bias on cosmological parameters across the survey configurations examined here, and, therefore, the additional terms that arise in the two-point function must be taken into account. Additionally, we showed that the resulting shift in cosmological parameters from ignoring relativistic corrections was strongly dependent on the magnification bias of the survey, and, as a result, the flux limit and luminosity function of the survey.

We include various tests of the \textsc{PowerFull} code in the appendix. The original \textsc{2-FAST} package for fast computation of spherical Bessel function integrals is available to the public at \href{https://github.com/hsgg/TwoFAST.jl}{https://github.com/hsgg/TwoFAST.jl}; the \textsc{PowerFull} extension introduced in this work is publicly available at \href{https://github.com/greglukens/PowerFull.jl}{https://github.com/greglukens/PowerFull.jl}. Additionally, we plan to introduce a mechanism for computing the tree-level bispectrum including second order relativistic effects; this will also be helpful for constraining primordial non-Gaussianity in surveys like SPHEREx. In a forthcoming paper, we will also address how the full-sky angular power spectrum can be used for geometrical and dynamical measurements of dark energy.

\section*{Acknowledgements}

We would like to thank Henry Gebhardt for his assistance in computational matters and valuable discussion. Additionally, we thank the Korea Institute for Advanced Study (KIAS) for their hospitality while working on this project. GL and DJ were supported by NSF grants AST-2307026 and AST-2407298 at the Pennsylvania State University (PSU). DJ was also supported by a KIAS Individual Grant, PG088301. This work was also supported by the Center for Advanced Computation at KIAS. 


\appendix

\section{Detailed derivation of the redshift-space kernel}\label{app:delta_s_derivation}
We derive the redshift-space kernel $F_\ell(k,r)$ introduced in \refsec{newtonian}. Substituting the TAM expansions of $\delta$, $v_r$, and $\partial_r v_r$ [\refeqs{delta_tam}{v_r_tam_der}] into \refeq{delta_s_tam} gives
\be
\label{eq:app_delta_s_tam}
    \delta_s(\bfx) =D(r) \sum_{\ell m} \int \frac{k^2 \dd k}{(2\pi)^3}\delta^k_{\lm}  4\pi i^\ell \left[b_g j_\ell(kr) - f \frac{\alpha}{r}\frac{j_\ell'(kr)}{k} - f j''_\ell(kr)\right]  Y_{\lm}(\nhat) \,.
\ee
To identify the relation between $\delta_{s,\lm}^k$ and $\delta_{\lm}^k$, we project onto the TAM basis using \refeq{phi_lm}:
\ba
\label{eq:delta_conversion_derivation1}
    \delta_{s,\lm}^k &\,= \int \dd^3 x \, \delta_s(\bfx) \left[4\pi i^\ell \Psi^k_{\lm}(\bfx) \right]^* \vs
    &\,= \int \dd^3 x \, \sum_{\ell'm'}\int \frac{k'^2 \dd k'}{(2\pi)^3} \delta^{k'}_{\ell' m'} 4\pi i^{\ell'}D(r) F_{\ell'}(k',r) Y_{\ell'm'}(\nhat) \left[4\pi i^\ell \Psi^k_{\lm}(\bfx) \right]^*\,,
\ea
where $F_\ell(k,r)$ is the redshift-space kernel defined in \refeq{redshift_space_kernel}. Separating the volume integral into radial and angular parts,
\ba
\label{eq:delta_conversion_derivation2}
    \delta_{s,\lm}^k = \int \dd r \,r^2  \int \dd^2 \nhat \sum_{\ell'm'}\int \frac{k'^2 \dd k'}{(2\pi)^3} \delta^{k'}_{\ell'm'} 4\pi i^{\ell'}D(r) &\, F_{\ell'}(k',r) Y_{\ell'm'}(\nhat)  \vs
    &\, \times 4\pi (-i)^\ell j_\ell(kr)Y^*_{\ell m}(\nhat)\,,
\ea
and applying the orthogonality relation
\be
\label{eq:spherical_harmonics_relation_pt2}
    \int \dd^2 \nhat \, Y_{\ell' m'}(\nhat) Y^*_{\lm}(\nhat) = \delta_{\ell\ell'}\delta_{mm'}\,,
\ee
we obtain
\ba
\label{eq:delta_conversion_derivation3}
    \delta_{s,\lm}^k &\,= \frac{2}{\pi}\int   \dd r \, r^2 \int \dd^2 \nhat \sum_{\ell'm'} \delta_{\ell\ell'}\delta_{mm'}\int k'^2 \dd k'\, \delta^{k'}_{\ell' m'} i^{\ell'-\ell}D(r) F_{\ell'}(k',r) j_\ell(kr) \vs
    &\,= \frac{2}{\pi}\int \dd k'\, k'^2  \,\delta^{k'}_{\lm} \int \dd r \,r^2  \, j_\ell(kr) D(r) F_\ell(k',r)\,,
\ea
which is consistent with \refeq{delta_s_tam_explicit}.

\section{Parametric comparison of Newtonian and relativistic RSD kernels}\label{app:newt_gr_comp}
The full Newtonian RSD expression in \refeq{redshift_space_kernel} contains, in addition to the usual Kaiser effect, the velocity term with a coefficient $\alpha(r)$. This term appears to be quite similar to the {\it velocity} term in the full relativistic expression in \refeq{F_ell_GR}. In this appendix, we compare the coefficients of the velocity term in these two expressions.

The coefficient $\alpha(r)$ in front of the velocity term [\refeq{alpha}] of the Newtonian treatment in \refeq{redshift_space_kernel} is
\be
\alpha(r) = \frac{\dd\ln[r^2f(r)D(r)a^3(r)\bar{n}_g(r)]}{\dd \ln r}\,.
\ee
With the evolution bias $ b_e=\dd \ln (a^3\bar{n}_g) / \dd \ln a$ defined in \refeq{def_be} and the identity $\dd \ln a/\dd\ln r = -aHr$, we may expand this coefficient as
\ba
\label{eq:alpha_coeff_expanded}
    \alpha(r)  &\, = -aHr \left( - \frac{2}{aHr} + \frac{\dd\ln (fD)}{\dd\ln a}    + b_e \right)\,.
\ea
In order to simplify further, we recall the definition of the linear growth rate $f=\dd \ln D/\dd\ln a$ [\refeq{linear_growth_rate}] to obtain 
\ba
\label{eq:dlnfdlna}
    \frac{\dd\ln (fD)}{\dd\ln a} &\, = \frac{a}{fD}\frac{\dd}{\dd a}\left(a\frac{\dd D}{\dd a}\right) 
    = 1 + \frac{a^2}{fD}\frac{\dd^2 D}{\dd a^2}\,.
\ea
We obtain the second derivative of the growth rate from the differential equation for the linear growth factor,
\ba
\label{eq:continuity_eq}
    \frac{\dd^2 D}{\dd a^2} + \left(\frac{\dd\ln H}{\dd a} + \frac{3}{a}\right) \frac{\dd D}{\dd a} - \frac{3}{2}\frac{\Omega_m}{a^2} D &\,= 0\,,
\ea
which yields
\ba
\label{eq:dlnfdlna_again}
    \frac{\dd\ln (fD)}{\dd\ln a} = -2 + \frac{3}{2}\frac{\Omega_m}{f} - \frac{\dd\ln H}{\dd\ln a}  \,.
\ea
If we plug this back into \refeq{alpha_coeff_expanded}, we are left with the final expression for the Newtonian coefficient:
\ba
\label{eq:alpha_coeff_expanded_simplified}
    -f\frac{\alpha}{r}  = aHf \left(- \frac{2}{aHr}  - 2 + \frac{3}{2}\frac{\Omega_m}{f}  - \frac{\dd\ln H}{\dd\ln a}    + b_e\right)\,.
\ea

Now, let us compare this to the relativistic coefficient present in the velocity term of the relativistic treatment in \refeq{B}. Ignoring magnification bias (i.e., $\mathcal Q = 0$), we find the coefficient to be
\be
\label{eq:B(Q=0)}
    aH \mathcal B(\mathcal Q = 0 ) = aHf \left(b_e - \frac{\dd\ln H}{\dd\ln a} - \frac{2}{aHr} - 1 \right)\,,
\ee
which is related to the Newtonian $\alpha$ term as
\ba
\label{eq:alpha_coeff_expanded_simplified_CDM}
    -f\frac{\alpha}{r}  = aH \mathcal B(\mathcal Q = 0) + aH \left(\frac{3}{2} \Omega_m - f\right)  \,.
\ea

Armed with this, we can rewrite the Newtonian expression for the redshift-space kernel \refeq{redshift_space_kernel} as
\be
\label{eq:newtonian_fell_new}
    F_\ell(k,r) = b_g j_\ell(kr) - f j_\ell''(kr) + aH\mathcal B(\mathcal Q = 0) \frac{j_\ell'(kr)}{k} + aH \left(\frac{3}{2} \Omega_m - f\right) \frac{j_\ell'(kr)}{k}
\ee

Note that the first three terms in \refeq{newtonian_fell_new} coincide with the first three terms from the relativistic treatment in \refeq{F_ell_GR} with no magnification bias; $\mathcal Q = 0$. In other words, the Newtonian derivation we have presented in this work is clearly a component of complete relativistic result. In particular, the Newtonian treatment captures all quantities that depend on the angle between line-of-sight direction, $\nhat$, and the Fourier wave vector, $\bfk$, which appears in the radial derivative of the spherical Bessel function. The Newtonian formalism, then, just lacks the relativistic terms that arise from the photon geodesics and gravitational potential.

On the other hand, the last term in \refeq{newtonian_fell_new} is only present in this expression; it is not in the relativistic treatment at all. Careful inspection shows that the term is missing in the relativistic formalism because of a subtle cancellation with the gravitational potential. Namely, the relativistic redshift perturbation, $\delta z$, given in \refeq{deltaz_GR}, contains effects from both the \textit{velocity}\footnote{Typically, the term $\partial_\parallel E'$ in the synchronous comoving gauge is interpreted as the Newtonian velocity; this is evident when comparing \refeq{radial_der_GR} with \refeq{v_r_tam}.} ($\partial_\parallel E'$) and gravitational potential ($E''$), so its radial derivative yields
\ba
    \frac{\partial}{\partial r} \left(\frac{\delta z}{aH} \right) &\,= \left[ 1 - \frac{1}{aH}\frac{\dd H}{\dd z} \right]\delta z + \frac{1}{aH} \left[ (\partial_\parallel - \partial_\tau)\partial_\parallel E' + (\partial_\parallel - \partial_\tau) E'' + \phi' + E'''\right] \vs
    &\,= \left[ 1 - \frac{1}{aH}\frac{\dd H}{\dd z} \right]\delta z + \frac{1}{aH} \partial^2_\parallel E' \,,
\ea
where the cancellation of the velocity and potential terms is apparent. Here, we use $\partial_r = \partial_\parallel - \partial_\tau$ and $\phi'=0$; recall that a prime indicates a derivative with respect to the conformal time, $\tau$. In the Newtonian expression, however, we have only the first term ($v_r = \partial_{\parallel} E'$), so the radial derivative becomes
\ba
    \frac{\dd}{\dd r}\left(\frac{v_r}{aH}\right) &\,= \left[ 1 - \frac{1}{aH}\frac{\dd H}{\dd z} \right]v_r + \frac{1}{aH} (\partial_\parallel - \partial_\tau)\partial_\parallel E' \vs
    &\,= \left[ 1 - \frac{1}{aH}\frac{\dd H}{\dd z} \right]\delta z + \frac{1}{aH} \partial^2_\parallel E' - \frac{1}{aH}\partial_\parallel E^{\prime\prime} ,
\ea
and the last term here is the same as the last term present in \refeq{newtonian_fell_new}.

\section{Review of the 2-FAST algorithm}\label{app:twofast_review}
The 2-FAST algorithm accelerates the computation of the overlapping integral involving one and two spherical Bessel functions 
\be
\label{eq:w_ellell'}
    w_{\ell,\ell'}(r,r') = \frac{2}{\pi} \int_0^\infty \dd k \, k^2 P(k) j_\ell(kr) j_{\ell'}(kr')\,,
\ee
by using the FFTLog transformation of the power spectrum $P(k)$ \cite{Hamilton::2000FFTlog}. 
Note that integrating these integrals is challenging because of the highly oscillatory nature of the spherical Bessel function.

The FFTLog performs a fast Fourier transform (FFT) of the power spectrum that has been uniformly sampled in logarithmic wavenumber space. To mitigate numerical artifacts---such as aliasing--- that can arise in this procedure, a bias parameter $q$ is introduced so that the function actually transformed by FFTLog is
\be
\label{eq:biased_pk}
    \left(\frac{k}{k_0}\right)^{3-q}P(k) = e^{(3-q)\kappa}\, P(k_0e^\kappa)\,,
\ee
where the logarithmic variable $\kappa$ is defined as
\be
\label{eq:kappa_fftlog}
    \kappa = \ln \left(\frac{k}{k_0} \right)\,,
\ee
with $k_0$ denoting a pivot wavenumber, typically chosen to be the midpoint of the logarithmic interval. The pertinent Fourier pair used in the 2-FAST algorithm is then
\ba
\label{eq:phi_q}
    \phi^q(t) &\, = \int_{-\infty}^\infty \frac{\dd\kappa}{2\pi} e^{i\kappa t} e^{(3-q)\kappa} \,P(k_0 e^{\kappa})\,,   \\
\label{eq:P(k)}
    P(k) &\, = e^{-(3-q)\kappa} \int_{-\infty}^\infty \dd t \,e^{-i\kappa t} \phi^q(t)\,.
\ea

The key insight for the fast computation is that, with the new variables $\rho=\ln(r/r_0)$ and $R=r'/r$,
\ba
\label{eq:r_rho_first}
w_{\ell,\ell'}(r,r')
=&
\frac{2k_0^3}{\pi}\int_{-\infty}^\infty
\dd \kappa \,  e^{3\kappa} P(k_0e^\kappa)j_{\ell}(k_0r_0 e^{\kappa+\rho} )j_{\ell'}(k_0r_0Re^{\kappa+\rho})
\vs
=&
\frac{2k_0^3}{\pi} e^{-q\rho} \int_{-\infty}^\infty
\dd \kappa \,  e^{(3-q)\kappa} P(k_0e^\kappa) e^{q(\kappa+\rho)} j_{\ell}(k_0r_0 e^{\kappa+\rho} )j_{\ell'}(k_0r_0Re^{\kappa+\rho})\,,
\ea
 we can transform the overlap integral into a convolution. 
That is, by using the Fourier transformation
\ba
e^{q\sigma} j_\ell(\alpha e^\sigma)j_{\ell'}(\beta e^{\sigma}) = \int\frac{\dd t}{2\pi} e^{it\sigma} M_{\ell\ell'}^q(t;\alpha,\beta)\,,
\label{eq:Mellellp}
\ea
and \refeq{P(k)}, the overlap integral reduces to
\ba
w_{\ell,\ell'}(r,r')
=&
\frac{2k_0^3}{\pi} e^{-q\rho} \int_{-\infty}^\infty
\dd \kappa \, \int \dd t' e^{-i\kappa t'}\phi^q(t')
\int\frac{\dd t}{2\pi} e^{it(\kappa+\rho)} M_{\ell\ell'}^q(t;k_0r_0,k_0r_0R)
\vs
=&
\frac{2k_0^3}{\pi} e^{-q\rho} 
\int \dd t' \phi^q(t')
\int \dd t \delta^D(t-t') e^{it\rho} M_{\ell\ell'}^q(t;k_0r_0,k_0r_0R)
\vs
=&
4k_0^3 e^{-q\rho} 
\int \frac{\dd t}{2\pi}\, e^{it\rho} \phi^q(t)  M_{\ell\ell'}^q(t;k_0r_0,k_0r_0R)
\,.
\ea
The last equation is the way that the 2-FAST algorithm computes the overlapping integral.

The last task is to compute the Fourier transformation defined in \refeq{Mellellp}: 
\be
M_{\ell\ell'}(t;\alpha,\beta) 
= \int \dd\sigma \, e^{-it\sigma +q\sigma} j_\ell(\alpha e^\sigma)j_{\ell'}(\beta e^{\sigma})\,,
\ee
which can be further simplified (here, $s=\alpha e^\sigma$), 
\be
M_{\ell\ell'}(t;,\alpha,\beta) 
= \int_0^\infty \frac{\dd s}{s}\left(\frac{s}{\alpha}\right)^{q-it}j_\ell(s)j_{\ell'}(Rs)
= \alpha^{it-q} \int_0^\infty \dd s \, s^{q-1-it}j_\ell(s)j_{\ell'}(Rs)\,.
\ee
Note that the integral can be expressed in terms of the Gauss hypergeometric function as
\ba
&\int_0^\infty \dd s\, s^n j_\ell(s)j_{\ell'}(Rs) 
\vs
=&\, 2^{n-2}R^{\ell'}\pi 
\frac{\Gamma\left(\frac{1+\ell+\ell'+n}{2}\right)}{\Gamma\left(\frac{2+\ell-\ell'-n}{2}\right)\Gamma\left(\frac32+\ell'\right)}
{}_2F_1\left(
\frac{-\ell+\ell'+n}{2},
\frac{1+\ell+\ell'+n}{2};
\frac{3}{2} + \ell' ; R^2
\right)\,,
\ea
for $|R|<1$, $\mathrm{Re}(n)<2$, and $\mathrm{Re}(n+\ell+\ell')>-1$. Furthermore, for $R = 1$, the Gauss hypergeometric function converges only if $\mathrm{Re}(1-n)>0$, which restricts the range for the biasing parameter to be $-\ell-\ell'<q<2$.

The 2-FAST implementation developed by Ref.~\cite{Gebhardt::2018twofast} uses the recursion relation of the Gauss hypergeometric function to compute $M_{\ell,\ell'}(t;\alpha,\beta)$. In particular, their strategy begins the recursion from a sufficiently high $\ell$ (typically more than 10,000) with an arbitrary initial condition and walks along the stable recursion direction ($\Delta\ell<0$) so that the result converges to the correct value of integration for the desired values of $\ell$. One could improve the algorithm by starting the recursion with better approximated initial conditions, but we find the current implementation sufficient for all practical purposes.

\section{The \textsc{PowerFull} package}\label{app:powerfull_details}
In this section, we present the \textsc{PowerFull} package that extends the 2-FAST implementation of the Kaiser effect to incorporate the full-relativistic galaxy power spectrum, including the local-type primordial non-Gaussianity effects.

\subsection{Extension of 2-FAST implementation}\label{app:twofast_extend}
\begin{table}[t!]
    \centering
    {\renewcommand{\arraystretch}{1.3}
    \begin{tabular}{|c|c|c|c|c|}
        \hline
         Terms & Integrals & $p_{\rm base}$ & $n$ & $q$ \\
         \hline \hline
         \footnotesize{Galaxy $\times$ (Galaxy, RSD), RSD $\times$ RSD} & $w_{\ell,00}^0\,,w_{\ell,02}^0\,,w_{\ell,20}^0\,,w_{\ell,22}^0$ & $0$ & $0$ & $1.3$ \\
         \hline
         \footnotesize{\begin{tabular}{@{}c@{}}(Galaxy, RSD) $\times$ \\ (Doppler, Potential, ISW, TD, Lensing) \end{tabular}} & 
         \begin{tabular}{@{}c@{}}$w_{\ell,10}^{-1}\,,w_{\ell,01}^{-1}\,,w_{\ell,12}^{-1}\,,w_{\ell,21}^{-1}$\\$w_{\ell,00}^{-2}\,,w_{\ell,02}^{-2}\,,w_{\ell,20}^{-2}\,,$ \end{tabular}
         &  \begin{tabular}{@{}c@{}}$0$ \\ $-2$ \end{tabular} 
         & $0$ 
         & \begin{tabular}{@{}c@{}} $1.5$ \\ $-0.2$ \end{tabular} \\
         \hline
         \footnotesize{\begin{tabular}{@{}c@{}}Doppler $\times$ (Doppler, Potential, ISW, TD, Lensing),\\ Potential $\times$ (Potential, ISW, TD, Lensing), \\ ISW $\times$ ISW, TD $\times$ TD, Lensing $\times$ Lensing\end{tabular}}& 
         \begin{tabular}{@{}c@{}}
         $w_{\ell,11}^{-2}$
         \\
         $w_{\ell,10}^{-3}\,,w_{\ell,01}^{-3}$
         \\
         $w_{\ell,00}^{-4}$
         \end{tabular}
         & \begin{tabular}{@{}c@{}}$-2$ \\ $-2$ \\ $-4$ \end{tabular}  
         & $0$ 
         & \begin{tabular}{@{}c@{}}$-0.2$ \\ $0.5$ \\ $-1.92$ \end{tabular} \\
         \hline
         \footnotesize{ PNG $\times$ (Galaxy, RSD)} & $u_{\ell,00}^{-2}\,,u_{\ell,20}^{-2}\,,u_{\ell,02}^{-2}$ & $-2$ & $-1$ & $0.5$\\
         \hline
         \footnotesize{ PNG $\times$ (Doppler, Potential, ISW, TD, Lensing)} 
         & \begin{tabular}{@{}c@{}}
         $u_{\ell,10}^{-3}\,,u_{\ell,01}^{-3}$
         \\
         $u_{\ell,00}^{-4}$
         \end{tabular}
         & \begin{tabular}{@{}c@{}}$-2$ \\ $-4$ \end{tabular}  
         & $-1$ & \begin{tabular}{@{}c@{}} $0.8$ \\ $-1.08$ \end{tabular}\\
         \hline
         \footnotesize{ PNG $\times$ PNG} & $v_{\ell,00}^{-4}$ & $-4$ & $-2$ & $0.02$\\
         \hline
    \end{tabular}
    }
    \caption{The $w_{\ell,jj'}^p$, $u_{\ell,jj'}^p$, $v_{\ell,jj'}^p$ functions we implement {\sc TwoFAST} for the fully-relativistic calculation of the angular power spectrum. The $p_{\rm base}$ is the base $p$ value for the {\sc TwoFAST} implementation when using the recursion relation in \refeqs{w01_twofast}{w21_twofast}. The last column shows the optimal values of the biasing parameter $q$ defined in \refeq{biased_pk}. The terms on the left provide the physical meaning behind the integrals, while $n$ denotes the power of the transfer function $T(k)$ used with the matter power spectrum inside the integrals. Here, ``TD" stands for time delay. The value of $q$ affects the convergence behavior of the integrals and, thus, requires careful selection. In \refapp{q_optimization}, we discuss the choice of $q$ for each case in more detail.}
    \label{tab:q_val_table}
\end{table}
The angular power spectrum that we want to compute is in the form of
\ba
\label{eq:cl_app}
    C_\ell(r,r') = D(r)D(r')\frac{2}{\pi} \int \dd k \, k^2 P(k) F_\ell(k,r) F_\ell(k,r')\,,
\ea
where the relativistic kernels $F_\ell(k,r)$ takes the form 
\refeq{F_ell_GR} for the Gaussian terms and the non-Gaussian corrections in \refeq{F_ell_modification}. 

The original 2-FAST algorithm in Ref.~\cite{Gebhardt::2018twofast} computed the overlapping integration needed for the computation of the simplest Kaiser formula with the kernel of \refeq{redshift_space_kernel} with $\alpha=0$. As a result, the original {\sf TwoFAST.jl} only outputs the four integrals in \refeq{kaiser_Cl}: $w_{\ell,00}^0$, $w_{\ell,02}^0$, $w_{\ell,20}^0$, and $w_{\ell,22}^0$, where 
\be
\label{eq:wljj'_app}
    w_{\ell,jj'}^p(r,r') = \frac{2}{\pi}\int \dd k\,k^{2+p} P(k) j_\ell^{(j)}(kr) j_\ell^{(j')}(kr')\,.
\ee
There, the integrals $w_{\ell,jj'}^p$ defined in \refeq{wljj'_app} are computed in terms of $w_{\ell,\ell'}(r,R)$ by exploiting the recursion relation for derivatives of spherical Bessel functions. Specifically, the second derivative of the spherical Bessel function may be written as
\be
\label{eq:j_ell''}
    j_\ell''(x) = f_{-2} \,j_{\ell-2}(x) + f_0 \,j_\ell(x) + f_2 \,j_{\ell+2}(x)\,,
\ee
with
\be
\label{eq:f0,f2,fneg2}
    f_{-2} = \frac{\ell(\ell-1)}{(2\ell - 1)(2\ell + 1)}\,, \qquad f_0 = -\frac{2\ell^2 + 2\ell -1}{(2\ell - 1)(2\ell +3)}\,, \qquad f_2 = \frac{(\ell + 1)(\ell + 2)}{(2\ell+1)(2\ell+3)}\,,
\ee
which allows the recast of $w_{\ell,00}^p$, $w_{\ell,20}^p$, and $w_{\ell,02}^p$ as
\ba
\label{eq:w_ell_00}
    w_{\ell,00}^p(r,R) &\,= w_{\ell,\ell}^p(r,R)\,, \\
\label{eq:w_ell_02}
    w_{\ell,02}^p(r,R) &\,= \begin{pmatrix} f_{-2} & f_0 & f_2 \end{pmatrix} \begin{pmatrix} w_{\ell,\ell-2}^p(r,R) \\ w_{\ell,\ell}^p(r,R) \\ w_{\ell,\ell+2}^p(r,R) \end{pmatrix}\,, \\[5pt]
\label{eq:w_ell_20}
    w_{\ell,20}^p(r,R) &\,= \begin{pmatrix} f_{-2} & f_0 & f_2 \end{pmatrix} \begin{pmatrix} w_{\ell-2,\ell}^p(r,R) \\ w_{\ell,\ell}^p(r,R) \\ w_{\ell+2,\ell}^p(r,R) \end{pmatrix}\,, \\[5pt]
\label{eq:w_ell_22}
    w_{\ell,22}^p(r,R) &\,= \begin{pmatrix} f_{-2} & f_0 & f_2 \end{pmatrix} \begin{pmatrix} w_{\ell-2,\ell-2}^p(r,R) & w_{\ell-2,\ell}^p(r,R) & w_{\ell-2,\ell+2}^p(r,R) \\ w_{\ell,\ell-2}^p(r,R) & w_{\ell,\ell}^p(r,R) & w_{\ell,\ell+2}^p(r,R) \\ w_{\ell+2,\ell-2}^p(r,R) & w_{\ell+2,\ell}^p(r,R) & w_{\ell+2,\ell+2}^p(r,R) \end{pmatrix} \begin{pmatrix} f_{-2} \\ f_0 \\ f_2 \end{pmatrix}\,.
\ea

In addition to these, the implementation of the angular power spectrum for the full relativistic kernels demands the computation of $w_{\ell,jj'}^p$ with various combinations of $j,j'\in\{ 0,1,2\}$, as well as $p=\{0,-2,-4\}$ as listed in the first three rows of Table.~\ref{tab:q_val_table}. For the computation of the first derivative of the spherical Bessel functions, we use the following recursion relations:
\ba
\label{eq:jl_recursion_sum}
&j_{\ell-1}(z) + j_{\ell+1}(z) = \frac{2\ell+1}{z} j_\ell(z)\,,
\\
\label{eq:jl_recursion_der}
&\ell \,j_{\ell-1}(z) - (\ell+1) j_{\ell+1}(z) = (2\ell+1)\,j_{\ell}'(z)\,.
\ea
The derivative relation \refeq{jl_recursion_der} expresses $j_\ell'$ as a linear combination of $j_{\ell\pm1}$, which yields the following identities for the odd-derivative overlap integrals:
\ba
\label{eq:w01_rec}
w_{\ell,01}^{p}
&\,=\frac{1}{2\ell+1}\Big[\ell \,w_{\ell,\ell-1}^{p} -(\ell+1) w_{\ell,\ell+1}^{p} \Big] \,,
\\
\label{eq:w10_rec}
w_{\ell,10}^{p}
&\,=\frac{1}{2\ell+1}\Big[\ell \,w_{\ell-1,\ell}^{p} -(\ell+1) w_{\ell+1,\ell}^{p}\Big]\,,
\\
\label{eq:w11_rec}
w_{\ell,11}^{p}
&\,= \frac{1}{(2\ell+1)^2}\Big[\ell^2 w_{\ell-1,\ell-1}^{p} -\ell(\ell+1) \left(w_{\ell-1,\ell+1}^{p}+w_{\ell+1,\ell-1}^{p}\right)+(\ell+1)^2w_{\ell+1,\ell+1}^{p} \Big]\,,
\\
\label{eq:w21_rec}
w_{\ell,21}^{p}
&\,=
\frac{1}{2\ell+1}\Big[
f_{-2}\big(\ell \,w_{\ell-2,\ell-1}^{p} -(\ell+1) w_{\ell-2,\ell+1}^{p}\big)+
f_{0}\big(\ell \,w_{\ell,\ell-1}^{p} -(\ell+1) w_{\ell,\ell+1}^{p}\big)\vs
&\, \qquad \qquad \quad+
f_{2}\big(\ell \,w_{\ell+2,\ell-1}^{p} -(\ell+1) w_{\ell+2,\ell+1}^{p}\big)\Big] \,,
\\
\label{eq:w12_rec}
w_{\ell,12}^{p}
&\,=
\frac{1}{2\ell + 1}\Big[
f_{-2}\big(\ell \,w_{\ell-1,\ell-2}^{p} -(\ell+1) w_{\ell+1,\ell-2}^{p}\big)+
f_{0}\big(\ell \,w_{\ell-1,\ell}^{p} -(\ell+1) w_{\ell+1,\ell}^{p}\big) \vs
&\, \qquad \qquad \quad +
f_{2}\big(\ell \,w_{\ell-1,\ell+2}^{p} -(\ell+1) w_{\ell+1,\ell+2}^{p}\big)\Big]
\,.
\ea
The right-hand sides of \refeqs{w01_rec}{w12_rec} still contain terms such as $w_{\ell\ell'}^p=w_{\ell,\ell\pm1}^p$, where $\Delta\ell=\ell-\ell'=\pm1$ is odd. Since the recursion relation of the 2-FAST implementation walks along the even-$\Delta\ell$ direction, we must convert them once more to even-$\Delta\ell$ quantities that are a part of 2-FAST computation. For that, we apply the sum rule \refeq{jl_recursion_sum}:
\ba
\label{eq:odd_to_even_1}
w_{\ell',\ell-1}^p
&\,= \frac{r'}{2\ell-1} \left(w_{\ell',\ell-2}^{p+1} + w_{\ell',\ell}^{p+1} \right)\,,
\qquad
w_{\ell-1,\ell'}^p
= \frac{r}{2\ell-1} \left(w_{\ell-2,\ell'}^{p+1} + w_{\ell,\ell'}^{p+1} \right)\,,
\\
\label{eq:odd_to_even_2}
w_{\ell',\ell+1}^p
&\,= \frac{r'}{2\ell+3} \left(w_{\ell',\ell+2}^{p+1} + w_{\ell',\ell}^{p+1} \right)\,,
\qquad
w_{\ell+1,\ell'}^p
= \frac{r}{2\ell+3} \left(w_{\ell+2,\ell'}^{p+1} + w_{\ell,\ell'}^{p+1} \right)\,.
\ea
Similar to the \refeqs{w_ell_00}{w_ell_22}, we define
\be
\label{eq:g_coeffs}
    g_{-2}=\frac{\ell}{(2\ell+1)(2\ell-1)}\,,\qquad
    g_{0}=\frac{1}{(2\ell-1)(2\ell+3)}\,,\qquad
    g_{2}=-\frac{\ell+1}{(2\ell+1)(2\ell+3)}\,,
\ee
so that the final expressions for the odd-derivative integrals in terms of the even-$\Delta\ell$ 2-FAST output become
\ba
\label{eq:w01_twofast}
w_{\ell,01}^{p}
&\,=
r'\left\{
g_{-2}\,w^{p+1}_{\ell,\ell-2}
+
g_{0}\, w^{p+1}_{\ell,\ell}
+
g_{2}\, w^{p+1}_{\ell,\ell+2}
\right\}\,,
\\
\label{eq:w10_twofast}
w_{\ell,10}^{p}
&\,=
r\left\{
g_{-2}\,w^{p+1}_{\ell-2,\ell}
+
g_{0}\, w^{p+1}_{\ell,\ell}
+
g_{2}\, w^{p+1}_{\ell+2,\ell}
\right\}\,,
\\
\label{eq:w12_twofast}
w_{\ell,12}^{p}
&\,=
r
\biggl\{
f_{-2}\left(g_{-2}\, w_{\ell-2,\ell-2}^{p+1} + g_0\, w_{\ell,\ell-2}^{p+1} + g_2\, w_{\ell+2,\ell-2}^{p+1} \right)
+
f_{0}\left(g_{-2}\, w_{\ell-2,\ell}^{p+1} + g_0\, w_{\ell,\ell}^{p+1} + g_2\, w_{\ell+2,\ell}^{p+1} \right)
\vs
&\,\qquad+
f_{2}\left(g_{-2}\, w_{\ell-2,\ell+2}^{p+1} + g_0\, w_{\ell,\ell+2}^{p+1} + g_2\, w_{\ell+2,\ell+2}^{p+1} \right)
\biggr\}\,,
\\
\label{eq:w21_twofast}
w_{\ell,21}^{p}
&\,=
r'
\biggl\{
f_{-2}\left(g_{-2}\, w_{\ell-2,\ell-2}^{p+1} + g_0\, w_{\ell-2,\ell}^{p+1} + g_2\, w_{\ell-2,\ell+2}^{p+1} \right)
+
f_{0}\left(g_{-2}\, w_{\ell,\ell-2}^{p+1} + g_0\, w_{\ell,\ell}^{p+1} + g_2\, w_{\ell,\ell+2}^{p+1} \right)
\vs
&\,\qquad+
f_{2}\left(g_{-2}\, w_{\ell+2,\ell-2}^{p+1} + g_0\, w_{\ell+2,\ell}^{p+1} + g_2\, w_{\ell+2,\ell+2}^{p+1} \right)
\biggr\}\,.
\ea
Therefore, we compute each odd-derivative integral $w_{\ell,jj'}^p$ as a linear combination of even-$\Delta\ell$ integrals $w_{\ell',\ell''}^{p+1}$, which can be computed with one additional call to the 2-FAST algorithm using the rescaled power spectrum $k^{p+1}P(k)$. This is the reason why the integrals with different $p$ values share the same $p_{\rm base}$ in the third column of Table~\ref {tab:q_val_table}.

Finally, for the computation of $u_{\ell,jj'}^p$ and $v_{\ell,jj'}^p$, whose integrands are more complex function in $k$ with a given power $p$ and transfer-function index $n$, we simply feed the rescaled spectrum $P(k)T^n(k)$ into the algorithm in place of $P(k)$ and use the same {\sc TwoFAST} function call.

In summary, the extended {\sc TwoFAST} computes the following integrals:
\be
    w_{\ell,jj'}^{p,n}(r,r') = \frac{2}{\pi}\int \dd k\,k^{2+p} P(k) \left[T(k)\right]^n j_\ell^{(j)}(kr) j_\ell^{(j')}(kr')\
\ee
for the values listed in Table~\ref{tab:q_val_table}, where we call $n=0$, $-1$, $-2$ cases, respectively, $w^p_{\ell,jj'}(r,r')$, $u^p_{\ell,jj'}(r,r')$, $v^p_{\ell,jj'}(r,r')$. The table also lists the base $p$ values for the {\sc TwoFAST} function call as well as the optimal bias parameter $q$ minimizing the numerical effects coming from the discrete FFTlog transformation.

\subsection{Optimization of the FFTLog Bias Parameter $q$}
\label{app:q_optimization}

The bias parameter $q$ in the FFTLog decomposition controls the power-law weighting $k^{-q}$ applied before the logarithmic Fourier transform.  While the exact result is independent of $q$, the finite discrete transform ($N_r$ samples in $\ln k$) introduces $q$-dependent truncation and aliasing errors; a poor choice of $q$ can degrade accuracy by orders of magnitude.  In the extended \textsc{TwoFAST} algorithm, the optimization is further complicated by the fact that $q$ enters not only the FFTLog decomposition of $P(k)$ but also the numerical stability of the ${}_2F_1$ hypergeometric functions that replace the analytic kernel of the original algorithm.  In this section, we describe a systematic optimization of $q$ for the extension of the \textsc{TwoFAST} algorithm.  The final result is summarized in the last column of Table~\ref{tab:q_val_table}.  Note that the original \textsc{TwoFAST} uses $q=1.1$ for $w_{\ell,jj'}^0$ in the first row of Table~\ref{tab:q_val_table}; we have revised the value to $1.30$ from the procedure outlined below.  Taking advantage of the fast \textsc{TwoFAST} function call, we adopt the optimal $q$ for each individual $(p,n)$ pair, instead of fixing $q$ across all cases with a common $p_{\rm base}$ and $n$.

We determine optimal $q$ values by comparing \textsc{TwoFAST} output against high-precision reference integrals computed via the method of \cite{LUCAS1995269}\footnote{The integrand is split into non-oscillatory combinations $h_1$ and~$h_2$ of spherical Bessel functions of the first and second kinds.  The non-oscillatory region is integrated with adaptive Gauss--Kronrod quadrature (\texttt{QuadGK}, \texttt{order}$\,=511$), and the oscillatory tails with a specialized oscillatory quadrature (\texttt{QuadOsc}).  The relative tolerance is set to $10^{-10}$. For more details of the implementation, see Ref.~\cite{Gebhardt::2018twofast}.}, which evaluates Eq.~\eqref{eq:wljj'_app} by direct numerical integration of the spherical Bessel functions.  Because this method avoids the FFTLog decomposition entirely, it serves as an independent benchmark.

For the test along the radial direction, we generate reference data at multipoles $\ell \in \{2, 42, 100, 500\}$, spanning the range $r\in[1,10^5]\;\mathrm{Mpc}/h$, which encloses the distance range used for the cosmological analysis of this paper.  For the test along the multipole direction, we generate reference data at $\ell\in\{2,3,\ldots,100\}\cup\{112,125,150,200,300,\ldots,1000,1200\}$ for $r=2303\;\mathrm{Mpc}/h$.  For both cases, we use 29 scale parameters $R \in \{0.1, \ldots, 0.9,\, 0.91, \ldots, 1.09, 1.1\}$.  The dense sampling near $R\sim1$ is necessary because the integrand drops quickly away from $R=1$ at large $\ell$.

The benchmark calculation was distributed across two HPC clusters equipped with dual-socket Intel Xeon E5-2660\,v2 (2.20\,GHz) and E5-2690\,v2 (3.00\,GHz) processors, respectively.  A total of 145 independent tasks were parallelized using SLURM job arrays with 10 cores per task, consuming approximately 18,000 core-hours.  The computational cost is dominated by the $\ell=500$ case ($\sim$12,000~core-hours, 68\%), followed by $\ell=100$ ($\sim$3,100~core-hours, 17\%), $\ell=42$ ($\sim$2,200~core-hours, 12\%), the fixed-$r$ run ($\sim$500~core-hours, 3\%), and $\ell=2$ ($\sim$90~core-hours, ${<}1$\%), reflecting the rapid increase in quadrature cost with multipole order.

We compare the reference data against \textsc{TwoFAST} results at $N_r = 4096$ with $k \in [10^{-5},\, 10^{3}]\;h/\mathrm{Mpc}$, the same set of parameters we use for the main analysis of the paper.  For the along-$r$ comparison, \textsc{TwoFAST} output is restricted to $10 \leq r \leq 6000\;\mathrm{Mpc}/h$, the radius range relevant for the analysis in the paper.

For each $(p,n)$ combination, we perform a systematic scan over 396 values of $q$ from $-1.99$ to $1.99$ in steps of $0.01$, excluding all integer values ($q = 0,\,\pm 1$) where the FFTLog kernel is singular.  Each $q$ evaluation covers all 29 values of $R$ and all $\ell$ values.  A single evaluation requires approximately 0.18 core-hours, representing a speedup of $\sim$70,000 over the direct integration, with the caveat that the \textsc{TwoFAST} calculation spans the full range $\ell\in[2,\,500]$ instead of the four specific $\ell$ values.  The full $q$ scan of 396 values was completed in $\sim$91 core-hours, distributed across three HPC clusters sharing a common filesystem, which is less than 1\% of the cost of the single direct-integration reference calculation.

For a single configuration $(p,n,j,j',R,\ell)$ and a given $q$, \textsc{TwoFAST} returns $w^{\TF}$ on a grid of $r=r_0e^\rho$ [see \refeq{r_rho_first}], which we interpolate to the radii for which reference calculation is available.  The comparison is performed on five channels per $(p,n,j,j')$: four fixed-$\ell$ channels ($\ell\in\{2,42,100,500\}$, varying $r$) and one fixed-$r$ channel ($r=2303\;\mathrm{Mpc}/h$, varying $\ell$).  Because the resulting $w(x)$ functions are highly oscillatory and their amplitude drops rapidly away from $R\sim 1$, a pointwise comparison is ill-suited for quantifying overall accuracy, because a small phase shift can produce large pointwise residuals even when the global shape is well reproduced.  We therefore adopt the relative $L_2$ error, which measures the integrated agreement over each channel:
\begin{equation}
  \epsilon_{L_2}
  = \frac{
      \displaystyle\sqrt{\int
        \bigl[w^{\TF}(x)-w^{\Luc}(x)\bigr]^2\,\dd x}
  }{
      \displaystyle\sqrt{\int
        \bigl[w^{\Luc}(x)\bigr]^2\,\dd x}
  }\,,
  \label{eq:epsL2}
\end{equation}
where $x=r$ for the fixed-$\ell$ channels (integrated over $r\in[10,\,6000]\;\mathrm{Mpc}/h$) and $x=\ell$ for the fixed-$r$ channel (integrated over $\ell\in[2,\,500]$).  The integrals are evaluated with the trapezoidal rule on the discrete grid.

\begin{figure}[t]
  \centering
  \includegraphics[width=\textwidth]{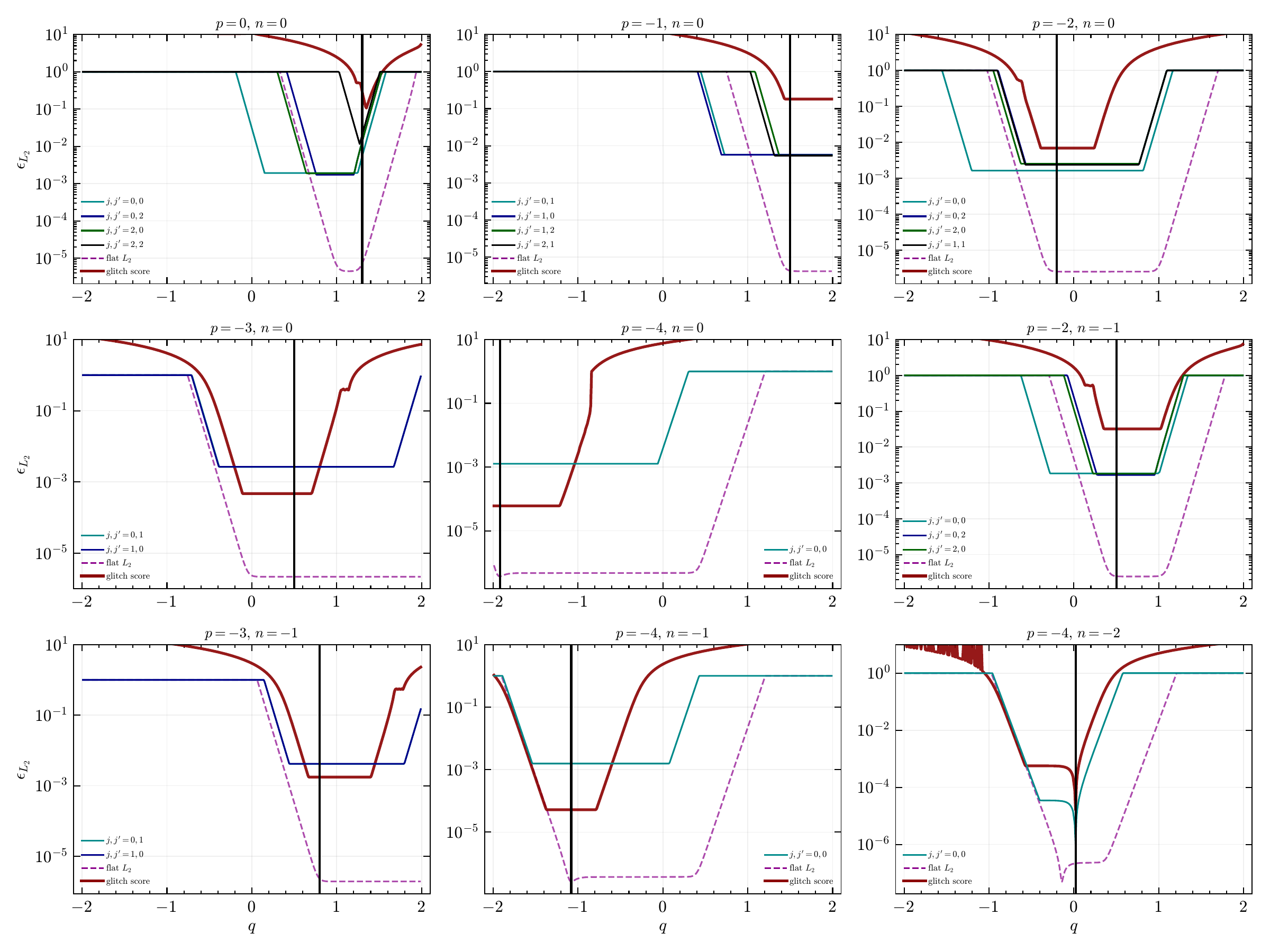}
  \caption{Error metrics as a function of the FFTLog bias parameter~$q$ for all nine $(p,n)$ pairs ($N_r=4096$).  Solid coloured curves: max metric $\epsilon_{\max}^{(j,j')}$ per $(j,j')$ sub-block.  Dashed dark magenta curve: flat metric $\epsilon_{\mathrm{flat}}$.  Solid dark red curve: glitch score $G(q)$.  The vertical black line marks the chosen $q^{\star}$.  The glitch score reveals structure within the $L_2$ plateau regions, enabling selection of $q$ values that avoid spurious artefacts in low-signal regimes.}
  \label{fig:qscan}
\end{figure}

For a given $(p,n)$ pair, we must select a single $q$ that works well across all $(j,j',R,\ell)$.  We define two complementary aggregate metrics.  The first is the worst-case (max) error per $(j,j')$ sub-block,
\begin{equation}
  \epsilon_{\max}^{(j,j')}(q;\,p,n)
  = \max_{R,\,\ell}\;\epsilon_{L_2}(q;\,p,n,j,j',R,\ell),
  \label{eq:eps_max}
\end{equation}
which identifies which derivative order is the limiting factor at each $q$ (solid curves in Fig.~\ref{fig:qscan}).  The second is a flat metric that combines all $(j,j',R,\ell)$ contributions into a single scalar:
\begin{equation}
  \epsilon_{\mathrm{flat}}(q;\,p,n)
  = \frac{
      \displaystyle\sqrt{\sum_{i}\Delta_i^2}
  }{
      \displaystyle\sqrt{\sum_{i}L_i^2}
  }\,,
  \label{eq:eps_flat}
\end{equation}
where $i$ runs over all valid $(j,j',R,\ell)$ combinations, $\Delta_i \equiv \sqrt{\int [w^{\TF}_i - w^{\Luc}_i]^2\,\dd x}$ is the absolute $L_2$ error, and $L_i \equiv \sqrt{\int [w^{\Luc}_i]^2\,\dd x}$ is the signal norm.  The flat metric treats all error vectors as components of a single vector in a high-dimensional space, making it less sensitive to a single outlier configuration than the max metric (dashed dark magenta curve in Fig.~\ref{fig:qscan}).

Within each channel, the {\sc TwoFAST} integration results from the 29~$R$ values span a wide dynamic range in signal amplitude; for some $R$ the signal norm $L_i$ is negligibly small, causing the relative error~\eqref{eq:epsL2} to diverge spuriously.  To filter these out, we group all data points by their $(p,n,j,j',\ell)$ quintuple (i.e.\ per channel) and within each group find the largest signal norm across all $R$ and all scanned $q$ values:
\begin{equation}
  L^{\max}_{(p,n,j,j',\ell)}
  = \max_{q,\,R}\;L_i(q,R).
\end{equation}
A data point is retained only if its signal norm exceeds 0.01\% of this maximum:
\begin{equation}
  L_i(q,R) > 10^{-4}\times L^{\max}_{(p,n,j,j',\ell)}.
  \label{eq:threshold}
\end{equation}

Finally, the $L_2$-based metrics integrate over the full $x$~range and are dominated by regions of high signal amplitude.  They can therefore miss localised artefacts in low-signal regions, precisely the regime where FFTLog truncation errors introduce spurious power-law contributions that are clearly visible in log-scale comparison plots but carry negligible weight in the $L_2$ norm.
To capture these artefacts, we introduce a glitch score that directly measures the visual separation between \textsc{TwoFAST} and the reference in log space:
\begin{equation}
  g_i(x) = \bigl|\log_{10}|w^{\TF}_i(x)|
            - \log_{10}|w^{\Luc}_i(x)|\bigr|.
  \label{eq:logdiff}
\end{equation}
This quantity equals the vertical distance (in dex) between the two curves as they appear in the comparison plots of the next paragraph, making it a natural proxy for visually identifiable discrepancies.
For each $(q,p,n,j,j',R,\ell)$-configuration data, the pointwise log-difference~\eqref{eq:logdiff} is evaluated only at grid points where at least one of $|w^{\TF}|$ or $|w^{\Luc}|$ exceeds $10^{-8}\times\max|w^{\Luc}_i|$, extended dynamic range than $L_2$-based comparison in order to catch the failure at small values.
To avoid the isolated single-point spikes $g_i$, possibly generated from the zero-crossings of $w^{\Luc}$, we apply a running-median filter (window width $W=21$ grid points) before taking the maximum:
\begin{equation}
  G_i(q) = \max_x\;\operatorname{median}_{W}\!\bigl[g_i(x)\bigr].
  \label{eq:glitch_file}
\end{equation}

Combining all three criteria, we determine the per-$(p,n)$ aggregate by taking the maximum of $G_i$ over all configurations whose signal is prominent enough to be plotted:
\begin{equation}
  G(q;\,p,n) = \max_{\substack{j,j',R,\ell \\[2pt]
      \max|w^{\Luc}_i| > 10^{-2}\,
      \max|w^{\Luc}|_{(p,n,j,j',\ell)}}} G_i(q).
  \label{eq:glitch_agg}
\end{equation}
Here, the denominator in the signal filter is the largest reference amplitude across all $q$ and~$R$ for a given $(p,n,j,j',\ell)$ channel.  This ensures that only files whose signal is within two orders of magnitude of the channel maximum contribute to the aggregate score (solid dark red curve in Fig.~\ref{fig:qscan}).

The selection proceeds in two stages.  First, for each $(p,n)$ we identify the candidate range $Q(p,n)$ by collecting all $q$ values satisfying
\begin{equation}
  \epsilon_{\mathrm{flat}}(q) \le \alpha\,
    \epsilon_{\mathrm{flat}}^{\min}
  \quad\text{and}\quad
  \epsilon_{\max}(q) \le \alpha\,
    \epsilon_{\max}^{\min},
  \label{eq:candidate}
\end{equation}
with $\alpha=2$, where $\epsilon_{\max}\equiv\max_{j,j'}\epsilon_{\max}^{(j,j')}$.  This intersection selects the region where both the overall and worst-case $L_2$ errors remain within a factor of two of their respective minima.  The $L_2$-based metrics reliably identify the broad range of $q$ values yielding globally small errors, but for several $(p,n)$ pairs this plateau is very wide, spanning most of the $q\in[-2,+2]$ range; within it, the $L_2$ metrics are essentially flat and cannot distinguish between $q$ values.

The glitch score breaks this degeneracy by targeting the specific failure mode of the FFTLog at sub-optimal $q$: spurious power-law contributions that dominate the signal in low-amplitude regions of parameter space (typically at low~$r$ and high~$\ell$, or at~$R\approx1$).  While these artefacts carry negligible $L_2$ weight, they are clearly visible in log-scale comparison plots as smooth curves diverging from the oscillatory reference.  Within the candidate set, we select
\begin{equation}
  q^{\star}(p,n)
  = \operatorname*{arg\,min}_{q\in Q}\;
    G(q;\,p,n).
  \label{eq:qstar}
\end{equation}

In practice, three situations arise.  When both $L_2$ and $G$ have well-defined minima, the two-stage procedure returns a sharp optimum; we adopt the automatically selected $q^{\star}$ (e.g.\ $(0,0)$, $(-4,-2)$).  When $G(q)$ is nearly constant across the candidate set while $L_2$ still varies, any candidate $q$ produces visually indistinguishable comparison plots; we round $q^{\star}$ to the nearest half-integer for simplicity (e.g.\ $(-1,0)$, $(-3,-1)$).  When both metrics are degenerate, we adopt the $\epsilon_{\mathrm{flat}}$ argmin (e.g.\ $(-4,0)$, $(-4,-1)$). The final $q^{\star}$ values are listed in the last column of the Table~\ref{tab:q_val_table}.

\begin{figure}[p]
  \centering
  \includegraphics[width=0.48\textwidth]{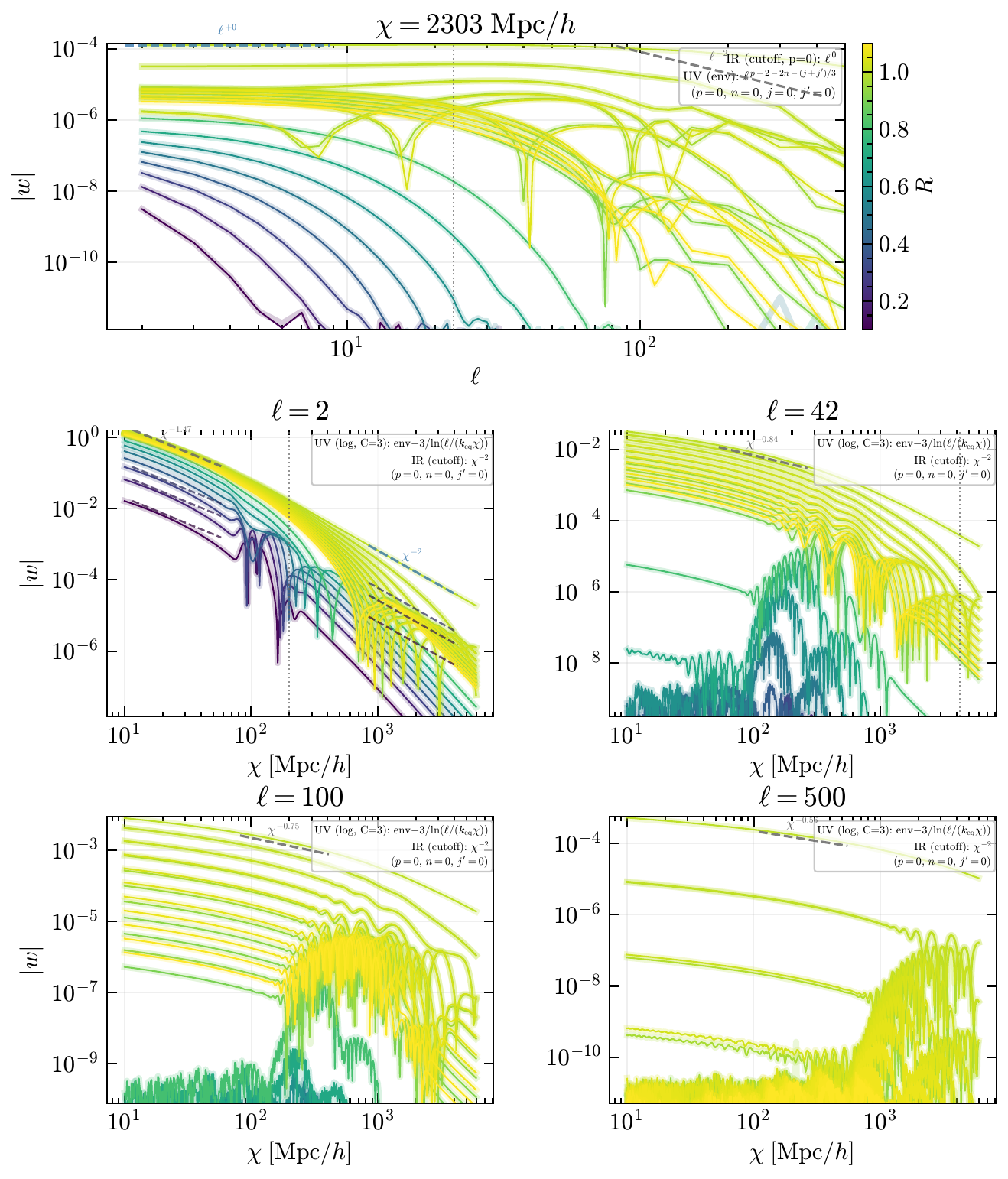}
  \hfill
  \includegraphics[width=0.48\textwidth]{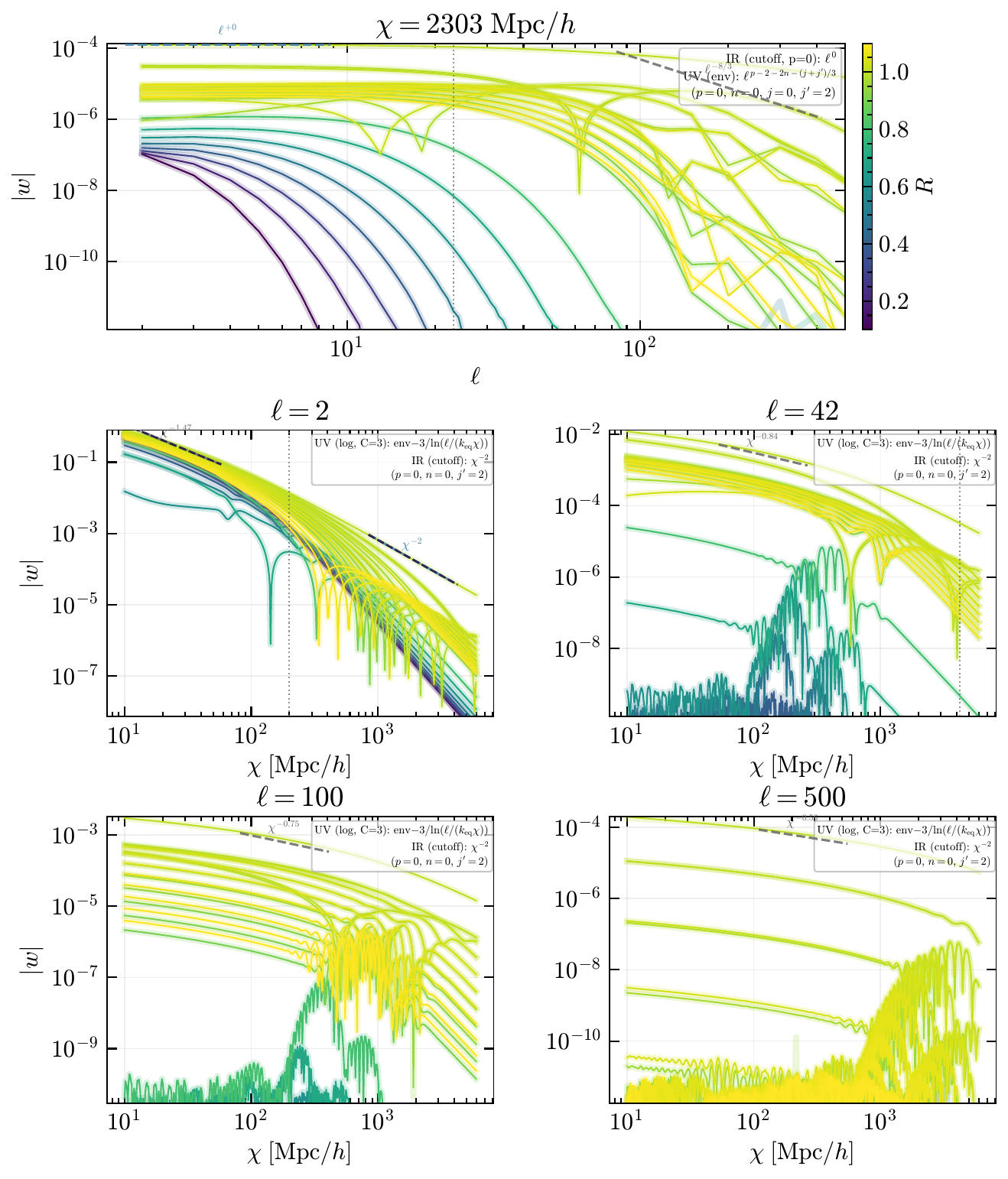}
  \\[6pt]
  \includegraphics[width=0.48\textwidth]{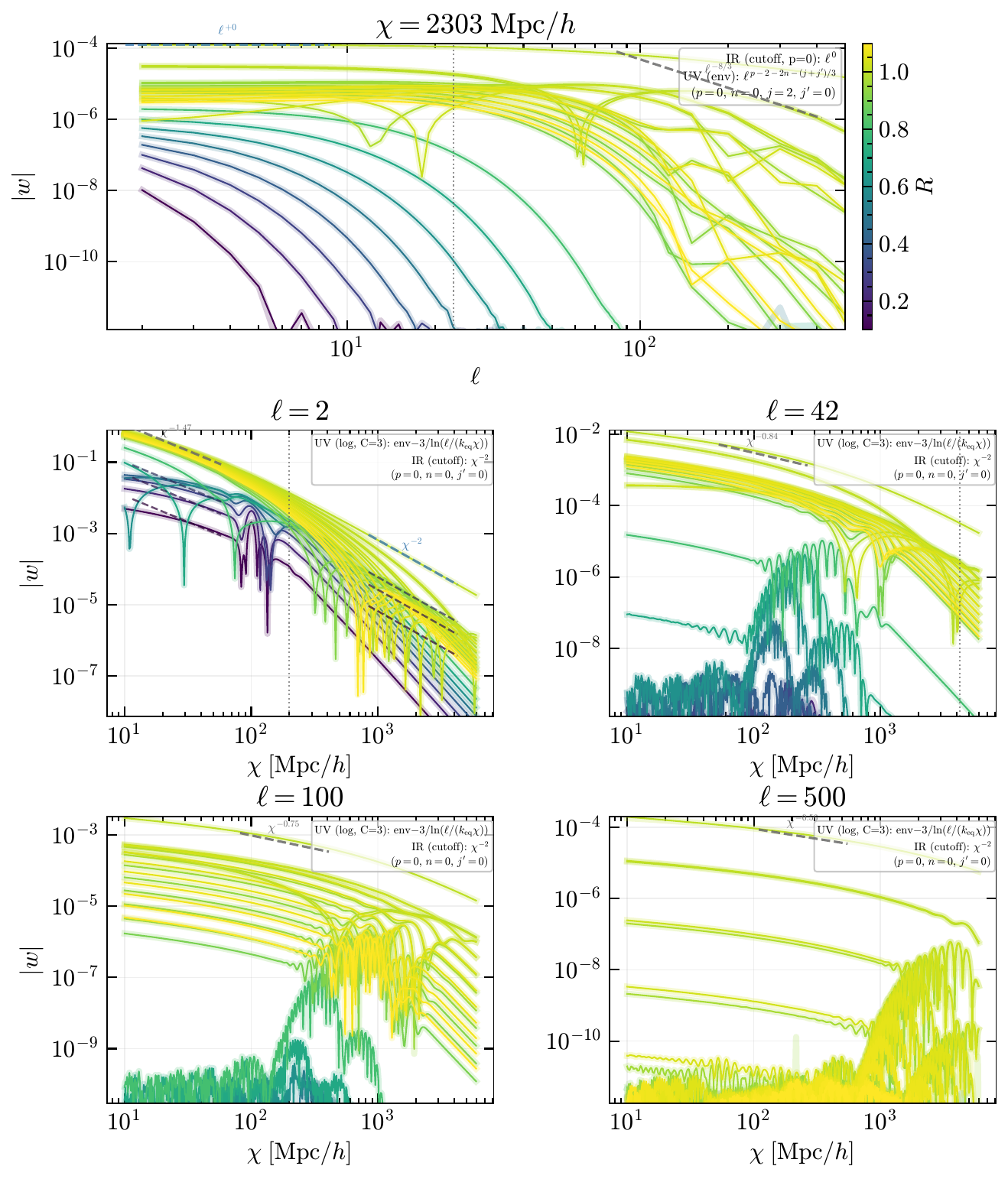}
  \hfill
  \includegraphics[width=0.48\textwidth]{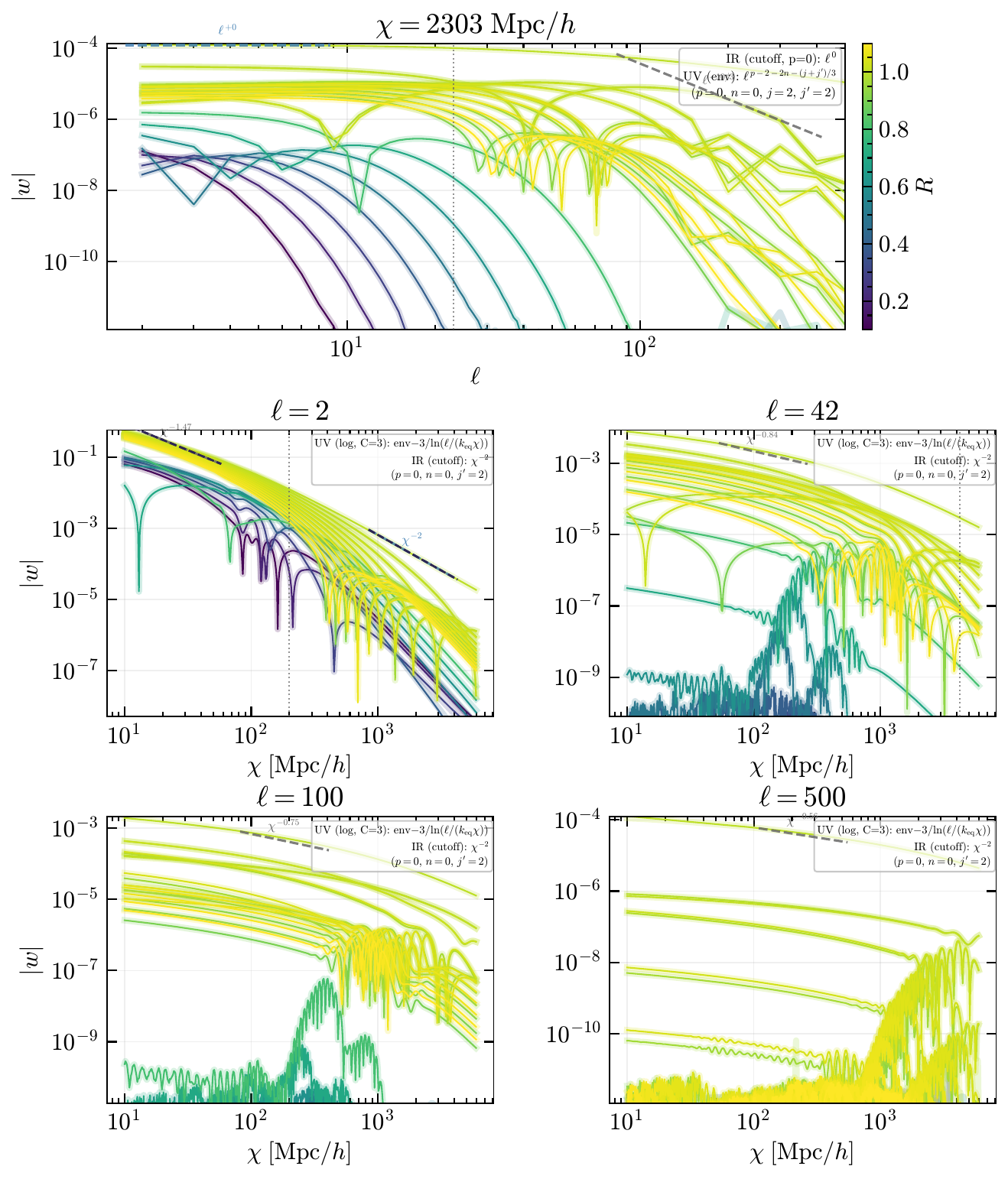}
  \caption{Direct comparison of \textsc{TwoFAST} ($N_r=4096$, $q^{\star}=+1.30$) against the Lucas reference for $(p,n)=(0,0)$.  Each sub-figure shows a different $(j,j')$ combination: $(0,0)$ [top left], $(0,2)$ [top right], $(2,0)$ [bottom left], and $(2,2)$ [bottom right].  See the text for panel descriptions.}
  \label{fig:comparison_p0n0}
\end{figure}

To verify that the chosen $q^{\star}$ values produce accurate results across the full parameter space, particularly for the $(p,n)$ pairs where all metrics are nearly degenerate, we compare \textsc{TwoFAST} ($N=4096$) against the Lucas reference for all 22 $(p,n,j,j')$ combinations at the selected~$q$. All comparison plots are in the {\sf PowerFull.jl/figs/comparison} folder of the package. Here, we show one example plot in Fig.~\ref{fig:comparison_p0n0} showing the comparison for $n=0$, $p=0$ cases with four $(j,j')$ combinations. The figure contains five sub-panels.  The top panel fixes $r=2303\;\mathrm{Mpc}/h$ and plots $|w|$ as a function of~$\ell$ (on a logarithmic scale, $\ell\leq 500$).  The four bottom panels fix $\ell\in\{2,42,100,500\}$ and plot $|w|$ as a function of~$r$ (log scale, $r\in[10,\,6000]\;\mathrm{Mpc}/h$).  In all panels, the Lucas reference is shown as thick, semi-transparent curves and the \textsc{TwoFAST} result as thin, opaque curves overlaid on top; the colour encodes the distance ratio~$R$ via a viridis colourmap spanning $R\in[0.1,\,1.1]$. For all four combinations, the \textsc{TwoFAST} curves trace the Lucas reference to within the line width, confirming sub-percent accuracy over the full range of $R$, $\ell$, and~$r$.  The same is true for the other 18 combinations, which can be found in the git repository.

\subsection{Scaling of the \textsc{TwoFAST} base integrals}
\label{app:scaling}

The \textsc{TwoFAST} base integrals $w_{\ell,jj'}^p(r,Rr)$, $u_{\ell,jj'}^p(r,Rr)$, and $v_{\ell,jj'}^p(r,Rr)$ share the same qualitative behavior as the example in Fig.~\ref{fig:comparison_p0n0}.  At fixed $r$, the amplitudes decrease with increasing $\ell$.  At fixed $\ell$, they are largest near the diagonal $R=r'/r\simeq 1$ and decrease away from it; the falloff
steepens with $\ell$.  This motivates the two-tier strategy in the \textsc{PowerFull} implementation.

In this appendix, we summarize the approximate power-law scaling of the base integrals with $\ell$, $r$, and $R$.  The goal is not to derive a uniform asymptotic expansion, but to explain the local power-law envelopes used in the comparison plots.  In regular regimes, the overlays follow from turning-point scaling of the spherical Bessel functions. In marginal regimes, especially for $p=-1$, the leading envelope is not sufficient, and we use empirical local fits.

For these estimates, we assume a Harrison-Zel'dovich-Peebles initial spectrum with $n_s=1$, and use the BBKS transfer-function asymptote $T(k)\simeq k^{-2}\ln k$ for $k\gg k_{\rm eq}$, with $k_{\rm eq}=0.01\,h/{\rm Mpc}$.  We write the base integrals schematically as
\begin{equation}
w^{p,n}_{\ell,jj'}(r,r')
=
\int \dd k\,
k^{2+p}\,[T(k)]^n\,P_{\rm L}(k)\,
j_\ell^{(j)}(kr)\,j_\ell^{(j')}(kr') ,
\label{eq:wpn}
\end{equation}
where $j_\ell^{(j)}$ denotes the $j$-th derivative of the spherical Bessel function with respect to its argument. Since $P_{\rm L}(k)\propto kT^2(k)$ for $n_s=1$, the effective integrand scales as
\begin{equation}
k^{3+p}\,T^{n+2}(k)\,
j_\ell^{(j)}(kr)\,j_\ell^{(j')}(kr') .
\label{eq:effective_integrand}
\end{equation}

\subsubsection{Turning-point estimate and Airy-region derivatives}

The leading envelope follows from the localization of $j_\ell(x)$ near its turning point $x\simeq \ell$.  In this region, the spherical Bessel function has the Airy form
\begin{equation}
j_\ell(x)
\sim
\ell^{-5/6}\,
{\rm Ai}\!\left(\frac{x-\ell}{\ell^{1/3}}\right),
\label{eq:airy_jl}
\end{equation}
up to order-unity constants and a shift between $\ell$ and $\ell+1/2$. The factor $\ell^{-5/6}$ is the Airy amplitude of the ordinary Bessel function, $\ell^{-1/3}$, multiplied by the spherical-Bessel prefactor $x^{-1/2}\sim \ell^{-1/2}$.

Each argument derivative pulls a factor $\ell^{-1/3}$ from the Airy argument $(x-\ell)/\ell^{1/3}$:
\begin{equation}
j_\ell^{(j)}(x)
\sim
\ell^{-5/6-j/3}\,
{\rm Ai}^{(j)}(y),
\qquad
y\equiv \frac{x-\ell}{\ell^{1/3}} ,
\label{eq:airy_derivative}
\end{equation}
and hence
\begin{equation}
j_\ell^{(j)}(x)\,j_\ell^{(j')}(x)
\sim
\ell^{-5/3-(j+j')/3}
\end{equation}
in the turning-point region.  This counting applies to the Airy form itself, not term by term in a recurrence.  Using $j_\ell'(x)=j_{\ell-1}(x)-(\ell+1)\,j_\ell(x)/x$, each term near the turning point is of order $\ell^{-5/6}$, but the two cancel at this order and leave a residual of order $\ell^{-7/6}$ set by the variation across the Airy width.  The same pattern extends to higher derivatives.
 
The Airy amplitudes Eq.~\eqref{eq:airy_derivative}, combined with the $k$-space turning-point width $\Delta k\sim \ell^{1/3}/r$, set the local turning-point envelope used below.  This estimate captures the amplitude envelope of the oscillatory integral, not the detailed phase cancellations or the BAO wiggles in the input power spectrum.

\subsubsection{Two contributions: turning point and equality}
\label{app:two_contrib}

The integrand of Eq.~\eqref{eq:wpn} has two structures in $k$: the Bessel pair $j_\ell^{(j)}(kr)\,j_\ell^{(j')}(kr')$, localized at the turning point $k_{\rm tp}\sim \ell/r$ with Airy width $\Delta k\sim \ell^{1/3}/r$, and the smooth factor $k^{3+p}T^{n+2}(k)$, a power-law profile with the transfer-function break at $k\sim k_{\rm eq}$.  Two scales, therefore, set the integral, and the envelope of $w$ follows the larger of the two corresponding contributions.
 
The turning-point contribution comes from evaluating the integrand at $k_{\rm tp}$ over the Airy width, using Eq.~\eqref{eq:airy_derivative} for the Bessel peak amplitude:
\begin{equation}
w_{\rm tp}
\simeq
\Delta k\,
k_{\rm tp}^{3+p}\,
T^{n+2}(k_{\rm tp})\,
j_\ell^{(j)}(k_{\rm tp}r)\,j_\ell^{(j')}(k_{\rm tp}r')
\propto
\frac{\ell^{1/3}}{r}
\left(\frac{\ell}{r}\right)^{3+p}
T^{n+2}\!\left(\frac{\ell}{r}\right)
\ell^{-5/3-(j+j')/3} .
\label{eq:wtp_general}
\end{equation}
The transfer-function factor reduces to $T\simeq 1$ for $k_{\rm tp}\ll k_{\rm eq}$ and to $T(k)\sim k^{-2}\ln k$ for $k_{\rm tp}\gg k_{\rm eq}$.
 
The equality contribution comes from the region around $k\sim k_{\rm eq}$, of width $\Delta k_{\rm eq}\sim k_{\rm eq}$:
\begin{equation}
w_{\rm eq}
\simeq
\Delta k_{\rm eq}\,
k_{\rm eq}^{3+p}\,
\left|j_\ell^{(j)}(k_{\rm eq}r)\right|^2 ,
\label{eq:weq_general}
\end{equation}
specialised to $r=r'$, with the order-unity $T(k_{\rm eq})$ absorbed.  The Bessel envelope at $k_{\rm eq}r$ depends on which side of the turning point the equality scale lies:
\begin{align}
\left|j_\ell^{(j)}(k_{\rm eq}r)\right|^2
&\sim \frac{1}{(k_{\rm eq}r)^2}
&&\text{for } k_{\rm eq}r\gg \ell ,
\label{eq:bessel_eq_large}\\
\left|j_\ell^{(j)}(k_{\rm eq}r)\right|^2
&\sim \frac{(k_{\rm eq}r)^{2(\ell-j)}}{[(2\ell+1)!!]^2}
&&\text{for } k_{\rm eq}r\ll \ell ,
\label{eq:bessel_eq_small}
\end{align}
up to oscillatory factors and derivative-dependent order-unity terms.
 
When the equality scale lies above the turning point ($k_{\rm eq}>k_{\rm tp}$, or $\ell< k_{\rm eq}r$), Eq.~\eqref{eq:bessel_eq_large} gives
\begin{equation}
w_{\rm eq}\simeq \frac{k_{\rm eq}^{2+p}}{r^2} ,
\label{eq:weq_above}
\end{equation}
which is $\ell$-independent at fixed $r$ and scales as $r^{-2}$ at fixed $\ell$.  When the equality scale lies below the turning point ($k_{\rm eq}<k_{\rm tp}$, or $\ell > k_{\rm eq}r$), $w_{\rm eq}$ is power-suppressed in $\ell$ via Eq.~\eqref{eq:bessel_eq_small} and the envelope reduces to $w_{\rm tp}$ alone.
 
Note that $w_{\rm tp}$ and $w_{\rm eq}$ are two estimates of the same integral from distinct $k$-regions, not additive contributions.  The envelope of $w$ tracks whichever is larger.  The transition occurs when the Bessel turning point samples the equality scale, $k_{\rm tp}=k_{\rm eq}$:
\begin{equation}
\ell_* \equiv k_{\rm eq}r ,
\qquad
r_* \equiv \frac{\ell}{k_{\rm eq}} ,
\end{equation}
so that $\ell_*$ marks the transition for fixed-$r$ curves as a function of $\ell$ and $r_*$ marks it for fixed-$\ell$ curves as a function of $r$.  In Fig.~\ref{fig:comparison_p0n0}, $r=2303\,{\rm Mpc}/h$ gives $\ell_*\simeq 23$; for the fixed-$\ell$ panels, $r_*=200,\,4200,\,10^4,\,5\times 10^4\,{\rm Mpc}/h$ at $\ell=2,\,42,\,100,\,500$.  The four scaling regimes below follow from applying this framework to fixed-$r$ and fixed-$\ell$ slices.  For $p=-1$, $w_{\rm tp}$ and $w_{\rm eq}$ match in scaling up to logarithms, and the larger-of-two rule becomes marginal.  We treat the resulting logarithmic envelopes case by case.

\subsubsection{Scaling with $\ell$ at fixed $r=r'$}

At fixed $r=r'$, the equality scale lies above the turning point for $\ell\ll \ell_*$ and below it for $\ell\gg \ell_*$, where $\ell_*=k_{\rm eq}r$ is the multipole corresponding to matter-radiation equality.
 
For $\ell\ll \ell_*$, $T(k_{\rm tp})\simeq 1$ in Eq.~\eqref{eq:wtp_general}, giving
\begin{equation}
w_{\rm tp}\;\propto\;\ell^{\,p+2-(j+j')/3} ,
\label{eq:topIRenv}
\end{equation}
where the $r$-dependent factors are irrelevant for the fixed-$r$ slope.  The equality contribution Eq.~\eqref{eq:weq_above} is $\ell$-independent:
\begin{equation}
w_{\rm eq}\;\propto\;\ell^0
\qquad
\hbox{(equality-dominated).}
\label{eq:topIReq}
\end{equation}
The envelope follows the larger of the two.  The $p=0$ cases sit in the equality-dominated regime, as do the marginal $p=-1$ cases for which the derivative order is not large enough to restore turning-point dominance.
 
The marginal case is $p=-1$ with $j+j'\ge 1$.  Here the larger-of-two rule gives the leading power but misses a logarithmic enhancement that exceeds both $w_{\rm tp}$ and $w_{\rm eq}$ at small $\ell$.  Integration by parts on the derivatives converts the integrand to a $j_\ell^2$-form with the smooth $k$-weight reduced by powers of $k$; for $p=-1$ the reduced weight, combined with the large-argument average $j_\ell^2(x)\sim 1/(2x^2)$, gives an effective $ \dd k/k$ integrand over the range $k_{\rm tp}\lesssim k\lesssim k_{\rm eq}$.  The integral therefore runs logarithmically across this range, with the Bessel small-argument suppression at $kr\sim \ell$ setting the lower boundary and the transfer-function break at $k_{\rm eq}$ setting the upper boundary:
\begin{equation}
w(r,\ell)
\sim
-\frac{\ln(k_{\rm eq}r/\ell)}{r^3} .
\end{equation}
The local $\ell$-slope is
\begin{equation}
\frac{ \dd \ln |w|}{\fixed{\dd}\ln \ell}
=
-\frac{1}{\ln(k_{\rm eq}r/\ell)} ,
\label{eq:topIRlog}
\end{equation}
evaluated at the local anchor point used for the overlay.
 
For $\ell\gg \ell_*$, the equality scale lies below the turning point, $w_{\rm eq}$ is power-suppressed by Eq.~\eqref{eq:bessel_eq_small}, and $w_{\rm tp}$ alone sets the envelope.  Using $T^{n+2}(k_{\rm tp})\sim (\ell/r)^{-2(n+2)}[\ln(\ell/r)]^{n+2}$ in Eq.~\eqref{eq:wtp_general} and retaining only the power-law part,
\begin{equation}
w_{\rm tp}\;\propto\;\ell^{\,p-2-2n-(j+j')/3} .
\label{eq:topUVenv}
\end{equation}

\subsubsection{Scaling with $r$ at fixed $\ell$ and $R=1$}

We next fix $\ell$ and set $r'=r$.  The equality scale lies below the turning point for $r\ll r_*$ and above it for $r\gg r_*$, where $r_*=\ell/k_{\rm eq}$.
 
For $r\ll r_*$, $w_{\rm eq}$ is power-suppressed by Eq.~\eqref{eq:bessel_eq_small} and $w_{\rm tp}$ alone sets the envelope.  Using $T^{n+2}(k_{\rm tp})\sim k_{\rm tp}^{-2(n+2)}[\ln k_{\rm tp}]^{n+2}$ in Eq.~\eqref{eq:wtp_general} and retaining only the power-law part,
\begin{equation}
w_{\rm tp}\;\propto\;r^{-p+2n} ,
\label{eq:botUVnaive}
\end{equation}
where the $\ell$-dependent factors are fixed.
 
The BBKS logarithm modifies the local slope.  We parameterize this correction as
\begin{equation}
\frac{\dd \ln |w|}{\dd \ln r}
=
(-p+2n)
-
\frac{C(p,n)}
{\ln\!\left[\ell/(k_{\rm eq}r)\right]} .
\label{eq:botUV}
\end{equation}
The coefficient $C(p,n)$ is set by the logarithmic power in $T^{n+2}$, with a small empirical adjustment for the marginal $p=0$ case:
\begin{table}[t!]
  \centering
\begin{tabular}{lll}
\toprule
$p$ & $C(p,n)$ & $C$ for $n=0,-1,-2$ \\
\midrule
$0$      & $2(n+2)-1$ & $3,\,1,\,-1$ \\
$\le -2$ & $2(n+2)$   & $4,\,2,\,0$ \\
$-1$     & special marginal form & see Eq.~\eqref{eq:pminus1uv} \\
\bottomrule
\end{tabular}
\caption{Coefficients $C(p,n)$ in the logarithmic radial-slope approximation of Eq.~\eqref{eq:botUV}.}
\label{tab:radial_slope_coefficients}
\end{table}
The base value $2(n+2)$ tracks the logarithmic power in the transfer function.  The $-1$ shift for $p=0$ accounts for the milder local slope in this marginal, UV-sensitive case.
 
For $p=-1$, the single $1/u$ correction in Eq.~\eqref{eq:botUV} does not capture the slope variation: the measured slope spans more than two units across the four fixed-$\ell$ panels, and a one-parameter fit leaves systematic residuals.  We extend the ansatz to two parameters,
\begin{equation}
\frac{\dd \ln |w|}{\dd\ln r}
=
1
-
\frac{C_1}{u}
-
\frac{C_2}{u^2},
\qquad
u\equiv
\ln\!\left[\frac{\ell}{k_{\rm eq}r}\right],
\label{eq:pminus1uv}
\end{equation}
and fit by least squares to the deep-UV (small-$r$) sub-range across the four panels, with $C_2$ allowed to depend linearly on $j+j'$:
\begin{equation}
C_1=1.5,
\qquad
C_2=6.3+2.5(j+j') ,
\label{eq:pminus1_coeffs}
\end{equation}
giving $C_2=8.8$ for $j+j'=1$ and $13.8$ for $j+j'=3$.  At representative points within the fit range, Eq.~\eqref{eq:pminus1uv} matches the measured slope to within about $0.05$ at intermediate $\ell$:
\begin{table}[t!]
  \centering
\begin{tabular}{cccc}
\toprule
$(j,j')$ & $\ell$ & fitted & measured \\
\midrule
$(0,1)/(1,0)$ & 2   & $-0.82$ & $-0.86$ \\
$(0,1)/(1,0)$ & 500 & $+0.68$ & $+0.66$ \\
$(1,2)/(2,1)$ & 2   & $-1.53$ & $-1.53$ \\
$(1,2)/(2,1)$ & 500 & $+0.61$ & $+0.67$ \\
\bottomrule
\end{tabular}
\caption{Comparison of fitted and measured radial slopes for representative derivative orders and multipoles.}
\label{tab:radial_slope_fits}
\end{table}
A residual deviation of order $0.4$ remains for $j+j'=3$ at $\ell=500$, reflecting an additional $\ell$ dependence of the marginal coefficient that the two-parameter form does not capture.
 
For $r\gg r_*$, $T(k_{\rm tp})\simeq 1$ in Eq.~\eqref{eq:wtp_general} and
\begin{equation}
w_{\rm tp}\;\propto\;r^{-(4+p)} ,
\label{eq:botIRenv}
\end{equation}
while the equality contribution Eq.~\eqref{eq:weq_above} gives
\begin{equation}
w_{\rm eq}\;\propto\;r^{-2} .
\label{eq:botIReq}
\end{equation}
For $p\le -2$, $w_{\rm tp}$ dominates and the envelope follows Eq.~\eqref{eq:botIRenv}.  For $p=0$, $w_{\rm eq}$ dominates and the envelope follows Eq.~\eqref{eq:botIReq}.  For $p=-1$ with $j+j'\ge 1$, Eqs.~\eqref{eq:botIRenv} and \eqref{eq:botIReq} are marginal in their relative weight; integration by parts gives
\begin{equation}
w(r)
\sim
\frac{\ln(k_{\rm eq}r/\ell)}{r^3},
\end{equation}
and therefore
\begin{equation}
\frac{\dd\ln |w|}{\dd\ln r}
=
-3
+
\frac{1}{\ln(k_{\rm eq}r/\ell)} .
\label{eq:botIRlog}
\end{equation}
Table~\ref{tab:radial_slope_large_r} summarizes the fixed-$\ell$ slopes for $r\gg r_*$.
\begin{table}[t!]
\centering
\begin{tabular}{ll}
\toprule
$p$ & slope in $r$ \\
\midrule
$0$ & $-2$ \quad equality-dominated \\
$-1$ with $j+j'\ge 1$ &
$-3+1/\ln(k_{\rm eq}r/\ell)$ \quad logarithmic marginal case \\
$\le -2$ & $-(4+p)$ \quad turning-point envelope \\
\bottomrule
\end{tabular}
\caption{Summary of asymptotic radial slopes at fixed multipole and large radius.}
\label{tab:radial_slope_large_r}
\end{table}
 
\subsubsection{$R$ dependence away from the diagonal}
 
The slope analysis above uses the diagonal curve $R=1$.  For the
off-diagonal curves at small $R$, we use the small-argument limit of the
second Bessel function,
\begin{equation}
j_\ell^{(j')}(kRr)
\simeq
\frac{\ell!}{(\ell-j')!}\frac{(kRr)^{\ell-j'}}{(2\ell+1)!!},
\qquad
R\ll 1 .
\end{equation}
At fixed $r$, this gives the approximate ratio
\begin{equation}
\frac{w(r,Rr)}{w(r,r)}
\simeq
R^{\ell-j'} .
\label{eq:smallRratio}
\end{equation}
Therefore the small-$R$ overlay lines use the same local $r$-slope as the
$R=1$ curve, but are shifted vertically by $R^{\ell-j'}$.  In the comparison
plots we apply this prescription for
\begin{equation}
R\le R_{\rm small}=0.3 .
\end{equation}
The data span only $R\in[0.1,1.1]$, so the large-$R$ asymptotic regime is
not tested.

\subsection{The PowerFull implementation}\label{app:powerfull_nR}

For all the {\sc TwoFAST} function calls, we use $N_r=4096$ {\sc FFTLog} sampling points, and $k_{\rm min}=10^{-5}~h/{\rm Mpc}$, $k_{\rm max}=10^3~h/{\rm Mpc}$. The input linear power spectrum need not cover the full wavelength range, because we fill out the power spectrum values outside of the tabulated $k$-range by a power law extrapolation at each end.  We choose the pivot radius $r_0=1/k_{\rm max}=10^{-3}~{\rm Mpc}/h$, which sets the result array logarithmically spaced between $0.001~{\rm Mpc}/h$ and $99551~{\rm Mpc}/h$; we keep the output between $z=0.01$ ($r\simeq 30~{\rm Mpc}/h$) and $z=5$ ($r\simeq 5400~{\rm Mpc}/h$).  We reserve a larger dynamical range for the {\sc FFTlog} than actually needed in the calculation to mitigate various numerical effects, and the same strategy is applied when choosing the optimal bias parameter $q$ in the previous section.

As explained in \refapp{twofast_review}, the \twofast\, library computes $w^{(p)}_{\ell,jj'}(r,r')$ on a logarithmic grid of radii $r$ for a given set of values $R \equiv r'/r$.  We parameterise $R$ by the number of points $n_R$ and the logarithmic spacing $\dd\ln R$, so that the grid spans between $\exp(-N\,\dd\ln R)$ and $\exp(N\,\dd\ln R)$ with $N \equiv (n_R - 1)/2$.

The four panels in \reffig{comparison_p0n0} show that $w^{(p)}_{\ell,jj'}(r, R)$ sharply peaks at $R=1$.  We therefore choose odd $n_R$ so that $R=1$ is always included.

\begin{table}[ht]

  \centering
  \vspace{2mm}
  \begin{tabular}{llll}
    \toprule
    Symbol & Definition & Physical origin \\
    \midrule
    $f_s(r)$ & $a^3 H^3 \Omega_\mathrm{m}(f-1)\,D$
             & Integrated Sachs-Wolfe \\
    $f_t(r)$ & $a^2 H^2 \Omega_\mathrm{m}\,D$
             & Shapiro time delay \\
    $f_{l}(r,\bar{r})$ & $a^2 H^2 \Omega_\mathrm{m}\,D (1/r-1/\bar{r})$
             & weak-lensing convergence \\
    \bottomrule
  \end{tabular}
  \caption{Cosmological kernel functions used in the {\sc PowerFull} implementation.  Here
    $a$ is the scale factor, $H$ the Hubble parameter, $\Omega_\mathrm{m}$
    the matter density parameter, $f\equiv\dd\ln D/\dd\ln a$ the linear growth rate, $D$ the linear growth factor, and $r$ the comoving radius.} 
    \label{tab:kernels}
\end{table}

Since we fix the radial grid of \powerfull\, the same as the $r$ grid of \twofast\, and \powerfull\, uses the same grid for both directions $(r_1, r_2)$, the mapping from the \twofast\, output on the $(r, R)$ plane to the physical radius on the $(r_1, r_2)$ plane is done by one-dimensional interpolation.  For each row of $r_1$ we build a cubic spline of $w^{(p)}_{\ell,jj'}(r_1, R)$ over the $R$-grid and evaluate it at $R = r_2/r_1$ for every $r_2$ in the radial grid.  The radial grid $r \in [30,\, 5400]\ {\rm Mpc}/h$ implies that the dimensionless ratio $R = r'/r$ between any two requested grid points covers $[r_{\rm min}/r_{\rm max},\, r_{\rm max}/r_{\rm min}] \approx [0.006,\, 180]$.  Two competing requirements constrain $(n_R, \dd\ln R)$.  At low $\ell$ the spherical Bessel functions oscillate slowly and the integrand has broad support in $R$, demanding a wide $R$ range.  At high $\ell$ the rapid oscillations of $j_\ell(kr_1)\,j_\ell(kr_2)$ confine the integrand to $r_1 \approx r_2$, demanding fine $\dd\ln R$.  A single $(n_R, \dd\ln R)$ pair that satisfies both at uniform cost is wasteful.

We therefore adopt a three-tier strategy.  The low-$\ell$ tier covers $\ell \in [2, 50]$ with $n_R = 4097$ and $\dd\ln R = 0.002$, giving $R \in [0.0166,\, 60.34]$; the wide $R$ range captures the broad integrand support at the lowest multipoles.  The mid-$\ell$ tier covers $\ell \in [51, 200]$ with $n_R = 2049$ and $\dd\ln R = 0.001$, giving $R \in [0.36,\, 2.79]$.  The high-$\ell$ tier covers $\ell \in [201, 500]$ with $n_R = 2049$ and $\dd\ln R = 0.0005$, giving $R \in [0.60,\, 1.67]$ around the $R = 1$ peak.  As $\ell$ increases the integrand support narrows, so the $R$ range contracts and the spacing $\dd\ln R$ refines from tier to tier; the lensing projections, which carry the $\ell(\ell+1)$ weight, set the resolution requirement and motivate the finer $\dd\ln R$ at high $\ell$.  We set the values at points outside the chosen $R$ range to zero, dropping the associated integration contributions.

The integrated terms $s_{\ell,jj';r}^p$, $\mathcal{S}_{\ell,jj'}^p$, and their analogues defined in Eqs.~(\ref{eq:s_jj'})--(\ref{eq:Z_jj'}) live in $(r, r')$ space, so we compute all integrations there. Specifically, given the base array, in $(r,r')$, from the {\sc TwoFAST} array, in $(r,R)$, by the one-dimensional interpolation along $R=r'/r$ for a fixed $r$, \textsc{PowerFull} constructs each 1D integral
\begin{equation}
  \mathcal{I}^{(p)}_{\ell,X}(r_i, r_j)
  = \frac{3}{D(r_i)}
    \int_0^{r_i} \mathrm{d}r_1\;
    f_X(r_1)\;
    w^{p}_{\ell,00}(r_1, r_j)\,,
  \label{eq:prefix_1D}
\end{equation}
as a cumulative trapezoidal summation along $r_1$ on the logarithmic grid and each 2D integral
\begin{equation}
  \mathcal{J}_\ell(r, r')
  = \frac{9}{D(r)D(r')}
    \int_0^{r}\!\mathrm{d}\!\ln r_1
    \int_0^{r'}\!\mathrm{d}\!\ln r_2\;
    \widetilde{f}_1(r_1)\,\widetilde{f}_2(r_2)\;
    w^{(-4)}_{\ell,00}(r_1, r_2)\,.
  \label{eq:prefix_2D}
\end{equation}
as two successive axis-wise cumulative trapezoidal summations; the base values $w$ enter once
as pre-tabulated arrays. The employment of the trapezoidal rule is driven by the structure of the problem, where the line-of-sight integrals are required not at a single upper limit but at every pair of source distances $(r, r')$ on the grid. Both the $X_{;r}$ and $X_{;r'}$ integrals are computed as cumulative (prefix) sums in $\mathcal{O}(N_r^2)$ operations, with endpoint halving at both ends of every $;r$ and $;r'$ sum, including the diagonal $r=r'$ boundary, so that each partial sum equals the trapezoidal integral up to its own upper limit. The trapezoidal weight pattern (uniform on the interior, halved at the two endpoints) extends incrementally with the upper limit and is therefore compatible with a single forward sweep through the grid. Higher-order closed Newton-Cotes schemes such as Simpson's $1/3$ rule break this incremental structure: their alternating interior weights and even-interval requirement force any cumulative variant to $\mathcal{O}(N_r^3)$ work for no accuracy gain, since both rules are $\mathcal{O}(h^2)$ on these smooth integrands.

Finally, we exploit the symmetry
\begin{equation}
  \label{eq:integration_symmetry}
  \Xi^{(p)}_{\ell,jj';r}(r, r') = \Xi^{(p)}_{\ell,j'j;r'}(r', r),
\end{equation}
which renders the lensing-galaxy integral redundant once the galaxy-lensing one is in hand.  {\sc PowerFull} realizes Eq.~(\ref{eq:integration_symmetry}) by an array transpose: the asymmetric 1D integrals at $j \neq j'$ are computed once in the $;r$ orientation, the $;r'$ counterpart with $j$ and $j'$ swapped is obtained as the transpose of the resulting $(r_1, r_2)$ matrix, and the 2D cross terms $\mathcal{X}^{(-4)}_{00}$, $\mathcal{Y}^{(-4)}_{00}$, $\mathcal{Z}^{(-4)}_{00}$ are similarly computed once in the $(r, r')$ orientation and reused for $(r', r)$.  The two orientations share identical interpolation residuals by construction.

\subsubsection{Testing the choice of $n_R$, $\dd\ln R$}\label{app:powerfull_nR_test}

The two parameters $(n_R, \dd\ln R)$ act on the result through two distinct mechanisms: the product $\exp(N \dd\ln R)$ sets the $R$ range, and $\dd\ln R$ alone sets the local spline-interpolation resolution.  We validate the chosen $(n_R, \dd\ln R)$ for each tier with two orthogonal tests.  The $R$-range test holds $\dd\ln R$ fixed and increases $n_R$ above the baseline value, which widens the $R$ window; the residual measures the contribution of the $R$ tail discarded by the baseline.  The $\dd\ln R$ test holds the $R$ window fixed and refines $\dd\ln R$; the residual measures the sensitivity to the local interpolation resolution.

For each test we compute the per-array, per-$\ell$ relative $L_2$ residual on the $(r_1, r_2)$ grid, comparing each stored component before it is assembled into $C_\ell^{\rm GR}$,
\be\label{eq:l2_rel}
  \epsilon_\ell^A(I) \;\equiv\;
  \frac{\bigl[\sum_{ij}(I^A_{ij}(\ell) - I^B_{ij}(\ell))^2\bigr]^{1/2}}
       {\bigl[\sum_{ij}(I^B_{ij}(\ell))^2\bigr]^{1/2}}\,,
\ee
where $I$ runs over all 61 stored arrays (39 integrated quantities and 22 base functions), and $A$ and $B$ denote the two builds being compared.

\begin{figure*}[!ht]
  \centering
    \includegraphics[width=0.325\textwidth]{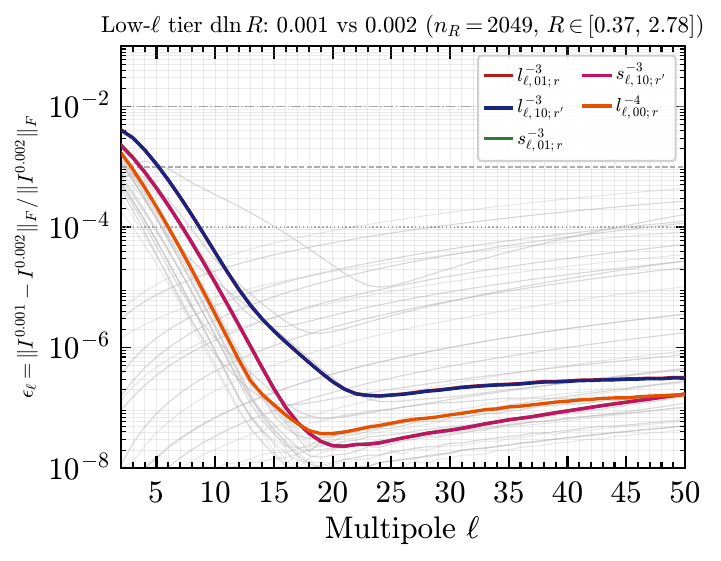}
  \includegraphics[width=0.325\textwidth]{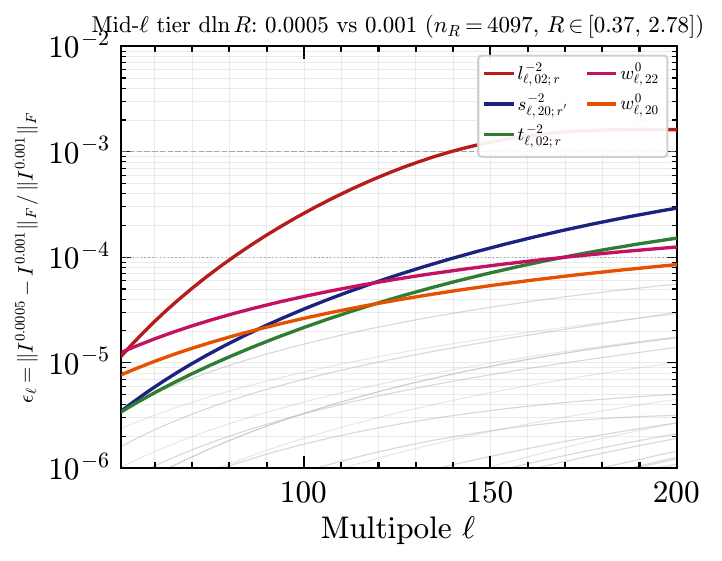}
  \includegraphics[width=0.325\textwidth]{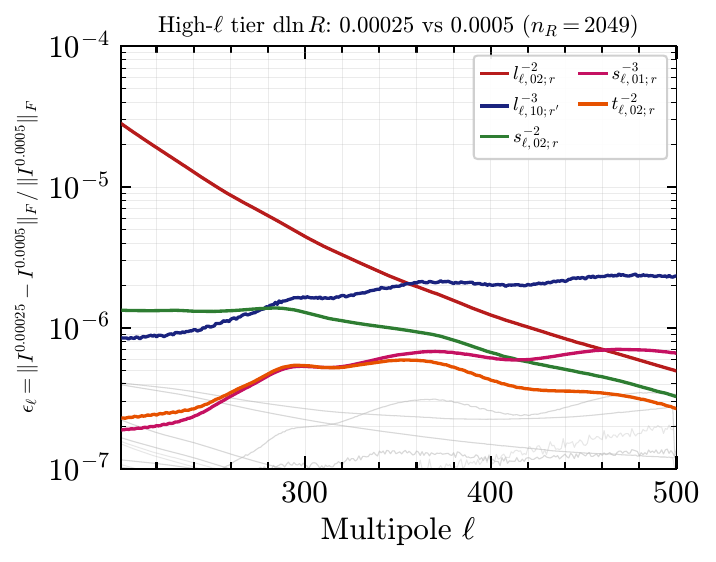}
  \includegraphics[width=0.325\textwidth]{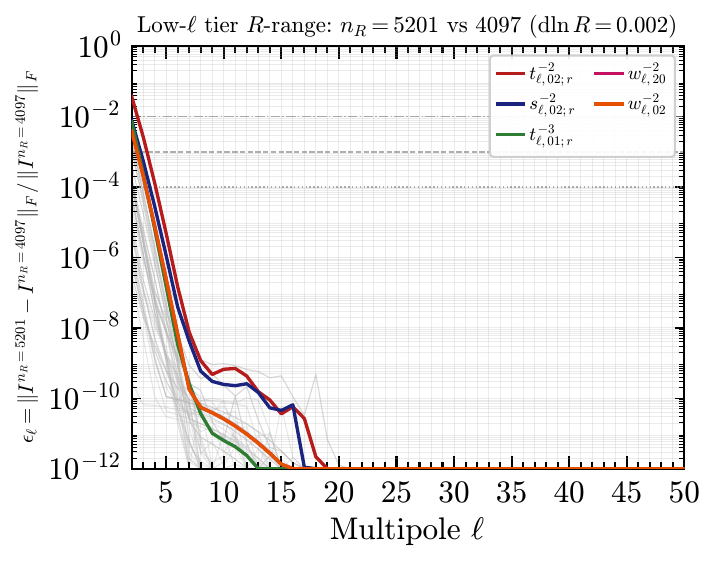}
 \includegraphics[width=0.325\textwidth]{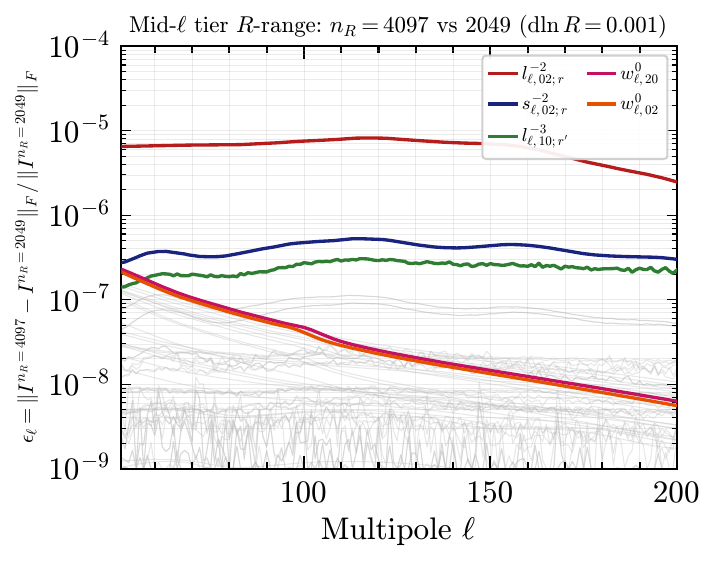}
 \includegraphics[width=0.325\textwidth]{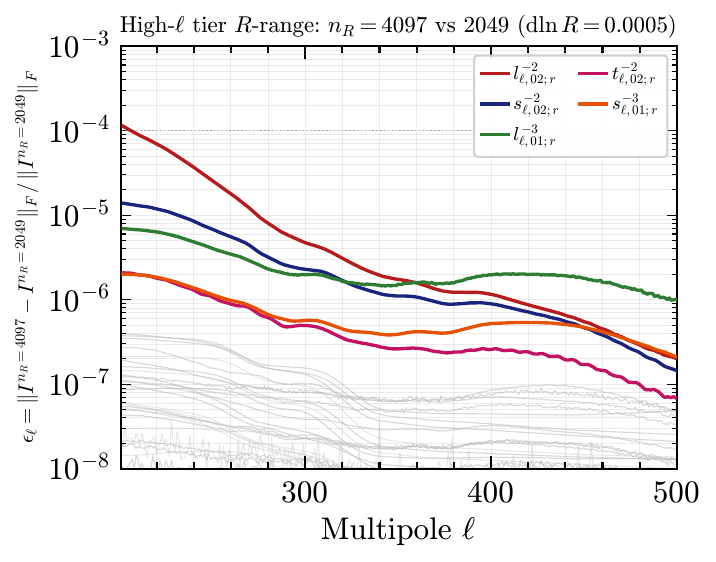}
  \caption{Convergence tests for the three production tiers, low-$\ell$ (left), mid-$\ell$ (middle), and high-$\ell$ (right).  Top row: $\dd\ln R$ tests, each comparing the baseline $\dd\ln R$ against a finer build at fixed $R$ window common to both builds.  Bottom row: $R$-range tests, each comparing the baseline $n_R$ against a wider build at fixed $\dd\ln R$.  The build parameters are given in each panel title.  Each panel shows the per-array, per-$\ell$ residual $\epsilon_\ell$ (\refeq{l2_rel}); the five arrays with the largest $\epsilon_\ell$ over the plotted $\ell$ range are coloured and labelled (transpose pairs deduplicated), and the remaining arrays form a grey envelope.  The horizontal lines mark $10^{-2}$ (dash-dot), $10^{-3}$ (dashed), and $10^{-4}$ (dotted).  The worst array is the lensing projection $l^{-2}_{\ell, 02; r}$ in every panel except the low-$\ell$ $R$-range test, where it is the companion $t^{-2}_{\ell, 02; r}$.}
  \label{fig:nR_test}
\end{figure*}

\paragraph{Low-$\ell$ tier:}
The low-$\ell$ baseline at $(n_R, \dd\ln R) = (4097, 0.002)$ gives $R \in [0.0166, 60.34]$.  For the $R$-range test (\reffig{nR_test}, bottom left), we compare against a wider build at $(n_R, \dd\ln R) = (5201, 0.002)$, with $R \in [0.0055, 181]$; both share $\dd\ln R$, so the residual isolates the contribution of the discarded $R$ tail.  The worst array $t^{-2}_{\ell, 02; r}$ and its transpose reach $3.8 \times 10^{-2}$ at $\ell = 2$ and fall to float64 underflow by $\ell = 20$; all remaining arrays sit below this value at $\ell = 2$.  The baseline $n_R = 4097$ captures the integrand support except at the very lowest multipoles.

For the $\dd\ln R$ test (\reffig{nR_test}, top left) we compare $\dd\ln R = 0.001$ against $\dd\ln R = 0.002$ on the $R$ window $[0.37, 2.78]$ common to both builds, with shared $n_R = 2049$.  With the $R$ window held fixed, the residual isolates the interpolation resolution.  The worst arrays are the $p = -3$ projections of $s$ and $l$ with $j + j' > 0$: $l^{-3}_{\ell, 01; r}$ and its transpose reach $4.1 \times 10^{-3}$ at $\ell = 2$, and $s^{-3}_{\ell, 01; r}$ and its transpose reach $2.2 \times 10^{-3}$; all four drop below $10^{-3}$ by $\ell = 5$ and to the float64 floor by $\ell = 20$.

\paragraph{Mid-$\ell$ tier.}
The mid-$\ell$ baseline at $(n_R, \dd\ln R) = (2049, 0.001)$ gives $R \in [0.36, 2.79]$, a narrower window than the low-$\ell$ tier.  For the $R$-range test (\reffig{nR_test}, bottom middle) we compare against $(n_R, \dd\ln R) = (4097, 0.001)$, with $R \in [0.13, 7.76]$.  The worst array $l^{-2}_{\ell, 02; r}$ and its transpose stay below $8.2 \times 10^{-6}$ across the tier; all remaining arrays sit below this value.  The narrow $R$ window therefore captures the mid-$\ell$ integrand support to better than $10^{-5}$.

For the $\dd\ln R$ test (\reffig{nR_test}, top middle) we compare $\dd\ln R = 0.0005$ against $\dd\ln R = 0.001$ on the $R$ window $[0.37, 2.78]$ common to both builds, with $n_R = 4097$.  The worst array $l^{-2}_{\ell, 02; r}$ and its transpose grow from $1.2 \times 10^{-5}$ at $\ell = 51$ to $1.6 \times 10^{-3}$ at the top of the tier; all remaining arrays sit below this value.  The lensing projection $l$ carries the $\ell(\ell+1)$ weight, and the near-cancellation in $l = (\ell(\ell+1)/2)(\tilde l - t/r)$ amplifies its interpolation residual relative to the other bases.

\paragraph{High-$\ell$ tier.}
The high-$\ell$ baseline at $(n_R, \dd\ln R) = (2049, 0.0005)$ gives $R \in [0.60, 1.67]$.  For the $R$-range test (\reffig{nR_test}, bottom right) we compare against a wider build at $(n_R, \dd\ln R) = (4097, 0.0005)$, with $R \in [0.36, 2.78]$.  The worst array $l^{-2}_{\ell, 02; r}$ and its transpose stay below $1.2 \times 10^{-4}$ across the tier; all remaining arrays sit below this value.  The rapid $j_\ell(kr) j_\ell(kr')$ oscillations confine the integrand to $R \sim 1$ at high $\ell$, and the baseline window $[0.60, 1.67]$ captures that support.

For the $\dd\ln R$ test (\reffig{nR_test}, top right) we compare $\dd\ln R = 0.00025$ against $\dd\ln R = 0.0005$ at $n_R = 2049$, on the $R$ window $[0.774, 1.292]$ common to both builds.  The worst array $l^{-2}_{\ell, 02; r}$ and its transpose stay below $2.8 \times 10^{-5}$ across the tier, and all remaining arrays sit below this value.  The baseline $\dd\ln R = 0.0005$ resolves the local spline at the $10^{-5}$ level across the high-$\ell$ tier.

The six panels show the robustness of the {\sc PowerFull} results against the change of $n_R$ and $\dd\ln R$. We have also performed the complementary accuracy check against a brute-force integration in \refapp{pkfull}.

\subsubsection{Sub-percent validation against brute-force integration}
\label{app:pkfull}
\begin{figure}
  \centering
  \includegraphics[width=0.8\textwidth]{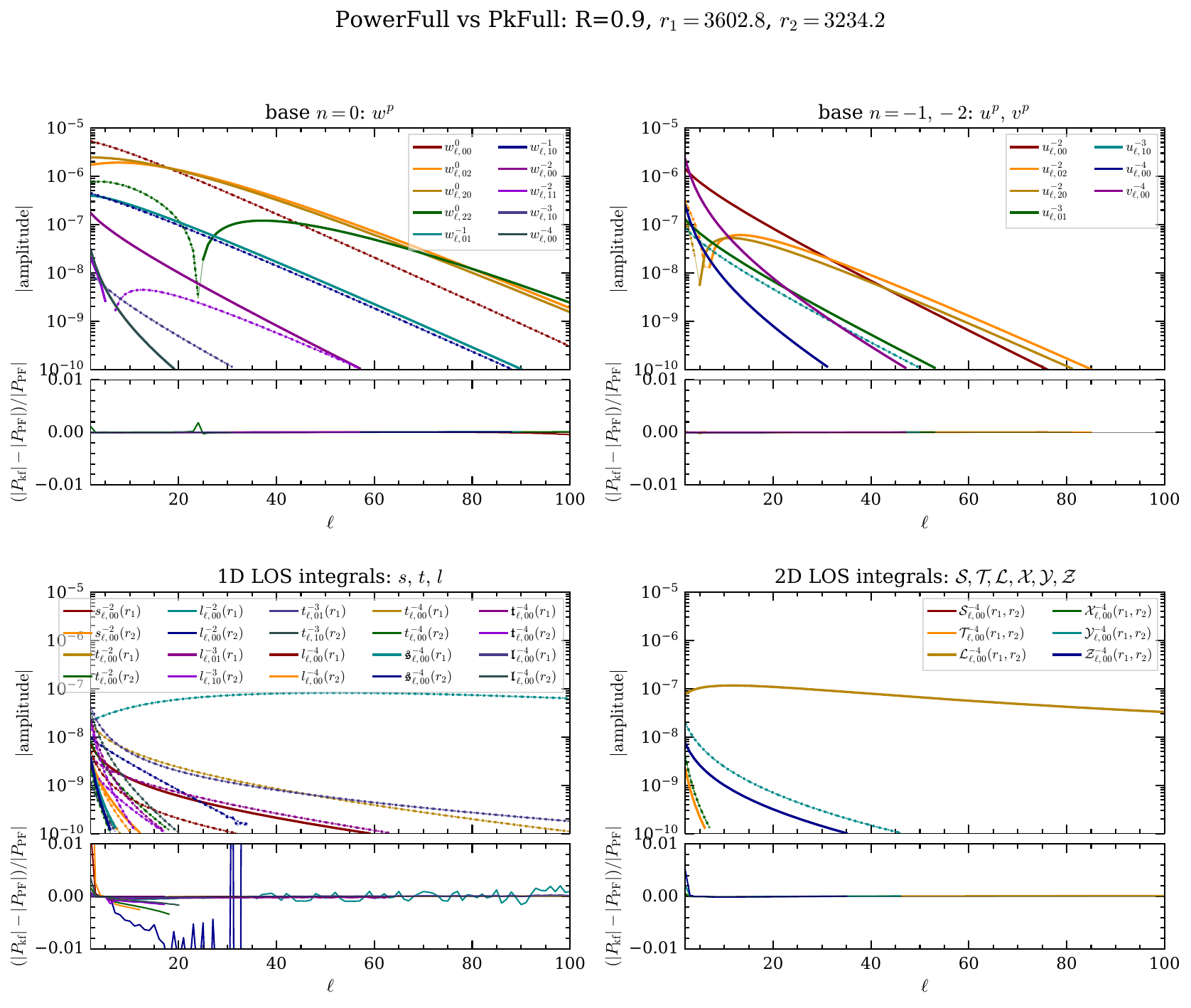}
  \includegraphics[width=0.8\textwidth]{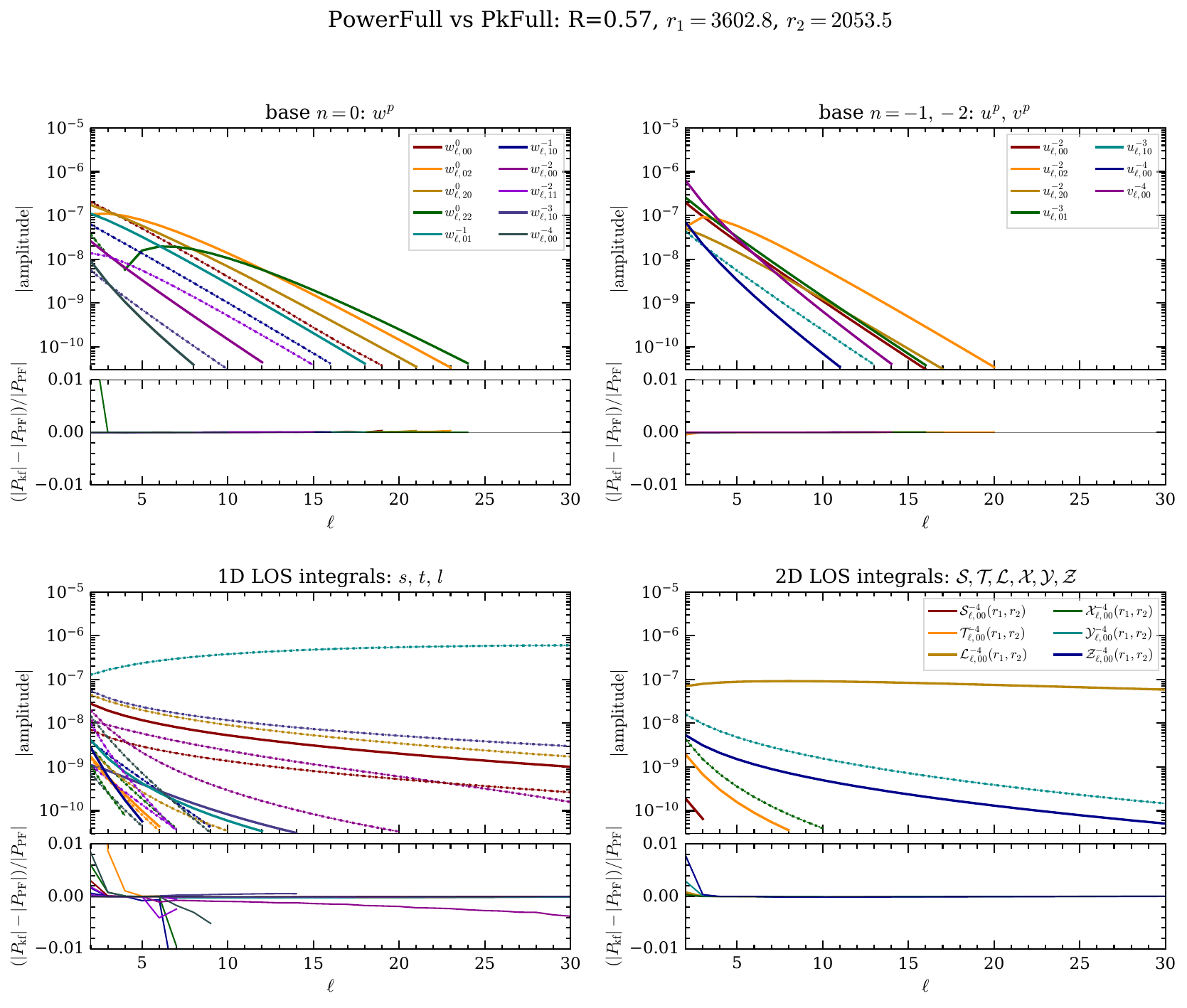}
  \caption{Per-base validation of \textsc{PowerFull} (dashed) at $R = 0.9$ (top) and $R = 0.57$ (bottom).  For the two-point bases (top row: $w$ at left, $u, v$ at right) the solid curve is an independent {\sc Julia} evaluation of the Lucas algorithm; for the line-of-sight projections (bottom row: 1D at left, 2D at right) it is \textsc{PkFull}'s {\sc FORTRAN} integral.  Residual panel window: $\pm 10^{-3}$.  The $w, u, v$ and $p = -2, -3$ bases agree at the $\sim 10^{-5}$ Lucas-vs-FFTLog floor; the lensing $l^{-4}$ at $\ell = 2$ exceeds the window on the $r_2$ side, reaching $3.6\%$ ($R = 0.9$) and $8.7\%$ ($R = 0.57$), with $t^{-4}$ a factor $15$--$50$ smaller.  Vertical spikes near $\ell \simeq 38$ are zero crossings of $l^{-3}$.}
  \label{fig:pkfull_base}
\end{figure}
\begin{figure}
  \centering
  \includegraphics[width=\textwidth]{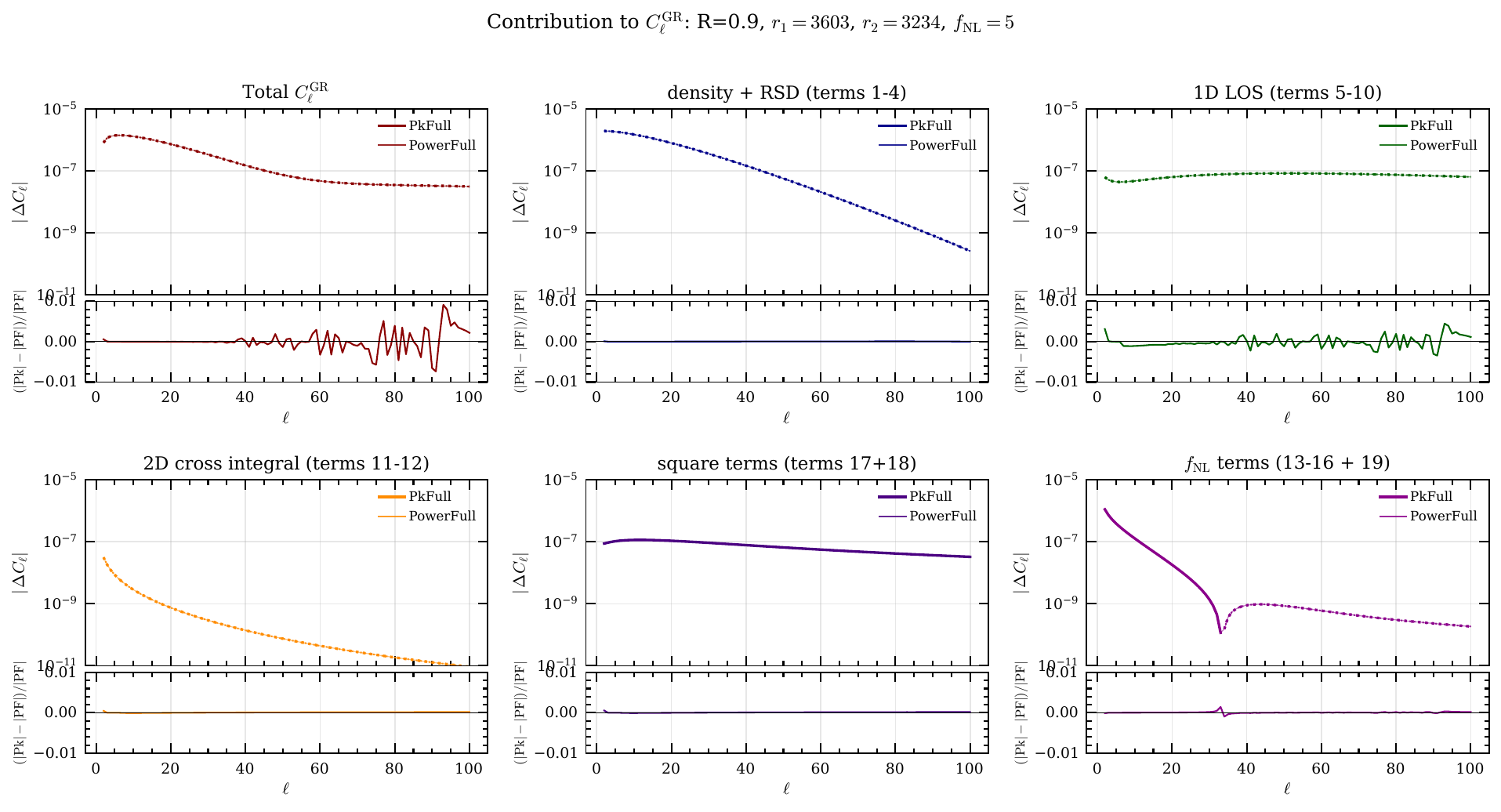}
  \includegraphics[width=\textwidth]{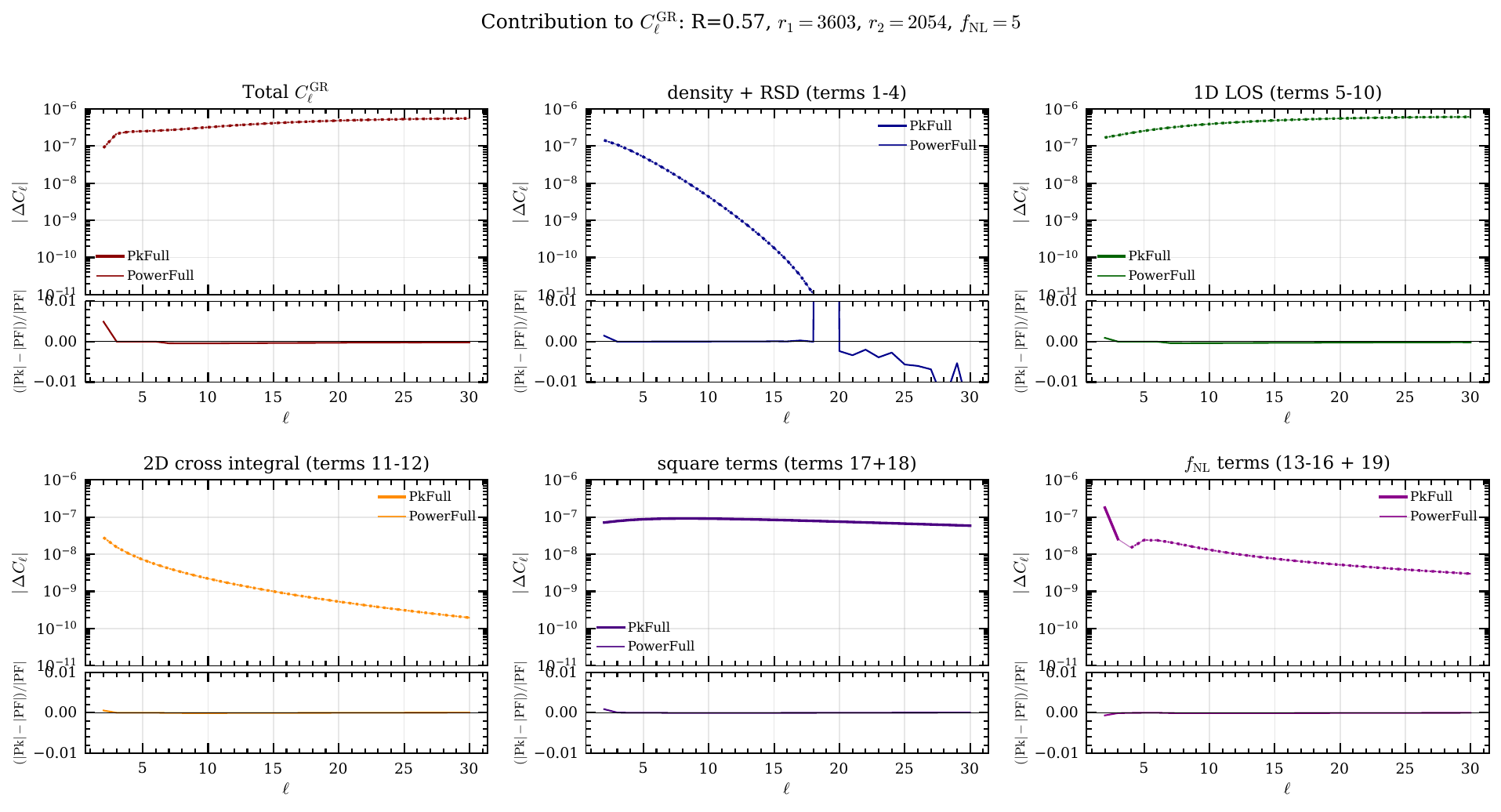}
  \caption{Contribution to $C_\ell^{\rm GR}$ at $f_{\rm NL} = 5$,
    $R = 0.9$ (left) and $R = 0.57$ (right).  Total plus the five line
    groups of \refeq{Cl_GR_expanded}.  The total $\ell = 2$ residual is
    $5.9 \times 10^{-4}$ ($R = 0.9$) and $4.9 \times 10^{-3}$
    ($R = 0.57$); the density$+$RSD and $f_{\rm NL}$ groups dominate the
    amplitude with opposite signs, so the total falls below either.
    The $\lesssim 9\%$ base alias on $l^{-4}$ of \reffig{pkfull_base} is
    suppressed by the $\mathcal{H}^{2}\mathcal{A}$ weighting.}
  \label{fig:pkfull_cl}
\end{figure}
As an independent cross-check of \textsc{PowerFull}, we compare its output against a brute-force numerical computation of the same integrals at two fixed target source pairs on a common radial grid.  The reference code \textsc{PkFull} is a single {\sc FORTRAN} solver that shares every cosmological and numerical input with \textsc{PowerFull}.  It computes the $22$ two-point bases ($w, u, v$) by the Lucas $h_1/h_2$ decomposition of Ref.~\cite{LUCAS1995269} for the $j = j' = 0$ members and by adaptive QUADPACK quadrature for the derivative members, and the $39$ line-of-sight projections ($24$ 1D LOS, $6$ PNG-coupled fraktur, $9$ 2D crosses) by a single Simpson $k$-loop that reverses the order of the nested $k$ and $r$ integrals of \textsc{PowerFull}.  Unlike the per-component ratios of \reffig{nR_test}, the comparisons in this section are all formed in the $C_\ell^{\rm GR}$ contribution that each term enters.

All cosmological background quantities ($z, a, H, \Omega_\mathrm{m}, f, D$) are read from a common tabulation and interpolated by identical cubic splines.  By construction, therefore, the integration kernel values $f_s, f_t, f_l$ (\reftab{kernels}) agree in both codes to machine precision on every grid point.  The linear matter power spectrum $P_L(k)$ uses the same cubic B-spline representation with power-law continuations at the low-$k$ and high-$k$ extrapolation boundaries.  The radial grid $\{r_j\}_{j=1}^{N_r}$ coincides with the \textsc{PowerFull} grid, filtered to the range $r \in [29.96, 5376.1]\,h^{-1}\mathrm{Mpc}$ with $N_r = 1155$ points, which we adopt for the {\sc PowerFull} implementation.  We fix $r_1 = r(z{=}2) = 3602.80\,h^{-1}\mathrm{Mpc}$ and test two source pairs, $r_2 = 3234.19\,h^{-1}\mathrm{Mpc}$ ($R \equiv r_2/r_1 = 0.8977$) and $r_2 = 2053.50\,h^{-1}\mathrm{Mpc}$ ($R = 0.5700$).  Spherical Bessel functions use the AMOS implementation of \citep{Amos1986} via \texttt{zbesj} and \texttt{zbesy}, the same routines that \textsc{PowerFull} invokes inside its FFTLog base-function construction.

The base panels and the line-of-sight panels of \reffig{pkfull_base} test two separate stages of \textsc{PowerFull}.  The base panels test whether the TwoFAST FFTLog output degrades when \textsc{PowerFull} maps it onto the $(r_1, r_2)$ grid by cubic-spline interpolation: we compare \textsc{PowerFull} against an independent {\sc Julia} evaluation of the Lucas algorithm at the exact $(r_1, r_2)$, using the stable $j_{\ell-1}$ derivative recurrence, so the residual is the interpolation degradation alone with no shared code path.  The line-of-sight panels test whether the cumulative trapezoidal scheme integrates accurately: we compare \textsc{PowerFull} against \textsc{PkFull}'s {\sc FORTRAN} integrals, which evaluate the same line-of-sight integrals with the $k$ and $r$ orders reversed.  At $R = 1$ the Lucas reference and \textsc{PowerFull} evaluate the analytic expression of \citep[Eq.~(35)]{Gebhardt::2018twofast} and agree at double-precision round-off.

The interpolation holds the bases at the $\sim 10^{-5}$ level in the ratio $P_{\rm Lucas}/P_{\rm PowerFull} - 1$ wherever the base magnitude stays well above the float64 floor: $\ell \lesssim 30$ at $R = 0.9$ and $\ell \lesssim 10$ at $R = 0.57$.  The ratio is sensitive to its denominator, so it grows wherever the base is small, either through a zero crossing or through the $j_\ell(k r_1) j_\ell(k r_2)$ cancellation that shrinks the integral at high $\ell$; the latter drives the ratio to a few $\times 10^{-4}$ by $\ell \sim 100$ at $R = 0.9$ and to the percent level by $\ell \sim 30$ at $R = 0.57$, the stronger cancellation there reaching the floor sooner.  This growth is in the ratio alone: the absolute interpolation error holds at the floor, and the $\mathcal{H}^2 \mathcal{A}$ weighting and internal cancellation of \refeq{Cl_GR_expanded} suppress it in the assembled $C_\ell^{\rm GR}$.  The residual is the interpolation degradation of the FFTLog reconstruction, not a coding mismatch.

While \textsc{PowerFull} constructs each 1D and 2D integral as cumulative trapezoidal summations, so that the base values $w$ enter once as pre-tabulated arrays, \textsc{PkFull} reverses this order: a single Simpson loop over $k$ evaluates the inner sum $I_X(k) = \sum_j \tilde f_X(r_j)\,j_\ell(k r_j)\,\dd\!\ln r$ at each $k$ and integrates over $k$, at $\mathcal{O}(N_k N_r)$ cost rather than $\mathcal{O}(N_r^2)$.  Interchange of the summation over $r$ and the integration over $k$ makes the two procedures mathematically equivalent on the same grid.
 
\textsc{PkFull} forms every base of \refeq{Cl_GR_expanded} individually: $22$ two-point bases ($w, u, v$), $24$ 1D LOS projections ($s, t, l$, four bases per symbol on each of the two source sides), $6$ PNG-coupled fraktur quantities ($\mathfrak{s}, \mathfrak{t}, \mathfrak{l}$ on both sides), and $9$ 2D crosses ($\mathcal{S, T, L}$ symmetric, $\mathcal{X, Y, Z}$ each in both argument orders).  Comparing the fields separately isolates the residual on each base from cancellations in the assembly.
 
\reffig{pkfull_base} compares the fields individually; the base panels show $17$ of the $22$ two-point bases, the five $\beta$-cousins absent from the {\sc Julia} reference set being omitted.  The two-point $w, u, v$ bases and the $p = -2$ LOS projections agree at the $\sim 10^{-5}$ floor described above over the low-$\ell$ range, with the ratio rising at high $\ell$ as the bases shrink toward the float64 floor.  The $p = -3$ pair $l^{-3}_{\ell,01}, l^{-3}_{\ell,10}$ is the same effect at a zero crossing: it passes through zero near $\ell \simeq 38$, and the ratio panel diverges there with no code mismatch.
 
The exception is the lensing projection $l^{-4}, t^{-4}$ at $\ell = 2$, in the 1D LOS panels where the reference is \textsc{PkFull}'s {\sc FORTRAN} integral.  The residual on $l^{-4}$ reaches $3.6\%$ at $R = 0.9$ and $8.7\%$ at $R = 0.57$ on the $r_2$ side, and $1.3\%$ and $0.30\%$ on the $r_1$ side, with $t^{-4}$ a factor $15$--$50$ smaller.  The $k^{-4}$ weighting concentrates the integrand at low $k$, where \textsc{PowerFull}'s FFTLog $k_{\min}$ cutoff aliases; \textsc{PkFull}'s direct $k$-integration carries no such floor.  The monotonic $r_2$ dependence follows: smaller $r_2$ samples larger physical scales, weighting $k \lesssim k_{\min}$ more heavily in $j_\ell(kr_2)$.  The paper observable $l = (\ell(\ell+1)/2)(\tilde l - t/r_s)$ amplifies this residual through the cancellation between $\tilde l$ and $t/r_s$, so the $l$ panels are more sensitive to the alias than the bare $\tilde l$.  The 2D quantities $\mathcal{T}^{-4}, \mathcal{L}^{-4}$ inherit the same aliasing through their $k^{-4}$ weighting; $\mathcal{X, Y, Z}$ stay at the floor.
 
The percent-level base residual at $\ell = 2$ does not propagate undamped to $C_\ell^{\rm GR}$.  The lensing projections enter \refeq{Cl_GR_expanded} with prefactor $\mathcal{H}^{2}\mathcal{A} \sim 10^{-7}$ at $z \simeq 2$, which suppresses the absolute mismatch by the same factor in the assembled spectrum.
 
\reffig{pkfull_cl} shows the residual of the five line groups of \refeq{Cl_GR_expanded} at $f_{\rm NL} = 5$.  The total $\ell = 2$ residual is $5.9 \times 10^{-4}$ at $R = 0.9$ and $4.9 \times 10^{-3}$ at $R = 0.57$.  At $f_{\rm NL} = 5$ the density$+$RSD and $f_{\rm NL}$ groups dominate the amplitude and carry opposite signs, so the total falls below either: at $R = 0.9$ the density$+$RSD group contributes $-1.93 \times 10^{-6}$ and the $f_{\rm NL}$ group $+1.12 \times 10^{-6}$, and at $R = 0.57$ the $f_{\rm NL}$ group ($+1.79 \times 10^{-6}$) is the largest while the density$+$RSD and 1D LOS groups together cancel most of it.  The relative residual is largest on the 1D LOS group, $3.1 \times 10^{-3}$ at $R = 0.9$ and $9.5 \times 10^{-4}$ at $R = 0.57$, tracking the $l^{-4}, t^{-4}$ base mismatch; every other group sits at the $10^{-4}$ level.
 
Across $\ell > 5$ every group stays at the sub-permille level, with one exception: at $R = 0.9$ the 1D LOS group and the total reach $\sim 9 \times 10^{-3}$ near $\ell = 93$, where the 1D LOS contribution changes sign and the cancellation amplifies the underlying per-base difference.  This residual sits well below cosmic variance at every $\ell$, so \textsc{PowerFull} reproduces the brute-force $C_\ell^{\rm GR}$ to sub-percent accuracy over the multipole range of interest.

\section{On the accuracy of Limber approximation}
\label{app:limber_validation}

The relativistic galaxy two-point statistics computed in this work are full-sky observables whose signal is dominated by large angular scales, that is, by the lowest multipoles.  The Limber approximation, which replaces the radial integral over a product of two spherical Bessel functions by its $\ell \to \infty$ delta-function limit, is the standard shortcut for projected statistics, but it is controlled only at high $\ell$.  In this section we quantify its accuracy across the multipole range we compute ($\ell \leq 500$) and show that it is not an admissible replacement for the full radial integration in this regime.  We form the ratio of each integral computed by \textsc{PowerFull} to its Limber prediction; the comparison both confirms that the full pipeline recovers the correct $\ell \to \infty$ limit, with the correct sign and order of magnitude, and measures the Limber error directly.  At the multipoles relevant for our calculation the sub-leading corrections to the approximation are at the $10^{-2}$ level on the non-lensing arrays (and at the $10^{-3}$ level on the lensing 2D diagonals), carry opposite signs between the 1D and non-lensing 2D blocks, and do not share a single $\ell$-scaling: three obstructions that together preclude a uniform Limber correction.  The Limber comparison therefore cannot itself validate \textsc{PowerFull} below the percent level; a distinct sub-percent check on the assembled $C_\ell^\mathrm{GR}$, based on direct brute-force integration rather than an asymptotic limit, is developed in \refapp{pkfull}.

\subsection{The Limber approximation}\label{app:limber}
The base functions $w^{p}_{\ell,jj'}(r, r')$ are computed via Hankel transforms of products of spherical Bessel functions weighted by the matter power spectrum.  When the function multiplied to the two spherical Bessel functions is smooth enough, the rapid oscillations of the Bessel functions localize the integrand, and expanding the identity 
\be
\frac{2}{\pi}\int k^2\dd k\, j_\ell(kr) j_\ell(kr') = \frac{\delta_D(r-r')}{r^2}
\ee
to the entire integral yields the Limber approximation \citep[e.g.,][]{LoVerde::2008limber}:
\be
\label{eq:limber_w}
  w^{p}_{\ell,00}(r, r')
  \;\xrightarrow{\;\ell \to \infty\;}
  \left(\frac{\nu}{r}\right)^{p}
  \frac{P_\mathrm{m}(\nu/r)}{r^2}\;
  \delta_\mathrm{D}(r - r')\,,
\ee
where $\nu \equiv \ell + 1/2$.  The appearance of $\nu$ rather than $\ell$ corresponds to the extended Limber approximation \citep{LoVerde::2008limber}.

The same derivation extends to the $n \neq 0$ base functions of {\sc TwoFAST} without modification.  The bases $u^p_{\ell,jj'}$ $(n=-1)$ and $v^p_{\ell,jj'}$ $(n=-2)$ differ from $w^p_{\ell,jj'}$ only by an extra factor $T(k)^n$ inside the Hankel transform, where $T(k) \equiv [P_\mathrm{m}(k)/P_\mathrm{prim}(k)]^{1/2}$ is the transfer function normalized against the primordial power-law extrapolation $P_\mathrm{prim}(k) \propto k^{n_s}$.  The Limber limit then follows from \refeq{limber_w} by the replacement $P_\mathrm{m}(\nu/r) \to P_\mathrm{m}(\nu/r)\,T(\nu/r)^n$ in the integrand, and the downstream formulas \refeqs{limber_1D}{limber_2D} below inherit the same factor.

For $j,j'>0$, we note that $j_\ell^{(j)}(kr) = k^{-j}\,\partial_r^j\, j_\ell(kr)$; the general base function therefore factorizes as
\be
\label{eq:wjj_derivative}
  w^p_{\ell,jj'}(r,r')
  = \partial_r^j\,\partial_{r'}^{j'}\,w^{p-j-j'}_{\ell,00}(r,r')\,.
\ee
Applying \refeq{limber_w} to the right-hand side gives
\be
\label{eq:limber_wjj}
  w^p_{\ell,jj'}(r,r')
  \;\xrightarrow{\;\ell \to \infty\;}
  \partial_r^j\,\partial_{r'}^{j'}\left[
  \left(\frac{\nu}{r}\right)^{p-j-j'}
  \frac{P_\mathrm{m}(\nu/r)}{r^2}\;
  \delta_\mathrm{D}(r - r')
  \right].
\ee
When this is folded into the line-of-sight integral against a weight function $f(r')$, the $\partial_{r'}^{j'}$ acts distributionally:
\be
\label{eq:limber_dist}
\int \dd r'\, f(r')\,\partial_{r'}^{j'}\delta_D(r-r') = (-1)^{j'} f^{(j')}(r)\,,
\ee
reducing the double integral to a single integral over $r$ with derivatives of the weight function.
 
We note that this direct-derivative route is the only valid path to the Limber limit for $j+j'>0$.  In the full numerical computation, the recursion relations in \refeqs{w01_twofast}{w21_twofast} express $w_{\ell,jj'}^p$ as linear combinations of even-$\Delta\ell$ base integrals $w_{\ell',\ell''}^{p+1}$; this representation is algebraically equivalent to \refeq{wjj_derivative} and numerically efficient within the {\sc TwoFAST} framework.  In the Limber limit, however, all off-diagonal terms $w_{\ell',\ell''}^{p+1}$ with $\ell'\neq\ell''$ vanish by orthogonality, and the finite-difference structure that encodes the distributional derivative $\delta_D^{(j')}$ is lost.  The direct-derivative form in \refeq{limber_wjj} preserves this structure by construction.

We compute the Limber approximation for the integration terms, therefore, by replacing $w_{\ell,00}^p$ using \refeq{limber_w}.  For the derivative terms with $j+j'>0$, although the Limber limit in \refeq{limber_wjj} is formally exact as $\ell\to\infty$, each derivative of $\delta_D(r-r')$ degrades the convergence by one power of $\nu$: the next-to-leading-order correction scales as $O(\nu^{-(2-j-j')})$ rather than $O(\nu^{-2})$.  At the multipoles relevant for our calculation ($\ell \lesssim 1000$), the $j'=1$ terms converge only at the $\sim 15\%$ level, while the $j'=2$ terms do not converge at all.  We therefore apply the Limber approximation only to the $j=j'=0$ base integrals and compute all derivative terms with the full {\sc TwoFAST} algorithm.

\subsection{Application to the radial integrals}
\label{app:limber_1D}

The one-dimensional (1D) integrals stored by \textsc{PowerFull} take the form
\begin{equation}
  \mathcal{I}^{(p)}_{\ell,X}(r_i, r_j)
  = \frac{3}{D(r_i)}
    \int_0^{r_i} \mathrm{d}r_1\;
    f_X(r_1)\;
    w^{p}_{\ell,00}(r_1, r_j)\,,
  \label{eq:limber_prefix_1D} 
\end{equation}
where $D(r)$ is the linear growth factor and $f_X(r)$ is one of the
cosmological kernel functions listed in Table~\ref{tab:kernels}.  Taking advantage of the logarithmic binning of $r$ in {\sc TwoFAST} for computing $w_{\ell,00}^p(r,R)$, we implement the integration by $f(r)\,\mathrm{d}r = r f(r)\,\mathrm{d}\!\ln r$, evaluated cumulatively in $r_1$ by the trapezoidal rule on the logarithmic radial grid, so that a single pass yields $\mathcal{I}^{(p)}_\ell(r_i, r_j)$ for every $r_i \geq r_j$.

Inserting the Limber limit~\eqref{eq:limber_w} into Eq.~\eqref{eq:limber_prefix_1D} and performing the $r_1$-integration, one obtains
\begin{equation}
  \mathcal{I}^{(p)}_{\ell,X}(r_i, r_j)
  \;\xrightarrow{\;\ell \to \infty\;}
  \frac{3}{D(r_i)}\;
  f_X(r_j)\,
  \left(\frac{\nu}{r_j}\right)^{p}
  \frac{P_\mathrm{m}(\nu/r_j)}{r_j^2}\,.
  \qquad (r_i \geq r_j)\,
  \label{eq:limber_1D}
\end{equation}
We compute the Limber approximation for $\mathfrak{s}_{\ell,jj'}^p, \mathfrak{t}_{\ell,jj'}^p, \mathfrak{l}_{\ell,jj'}^p$ using exactly the same procedure, replacing $w_{\ell,jj'}^p$ with $u_{\ell,jj'}^p$.

The two-dimensional (2D) integrals, used for the second-order terms
$\mathcal{S}^{-4}_{\ell,00}$, $\mathcal{T}^{-4}_{\ell,00}$, and $\mathcal{L}^{-4}_{\ell,00}$, are stored as
\begin{equation}
  \mathcal{J}_\ell(r, r')
  = \frac{9}{D(r)D(r')}
    \int_0^{r}\!\mathrm{d}\!\ln r_1
    \int_0^{r'}\!\mathrm{d}\!\ln r_2\;
    r_1 f_1(r_1)\,r_2 f_2(r_2)\;
    w^{(-4)}_{\ell,00}(r_1, r_2)\,.
  \label{eq:limber_prefix_2D}
\end{equation}
Applying the Limber limit collapses the double integral to a single sum via
$\delta_\mathrm{D}(r_1 - r_2)$:
\begin{equation}
  \mathcal{J}_\ell(r, r')
  \;\xrightarrow{\;\ell \to \infty\;}
  \frac{9}{D(r)D(r')}
  \Delta\!\ln r
  \sum_{r_a \leq \min(r,r')}
  f_1(r_a)\,f_2(r_a)\,
  \left(\frac{\nu}{r_a}\right)^{\!-4}
  \frac{P_\mathrm{m}(\nu/r_a)}{r_a}\,.
  \label{eq:limber_2D}
\end{equation}
The summand contains $P_\mathrm{m}/r_a$ rather than $P_\mathrm{m}/r_a^2$: the Limber delta function in the logarithmic measure gives $\delta_\mathrm{D}(r_1 - r_2) = \delta_\mathrm{D}(\ln r_1 - \ln r_2)/r_1$, adding $1/r_a$ to the $1/r_a^2$ from \refeq{limber_w} and bringing the denominator to $r_a^3$; the two kernel prefactors $r f_i$ from the logarithmic measure then supply $r_a^2$, yielding the net $1/r_a$.

The Limber approximations for $\mathcal X^{-4}_{\ell,00}$, $\mathcal Y^{-4}_{\ell,00}$, and $\mathcal Z^{-4}_{\ell,00}$ follow from Eq.~\eqref{eq:limber_2D}, where we use the corresponding pairs of radial kernel functions $f_1$ and $f_2$.

\subsection{Results}
\label{app:limber_results}

At each grid point $(r_i, r_j)$ we form the ratio $R_\ell \equiv \mathcal{I}_\ell^{\rm computed}/\mathcal{I}_\ell^{\rm Limber}$ between the numerically integrated value and the Limber prediction; for the 2D integrals we evaluate at the diagonal $r_i = r_j$.  The test covers all $24$ arrays at $j = j' = 0$: the $12$ one-dimensional $s$, $t$, $l$ integrals at $(p, n) = (-2, 0)$ and $(-4, 0)$ in both $;r$ and $;r'$ orientations; the $6$ one-dimensional $\mathfrak{s}, \mathfrak{t}, \mathfrak{l}$ integrals at $(p, n) = (-4, -1)$; and the $6$ two-dimensional $\mathcal{S}, \mathcal{T}, \mathcal{L}, \mathcal{X}, \mathcal{Y}, \mathcal{Z}$ integrals at $(p, n) = (-4, 0)$.  Derivative terms with $j + j' > 0$ are excluded for the reason discussed below \refeq{limber_wjj}.  Because the trapezoidal rule assigns the boundary point half-weight, the $;r$ integrals are evaluated at $i = j + 4$ and the $;r'$ integrals at $i = j - 4$, ensuring the Limber delta-function peak lies in the interior of the integration range.

\begin{figure*}
  \centering
  \includegraphics[width=\textwidth]{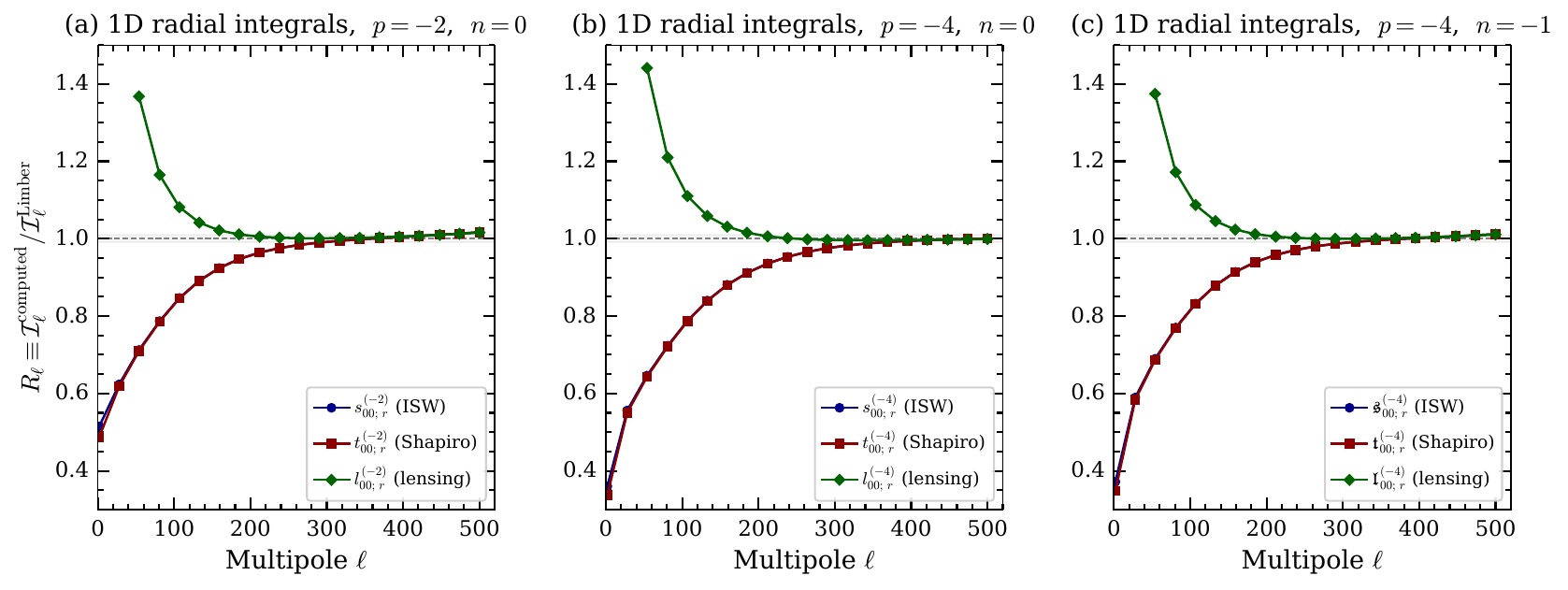}
  \caption{Limber-ratio test of the nine 1D radial integrals, evaluated
    at $r_j = 399.5\;\mathrm{Mpc}/h$ and $i = j + 4$.  The three panels
    split the arrays into $(p, n)$ groups: (a) $(p, n) = (-2, 0)$;
    (b) $(p, n) = (-4, 0)$; (c) $(p, n) = (-4, -1)$, corresponding to the
    $\mathfrak{s}, \mathfrak{t}, \mathfrak{l}$ arrays built on the
    $u$-base.  Each panel overlays the three cosmological kernels
    ($s$, $t$, $l$); within each group the three curves agree to
    $\sim 10^{-5}$ and sit on top of one another at the resolution of
    the figure.  The shaded band marks $|1 - R| < 1\%$.  All nine arrays
    converge monotonically toward $R = 1$ above $\ell \simeq 30$; at
    $\ell = 500$ the residuals separate into $R \simeq 1.016$ (a),
    $R \simeq 1.000$ (b), and $R \simeq 1.011$ (c).  The $;r'$ variants
    are not shown; they agree with $;r$ to three decimal places by the
    transpose construction.  The lensing kernel $f_\ell(r, r') \propto 1/r - 1/r'$
    vanishes on the diagonal $r = r'$, so the lensing ratio is degenerate at
    $\ell = 2$ (both the stored integral and its Limber image vanish in the
    asymptotic limit), leaving the green curves outside the plot frame at
    $\ell \lesssim 30$.  From $\ell \simeq 50$ onward the $l$ ratio rejoins the
    $s$, $t$ group ratio.}
  \label{fig:limber_1D}
\end{figure*}

\reffig{limber_1D} shows $R_\ell$ as a function of $\ell$ at a fixed reference point $r_j = 399.5\;\mathrm{Mpc}/h$ ($z = 0.138$) for the nine 1D $;r$ integrals.  At low $\ell$ the ratio departs from unity because the Limber approximation itself breaks down; for the non-lensing $s$ and $t$ kernels $R_{\ell=2}$ ranges from $0.34$ to $0.52$ across the six arrays, and the lensing-observable combination $l$ is degenerate at $\ell = 2$ because both the stored bracket and its Limber image are vanishingly small.  Above $\ell \simeq 30$ all arrays converge monotonically toward $R = 1$, and at $\ell = 500$ the residuals separate into three groups: the $(p, n) = (-2, 0)$ kernels sit at $R \simeq 1.016$, the $(p, n) = (-4, 0)$ kernels at $R \simeq 1.000$, and the $\mathfrak{s}, \mathfrak{t}, \mathfrak{l}$ kernels at $(p, n) = (-4, -1)$ at $R \simeq 1.011$.  Within each group the three cosmological kernels ($s$, $t$, $l$) agree to $10^{-3}$ for $\ell \gtrsim 50$, confirming that the residual is controlled by the Hankel-transform structure of the base function rather than by the kernel choice; the $;r'$ variants agree with the $;r$ results to three decimal places by the transpose construction.

\begin{figure}
  \centering
  \includegraphics[width=0.7\textwidth]{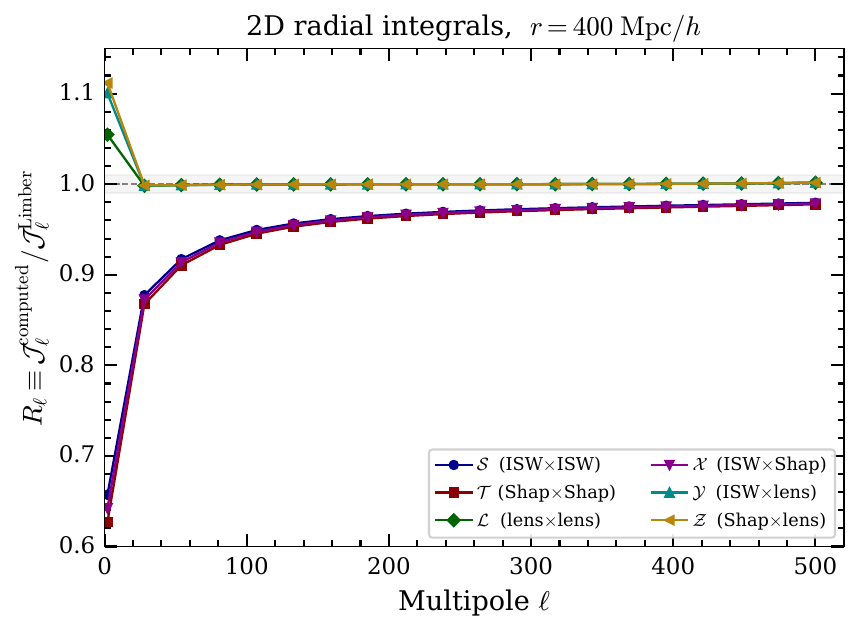}
  \caption{Limber-ratio test of the six 2D radial integrals at the
    diagonal $\mathcal{J}_\ell(r, r)$ with $r = 400\;\mathrm{Mpc}/h$.
    The three non-lensing auto- and cross-correlations
    ($\mathcal{S}, \mathcal{T}, \mathcal{X}$) sit at $R \simeq 0.978$ at
    $\ell = 500$, a $\simeq 2\%$ undershoot of the same magnitude and
    sign as the $(-2, 0)$ 1D group of \reffig{limber_1D}.  The three
    lensing diagonals ($\mathcal{L}, \mathcal{Y}, \mathcal{Z}$) sit at
    $R = 1.0017$, a $\simeq 0.2\%$ overshoot two orders of magnitude
    smaller than the non-lensing residual: the lensing kernel
    $f_\ell(r, r') \propto 1/r - 1/r'$ vanishes at the diagonal endpoint
    and suppresses the leading Limber correction
    \citep{LoVerde::2008limber}.}
  \label{fig:limber_2D}
\end{figure}

For the 2D integrals we test the diagonal elements $\mathcal{J}_\ell(r, r)$ for all six kernel combinations: the three auto-correlations $\mathcal{S}$ (ISW$\times$ISW), $\mathcal{T}$ (Shapiro$\times$Shapiro), $\mathcal{L}$ (lensing$\times$lensing), and the three cross-correlations $\mathcal{X}$ (ISW$\times$Shapiro), $\mathcal{Y}$ (ISW$\times$lensing), $\mathcal{Z}$ (Shapiro$\times$lensing).  \reffig{limber_2D} shows the $\ell$-convergence: all six arrays converge monotonically toward $R = 1$ above $\ell \simeq 30$, and split into two sub-groups at $\ell = 500$.  The three non-lensing diagonals $\mathcal{S}, \mathcal{T}, \mathcal{X}$ sit at $R = 0.978$--$0.979$, a $\simeq 2\%$ undershoot of the same magnitude and sign as the $(-2, 0)$ 1D group.  The three lensing diagonals $\mathcal{L}, \mathcal{Y}, \mathcal{Z}$ sit at $R = 1.0017$, a $\simeq 0.2\%$ overshoot, two orders of magnitude smaller than the non-lensing residual.  The lensing kernel $f_\ell(r, r') \propto 1/r - 1/r'$ vanishes at the diagonal endpoint $r_1 = r_2 = r$, and this suppresses the leading sub-leading Limber correction in the lensing diagonals, recovering the sub-percent convergence of the lensing observable known from the WL literature \citep{LoVerde::2008limber}.

\begin{figure*}
  \centering
  \includegraphics[width=\textwidth]{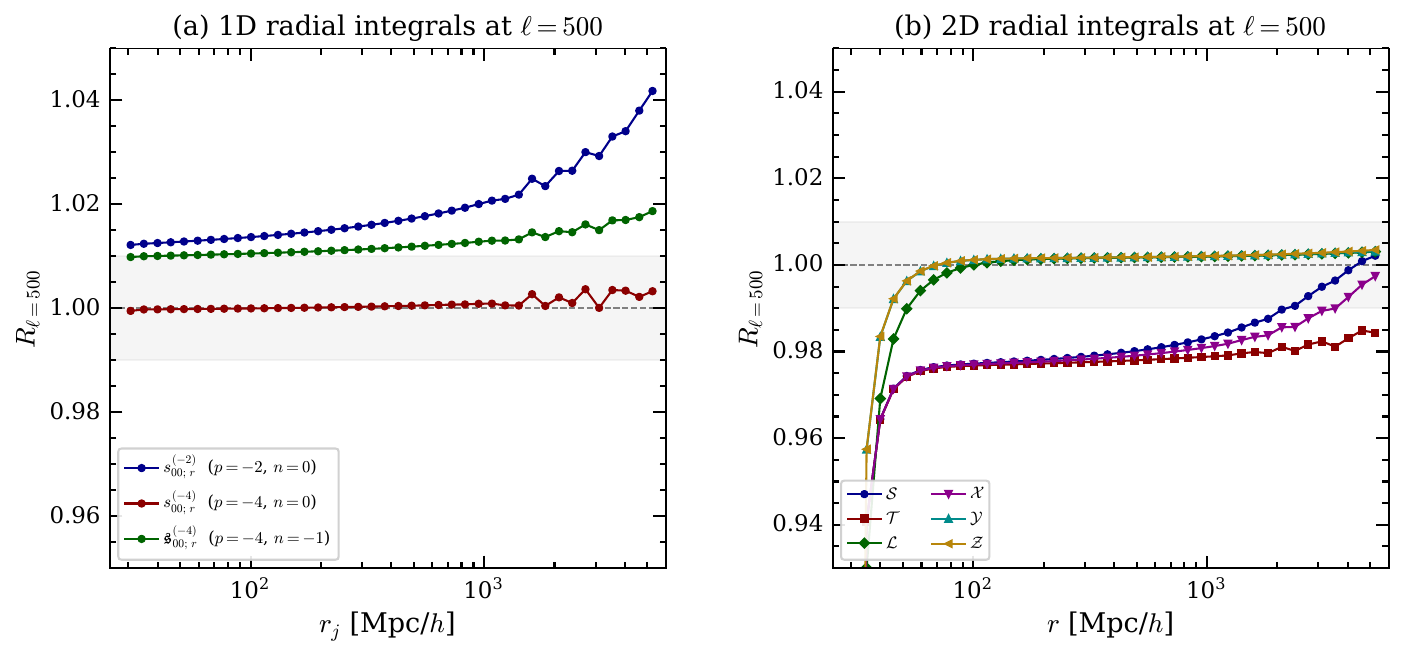}
  \caption{$r$-dependence of $R_{\ell=500}$ across the full radial grid.
    \textbf{(a)}~Three 1D arrays (one per $(p, n)$ group).  The $(-2, 0)$
    overshoot grows monotonically with $r$, consistent with NLO scaling
    $(r/\nu)^2$ at positive sign; the $(-4, 0)$ group sits closest to
    unity with small-amplitude oscillations about $R = 1$ and a single
    crossing near $r \simeq 120\;\mathrm{Mpc}/h$, reflecting a partial
    cancellation in the leading sub-leading coefficient.
    \textbf{(b)}~Six 2D arrays.  The non-lensing diagonals
    $\mathcal{S}, \mathcal{T}, \mathcal{X}$ undershoot at small $r$ and
    approach $R = 1$ from below as $r$ grows; the lensing diagonals
    $\mathcal{L}, \mathcal{Y}, \mathcal{Z}$ pass through $R = 1$ near
    $r \simeq 70$--$100\;\mathrm{Mpc}/h$ and overshoot only mildly at
    large $r$, reaching $R \simeq 1.003$ at $r \simeq 5000\;\mathrm{Mpc}/h$.
    The vanishing of the lensing kernel on the diagonal is what separates
    the two 2D sub-groups.}
  \label{fig:limber_rdep}
\end{figure*}

The $\ell$-convergence test fixes $r_j$ at a single reference point.  \reffig{limber_rdep} shows $R_{\ell=500}$ as a function of $r_j$ over the full radial range $30$--$5300\;\mathrm{Mpc}/h$.  The $r$-dependence differs qualitatively between the 1D and 2D arrays.  For the $(p, n) = (-2, 0)$ 1D integrals, $R$ rises monotonically from $\simeq 1.012$ at $r \simeq 30\;\mathrm{Mpc}/h$ to $\simeq 1.042$ at $r \simeq 5000\;\mathrm{Mpc}/h$: the overshoot above unity \emph{grows} with $r$, consistent with the next-to-leading-order correction scaling as $(r/\nu)^2$ at fixed $\ell$ \citep{LoVerde::2008limber} with the sign of this group.  For the non-lensing 2D diagonals $\mathcal{S}, \mathcal{T}, \mathcal{X}$, $R$ rises from $\simeq 0.94$ at $r \simeq 35\;\mathrm{Mpc}/h$ to $\simeq 1.00$ at $r \simeq 5000\;\mathrm{Mpc}/h$: the undershoot below unity \emph{shrinks} with $r$, the opposite $r$-trend to the 1D case.  The lensing 2D diagonals $\mathcal{L}, \mathcal{Y}, \mathcal{Z}$ sit substantially below unity at small $r$ (down to $R \simeq 0.56$ for $\mathcal{L}$ at $r \simeq 30$) and pass through unity near $r \simeq 70$--$100\;\mathrm{Mpc}/h$, settling to $R \simeq 1.003$ at $r \simeq 5000\;\mathrm{Mpc}/h$.  The 2D sub-leading correction therefore carries a different $r$-dependence than a single $(r/\nu)^2$ term would predict, reflecting the distinct delta-function-collapse structure of the 2D diagonal integrand in \refeq{limber_2D}.  The $(p, n) = (-4, 0)$ 1D group remains closest to unity across all $r$, oscillating between $R \simeq 0.9995$ and $\simeq 1.004$ with a zero crossing near $r \simeq 120\;\mathrm{Mpc}/h$; its small residual magnitude reflects a cancellation in the leading $(r/\nu)^2$ coefficient for this particular combination of $p$ and $n$, not a faster asymptotic convergence.

The residuals at $\ell = 500$ reflect sub-leading corrections to the Limber approximation, whose coefficients and $\ell$-scalings depend on the specific combination of shape derivatives that the $\delta$-function collapse picks out; different arrays carry different such combinations, and the numerical size of the residual varies by two orders of magnitude between groups.  Concretely, the 1D $(p, n) = (-4, 0)$ arrays reach $|1-R| \simeq 1 \times 10^{-4}$ at $\ell = 500$ as they pass through a zero crossing near $\ell \simeq 491$; the corresponding $(-2, 0)$ arrays cross near $\ell \simeq 357$ and overshoot to $R = 1.016$, a residual of $\simeq 1.6\%$ with the opposite sign from the low-$\ell$ deficit.  The $(p, n) = (-4, -1)$ arrays, built on the $u$-base with the $T(\nu/r)^{-1}$ weighting, sit at $|1-R| \simeq 1.1 \times 10^{-2}$ after a zero crossing near $\ell \simeq 382$; the transfer-function factor shifts the NLO coefficient relative to the pure $w$-base.  The three non-lensing 2D diagonals $\mathcal{S}, \mathcal{T}, \mathcal{X}$ sit at $|1-R| \simeq 2 \times 10^{-2}$ with $R < 1$, the same order of magnitude as the $(-2, 0)$ 1D residual.  The three lensing 2D diagonals $\mathcal{L}, \mathcal{Y}, \mathcal{Z}$ sit at $|1-R| \simeq 2 \times 10^{-3}$ with $R > 1$, two orders of magnitude smaller than the non-lensing residual: the lensing kernel $f_\ell(r, r') \propto 1/r - 1/r'$ vanishes on the diagonal and suppresses the leading sub-leading correction, reproducing the sub-percent convergence of the lensing observable known from the WL literature \citep{LoVerde::2008limber}.

Any attempt to characterize the approach to unity through an empirical power-law $|1-R| \propto \ell^{-\alpha}$ at $\ell \leq 500$ is complicated by the zero crossings noted above: the fitted slope depends strongly on whether the range straddles the crossing or lies entirely below or above it, and a single $\alpha$ therefore cannot summarize the whole sub-leading structure.  The clean data-level statement is that the stored integrals approach their Limber limits in sign and in absolute magnitude at every tested $\ell \geq 30$, with the numerical residuals listed above; the asymptotic $\nu^{-2}$ decay of Ref.~\cite{LoVerde::2008limber} has not yet set in for any of the 1D $(-2, 0)$ or non-lensing 2D arrays at the multipoles we compute.

In total, the test covers all $24$ stored arrays at $j = j' = 0$ across three cosmological kernels, three $(p, n)$ combinations, both integration directions, and both auto- and cross-correlations.  The stored integrals recover the Limber limit in sign and at the expected sub-leading magnitude at every tested $\ell \geq 30$.  The quantitative upper bound on the numerical accuracy of the stored integrals set by this test is therefore the Limber residual itself, $\sim 10^{-2}$ for the non-lensing arrays and $\sim 10^{-3}$ for the lensing diagonals: an asymptotic-limit comparison cannot discriminate pipeline errors below this floor, and a stronger check is required to probe the stored integrals beneath it.  We provide such a check against brute-force numerical integration in \refapp{pkfull}.

\subsection{Why Limber is inadequate at the percent level}
\label{app:limber_limits}

The residuals reported above are the asymptotic behavior of the stored integrals as $\ell \to \infty$; they are not numerical errors in \textsc{PowerFull}.  Two distinct conclusions follow, both with practical consequences for anyone tempted to substitute the Limber approximation for the full calculation.  The test does not set a precision bound on the stored integrals, because at the Limber residual floor the comparison is dominated by the approximation itself rather than by the numerical pipeline.  The Limber approximation also cannot serve as a sub-percent replacement for \textsc{PowerFull}: the same residuals that prevent the test from probing better than $\sim 10^{-2}$ are the errors one inherits by using the Limber formula in place of the full integral at the multipoles we compute.  Four structural features of the Limber limit make both conclusions unavoidable at $\ell \leq 500$.

\textbf{(i)~Asymptotic regime not reached.}  The leading-order Limber formula \refeq{limber_w} is the first term of a series in $\nu^{-1}$ whose next-to-leading correction for a smooth 1D integrand scales as $O(\nu^{-2})$ \citep{LoVerde::2008limber}.  For the $(p, n) = (-2, 0)$ 1D arrays and for the three non-lensing 2D diagonals, the residual at $\ell = 500$ sits at the $1$--$2\%$ level, and its empirical $\ell$-dependence is not yet a power law in $\nu^{-2}$: the $(-2, 0)$ residual crosses zero near $\ell \simeq 357$ and overshoots above unity; the non-lensing 2D residual is monotonic in $\ell$ but with an approach far slower than $\ell^{-2}$.  The numerical coefficient of the sub-leading term is therefore not small enough to be neglected, and the asymptotic decay that would make the remaining error shrink quickly with $\ell$ has not set in for these arrays at the multipoles we compute.  The Limber approximation is, in this regime, not a controlled expansion but a finite-order truncation of an incompletely converged series.

\textbf{(ii)~Opposite signs between 1D and non-lensing 2D structures.}  The 1D $(p, n) = (-2, 0)$ residuals at $\ell = 500$ sit at $R = 1.016$ (above unity), while the three non-lensing 2D diagonals $\mathcal{S}, \mathcal{T}, \mathcal{X}$ sit at $R = 0.978$--$0.979$ (below unity) at the same multipole on the same $r$ grid.  The 2D correction arises when the $\delta$-function collapse places $r_1 = r_2$ at the upper endpoint of the support and both kernel axes contribute a cross-derivative.  Because the 1D and non-lensing 2D corrections enter the final $C_\ell^{\mathrm{GR}}$ with opposite signs but at the same order of magnitude, the partial cancellations that bring the Limber prediction close to the full result for one array structure drive it away for the other, and no single multiplicative rescaling acts across both sets.  A Limber-based $C_\ell^{\mathrm{GR}}$ therefore inherits a $\sim 2\%$ bias whose sign depends on which term dominates the signal at a given $\ell$, with no simple recipe to remove it.

\textbf{(iii)~Kernel-dependent coefficients.}  The sub-leading coefficient depends on the $(p, n)$ structure of the integrand: the $(-2, 0)$, $(-4, 0)$, and $(-4, -1)$ 1D groups sit at $|1 - R| \simeq 1.6 \times 10^{-2}$, $1 \times 10^{-4}$, and $1.1 \times 10^{-2}$ respectively at $\ell = 500$, a spread of two orders of magnitude on the same grid and same multipole.  The split between non-lensing 2D ($|1-R| \simeq 2 \times 10^{-2}$) and lensing 2D ($|1-R| \simeq 2 \times 10^{-3}$) adds a third decade.  This kernel-dependence extends to the $r$-direction: on the same arrays the residual at $\ell = 500$ varies by a factor of a few across the full radial range $30$--$5000\;\mathrm{Mpc}/h$.  The bias inherited by a Limber-based predictor is therefore an \emph{unknown function} of the specific kernel combination and the $(r, r')$ pair, not a single scalar that could be calibrated once and reused.

\textbf{(iv)~Derivative terms break the expansion.}  The base functions with $j + j' > 0$ carry distributional derivatives $\delta^{(j')}_\mathrm{D}(r - r')$ in their Limber limits \refeq{limber_wjj}, and each derivative degrades the $\nu$-convergence by one power: the sub-leading correction scales as $O(\nu^{-(2 - j - j')})$ rather than $O(\nu^{-2})$.  For $j + j' = 1$ the residual is formally $O(\nu^{-1})$ and numerically remains at the $\sim 15\%$ level at $\ell = 500$; for $j + j' = 2$ the correction is formally $O(1)$ and does not converge within the multipoles we compute.  More than half of the $22$ base functions stored by \textsc{TwoFAST} fall into the derivative-containing category, and their Limber limits are therefore quantitatively meaningless at $\ell \leq 500$.  Any $C_\ell^{\mathrm{GR}}$ term that involves a radial derivative of a kernel, including the full Shapiro contribution and part of the velocity-gravity cross terms, is outside the regime where the Limber approximation is applicable at the multipoles relevant for current and next-generation galaxy surveys.

The Limber approximation thus remains a useful order-of-magnitude sanity check and a confirmation that the pipeline has the correct $\ell \to \infty$ limit, but the relativistic full-sky two-point statistics studied here live precisely where it fails.  Their signal is concentrated at large angular scales, the lowest multipoles, where the sub-leading Limber corrections reach the $1$--$2\%$ level with kernel-dependent signs and no uniform $\ell$-scaling; a Limber-based $C_\ell^\mathrm{GR}$ therefore carries a percent-level, sign-indefinite bias that cannot be calibrated away, and the derivative-coupled terms lie outside the controlled regime altogether.  The full radial integration is not an optional refinement but a requirement for these observables at the multipoles relevant to Stage-IV galaxy surveys such as Euclid, LSST, and DESI.  For a sub-percent check on the physically meaningful contributions to $C_\ell^\mathrm{GR}$ we develop in \refapp{pkfull} a reference that shares every input and grid point with \textsc{PowerFull} and evaluates the integrand by direct numerical quadrature.

\bibliographystyle{JHEP}
\bibliography{tamred}

\end{document}